\documentclass[notitlepage,tightenlines,aps,pra,reprint,twocolumn,amsmath,amssymb,citeautoscript,nofootinbib,floatfix, longbibliography]{revtex4-1}
\usepackage{float}
\usepackage{graphicx}
\usepackage{dcolumn}
\usepackage{bm}
\usepackage{latexsym,epsfig}
\usepackage{graphicx}
\usepackage{verbatim}
\usepackage{comment}
\usepackage{amsmath} 
\usepackage{amssymb}
\usepackage{stmaryrd}
\usepackage{color}
\usepackage{soul}
\usepackage{epstopdf}
\usepackage{grffile}
\usepackage{mathtools}
\usepackage{physics}
\usepackage{hyperref}
\hypersetup{
    colorlinks=true,      
    urlcolor=blue,
    citecolor=blue,
    linkcolor=blue
}

\DeclareGraphicsExtensions{.eps}
\newcommand{\beq}{\begin{equation}}
\newcommand{\eeq}{\end{equation}}
\newcommand{\bea}{\begin{eqnarray}}
\newcommand{\eea}{\end{eqnarray}}
\newcommand{\ben}{\begin{eqnarray*}}
\newcommand{\een}{\end{eqnarray*}}
\newcommand{\bfig}{\begin{figure}}
\newcommand{\efig}{\end{figure}}

\begin{document}
\title{Interaction driven topological phase transitions of hardcore bosons on a two-leg ladder}

 \affiliation{Homi Bhabha National Institute, Training School Complex, Anushaktinagar, Mumbai 400094, India}

\author{Rajashri Parida$^{1,2}$, Ashirbad Padhan$^{1,2}$ and Tapan Mishra$^{1,2,*}$}
\affiliation{$^1$School of Physical Sciences, National Institute of Science Education and Research, Jatni 752050, India\\
$^2$ Homi Bhabha National Institute, Training School Complex, Anushaktinagar, Mumbai 400094, India}
 \email{mishratapan@gmail.com}
\date{\today}

\begin{abstract}
We investigate the topological properties of hardcore bosons possessing nearest-neighbor repulsive interactions on a two-leg ladder. We show that by allowing nearest neighbour dimerized interactions instead of hopping dimerization, the system exhibits topological phases and phase transitions under proper conditions. First, by assuming uniform hopping throughout the ladder, we show that when interaction along the legs are dimerized and the dimerization pattern is different in the legs, a trivial rung-Mott insulator to a topological bond order phase transition occurs as a function of the dimerization strength. However, for a fixed dimerization strength, the system exhibits a topological to trivial phase transition with increase in the rung hopping. A completely different scenario appears when the rung interaction is turned on. We obtain that for a ladder with uniform hopping, the repulsive interaction either turns the topological phase into a trivial rung-Mott insulator or a charge density wave phase. Such topological features are absent when the dimerization pattern in the nearest neighbour interaction is considered to be identical in both the legs of the ladder. We numerically obtain the ground state properties and also show the signatures of topological phase transitions through Thouless charge pumping.

\end{abstract}

\maketitle

\section{Introduction}\label{introduction}

The discovery of the quantum Hall effect has ignited profound interest in the exploration of topological phase transitions in condensed matter systems~\cite{Klitzing,qhe_thouless}. The topological phases possess dual significance, embodying both fundamental understanding~\cite{Hassanreview} and technological applications~\cite{vonKlitzing2017, senthil_review,Fidkowski,Oshikawa,Xiao-Gang,sptinteract,Pachos_2014}. Symmetry-protected topological (SPT) phases represent a crucial class of topological states, that relies on the protection of specific underlying symmetries for their stability and features a finite bulk excitation gap while exhibiting intricately gapless behavior in their edge/surface modes~\cite{Asboth2016_ssh,senthil_review}. Any transition to another gapped phase due to strong perturbation is accompanied by a closure of the excitation gap. The one-dimensional Su-Schrieffer-Heeger (SSH) model is the simplest to exhibit such a gapped SPT phase and a transition to another trivial gapped phase~\cite{ssh_model}. While in the single particle limit the topological feature is characterized by the zero energy edge states and quantized Zak phase~\cite{Zak1989}, in the many-body limit the topological ground state at half-filling exhibits finite edge polarization, non-local correlations, and bond ordering in the bulk. Due to the simplicity of the model, the SSH model and its variants have been widely discussed in the literature and associated topological properties have been experimentally demonstrated in various systems such as cold atoms in optical lattices, photonic lattices, superconducting circuits, ion traps, mechanical systems, acoustic systems and electrical circuits~\cite{Atala2013,Takahashi2016pumping,Lohse2016,Mukherjee2017,Lu2014,ssh_expt_2,Kitagawa2012,Leder2016, floquet_ssh_ladder}.

While describing topology in non-interacting systems using symmetries of single-particle bands and topological invariant is straightforward, this classification becomes challenging in an interacting system due to the complexities arising from the competition between the symmetries, strong correlations, particle statistics, and lattice topology~\cite{rachel_review}. Additionally, in various instances, interactions are believed to disrupt the non-local order of the SPT phases through spontaneous symmetry breaking.  Furthermore, there are situations where topology persists under weak to moderate interactions, but strong interactions result in the loss of topological characteristics. Due to this, the study of stability of topological phases and phase transitions in interacting systems have recently gained enormous interest. Although such studies shed light on the competing effects of interaction and the underlying topological nature of the system, in low-dimensional systems strong correlations play important roles. 
In this context, a variety of studies have been performed in the framework of the interacting SSH model to investigate the role of local and/or non-local interactions among the particles on the topological phases and phase transitions in bosonic, spin, and fermionic systems ~\cite{fleishauer_prl, mondal_pumping, DiLiberto2016,DiLiberto2017,sjia, wu_vincet_liu,Taddia2017,fraxanet,juliafarre, mondal_sshhubbard}. Moving away from the one-dimensional chain, the topological properties in interacting quasi-one-dimensional and two-dimensional SSH models have also been investigated~\cite{Nersesyan,ssh_ladder_glide_symmetry, padhan, suman_adhip, batrouni,wu_vincet_liu}. Such systems of two or more connected SSH chains reveal interesting scenarios as one goes from one to two dimensions through couplings. Recently, several such interaction effects have been observed in various quantum simulators by utilizing appropriate equilibrium and non-equilibrium properties of the topological phases~\cite{browyes,bloch_haldane, ssh_expt_1,ssh_expt_2, ssh4_expt, suotang_jia_expt}.

While a great deal of studies have been performed to understand the effect of interactions in models with hopping dimerization, some recent studies have unveiled the existence of a topological phase driven purely by interactions in systems where the non-interacting limit is a gapless liquid ~\cite{Mondal_topology, Hetenyi_tvvp}. In particular, a 
recent study demonstrates that by considering nearest neighbor (NN) interacting hardcore bosons or spinless fermions in a one-dimensional lattice with uniform hopping, a finite dimerization in the interaction (bond alternating NN interactions) drives the system from the gapless superfluid (SF) phase to gapped bond order (BO) phases.
Depending on the choice of the dimerization pattern in the NN interaction, the BO phase was found to be either topological or trivial in nature. In other words, if the dimerization pattern in the NN interaction is of weak-strong-weak-strong (strong-weak-strong-weak) type then a topological (trivial) BO phase emerges. This results in a well-defined topological phase transition between the two BO phases through a gap-closing point as a function of the strength of dimerization. Although such topological phase transition resembles the one due to the hopping dimerization in the SSH model, there is a clear difference between the two cases in the limit of strong NN interaction where the topological phase transition between the two BO phase occurs through a gapped charge-density-wave (CDW) phase. Such topological phase transitions do not have any non-interacting counterparts, unlike other models where the non-interacting system exhibits some sort of topological character. Now the question is whether such topological phases driven solely by interaction is stable if two one-dimensional chains are coupled to each other in the form of a two-leg ladder. The important difference between a two-leg ladder and a one-dimensional chain is that the single particle spectrum of the former can be gapped depending on the strength of the rung hopping whereas for the latter it is always gapless. The difference is even more significant in the non-interacting many-body limit when hardcore bosons are considered in place of spinless fermions. In the case of hardcore bosons, the system is known to be a gapped rung-Mott insulator (RMI) for any finite value of the rung hopping~\cite{rmi_laflorencie}. At this point, a natural question arises, whether such interaction driven topological phase transitions can be established in a system of hard-core bosons on a two-leg ladder or not. 

Motivated by these, in this paper, we investigate the ground state properties of NN interacting hardcore bosons on a two-leg ladder with uniform NN hopping along the legs. We show that if the dimerized NN interactions in the legs are chosen in such a way that one of the legs favors the topological BO phase whereas the other favors a trivial BO phase according to Ref.~\cite{Mondal_topology} (which we call the staggered dimerization), then a well defined topological phase transition can be established due to the competing effect of the strength of dimerization and rung couplings.

We obtain a phase transition from the trivial RMI phase to a topological BO phase with an increase in the dimerization strength if hopping is uniform throughout the ladder. However, for fixed dimerization strength, we find a transition from the topological BO phase to the RMI phase as a function of the rung hopping. In addition to the rung hopping, if rung repulsion is turned on, a topological phase transition is favorable for some specific range of values of dimerized NN interaction. Moreover, for weaker dimerized NN interaction the system remains in the trivial gapped phase, and for stronger dimerization, there exists a phase transition from the topological phase to a CDW phase. We first characterize these features from the ground state properties and then we show the signatures of the topological phases through Thouless charge pumping~\cite{monika_review,Thouless1983}. 
In the end, we compare our results by considering uniform dimerized interaction i.e. identical dimerization pattern in the NN interaction that favors topological phases in both the legs.

\section{Model and approach}\label{modelmethod}
\begin{figure}[t]
\centering
\includegraphics[width=0.9\linewidth]{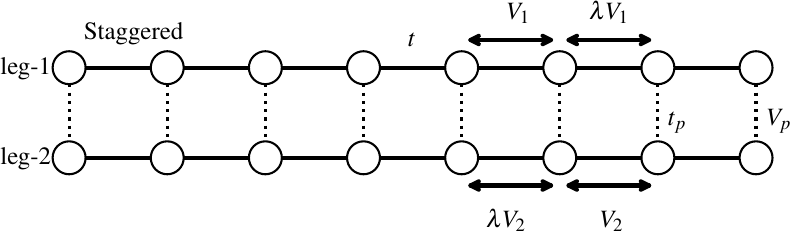}
\caption{Schematic diagram illustrating a two-leg ladder with dimerized NN interaction. The circles represent the sites and the hopping strengths along the legs and the rungs are denoted by $t$ (solid line) and $t_p$ (dotted line),  respectively. The NN interaction strengths are represented as $V_1$ and $V_2$ in leg-1 and leg-2, respectively, and we assume $V_1=V_2=1$. The bond-alternating dimerized NN interaction is achieved by introducing $\lambda V_1$ and $\lambda V_2$ with $\lambda\neq1$ in leg-1 and leg-2, respectively. $V_p$ denotes the NN interaction along the rungs. The choice of dimerization is called staggered if the particles in one of the legs are in the topological phase whereas they are in the trivial phase in the other leg.}
\label{fig:dimerizaed_ladder_staggered}
\end{figure}
The model describing the system of hardcore bosons on a two-leg ladder with staggered dimerization in the NN interaction is given by 
\begin{eqnarray}
 H&=-&t\sum\limits_{\alpha\in(1,2)}\sum\limits_{i}({a}^\dagger_{\alpha,i}{a}_{\alpha,i+1}+\text{H.c.}) \nonumber\\&+&\sum\limits_{\alpha\in(1,2)}\sum\limits_{i}V_{\alpha,i} \left(n_{\alpha,i}-\frac{1}{2}\right)\left(n_{\alpha,i+1}-\frac{1}{2}\right)\nonumber\\&-& t_p\sum\limits_{i}(a^\dagger_{1,i}a_{2,i}+\text{H.c.})\nonumber\\&+&V_p\sum\limits_{i}\left(n_{1,i}-\frac{1}{2}\right)\left(n_{2,i}-\frac{1}{2}\right)
 \label{eq:ham}
\end{eqnarray}
where $\alpha$ and $i$ represent the leg index and rung index of the ladder respectively. Here $a^\dagger_{1,i}$($a_{2,i}$) and $n_{1,i}$($n_{2,i}$) are the bosonic creation (annihilation) and onsite number operators at $i$th site of leg-1 (leg-2). The parameter $t$ signifies the hopping strength in both the legs, while $t_p$ represents the hopping strength between the legs or along the rungs of the ladder. $V_{\alpha,i}$ and $V_p$ are the NN interaction strengths in legs and rungs respectively. 

We define $V_{\alpha,i}$ as
\begin{equation}
    V_{\alpha,i} =  
    \begin{cases}
        V_\alpha,\alpha i\in\text{odd} \\
        \lambda V_\alpha,\alpha i\in\text{even},
    \end{cases}
\end{equation}
with $\lambda$ denoting the strength of dimerization such that a staggered pattern in dimerized NN interaction is achieved as depicted in Fig.~\ref{fig:dimerizaed_ladder_staggered}. In our studies, we assume $V_1=V_2=V=1$. The hardcore constraint is achieved by the condition ${a^\dagger}^2 = 0$ which ensures that not more than one boson can occupy a single site. Here $t = 1$ sets the energy unit which makes all the other parameters of the system dimensionless. 

Note that the model above can be equivalently mapped to a spin-$1/2$ ladder model with Heisenberg interactions along the legs ($J_1~\text{and}~J_2$) as well as the rungs ($J_p$) by associating the occupation state of hardcore bosons to $S_z$ states of spins by using the Holstein–Primakoff transformation~\cite{hp_transformation} defined as
\begin{equation}
    S_i^+=\sqrt{1-a_i^\dagger a_i}a_i,~S_i^-=a_i^\dagger\sqrt{1-a_i^\dagger a_i}~,S_i^z=\frac{1}{2}-a_i^\dagger a_i .
    \label{eq:hp_transform}
\end{equation}
Specifically, the bosonic occupation state $|1\rangle$ corresponds to the spin state $|\downarrow\rangle$, and $|0\rangle$ corresponds to $|\uparrow\rangle$.

The bosonic model above can be divided into three parts as follows.
\begin{equation}
    H = H_{\text{leg-1}}+H_{\text{leg-2}}+H_{\text{rung}}
    \label{eq:spin_hamiltonian}
\end{equation}
After using the transformation defined in Eq.~(\ref{eq:hp_transform}), the transformed Hamiltonian in spin language can now be written as
\begin{equation}
    H_{\text{leg}-\alpha}=J^{\perp}_{\alpha}\sum\limits_i(S_{\alpha,i}^xS_{\alpha,i+1}^x+S_{\alpha,i}^yS_{\alpha,i+1}^y)+\sum\limits_iJ_{\alpha,i}^{||} S_{\alpha,i}^zS_{\alpha,i+1}^z
\end{equation}
and
\begin{equation}
    H_{\text{rung}}=\sum\limits_i J_p^{\perp}(S_{1,i}^xS_{2,i}^x+S_{1,i}^yS_{2,i}^y)+\sum\limits_i J_p^{||} S_{1,i}^zS_{2,i}^z
\end{equation}
with $\alpha$ and $i$ indices representing the same as in the bosonic case, $J_\alpha^{\perp}=-2t, J_p^{\perp}=-2t_p, J_p^{||}=V_p$ and 
\begin{equation}
J_{\alpha,i}^{||} =  
    \begin{cases}
        V_\alpha,~\alpha i\in\text{odd} \\
        \lambda V_\alpha,~\alpha i\in\text{even}
    \end{cases}
\end{equation}
This mapping guarantees that the physics of the spin Hamiltonian corresponding to Eq.~(\ref{eq:spin_hamiltonian}) and the hard-core bosonic Hamiltonian defined in Eq.~(\ref{eq:ham}) share identical properties. In the present study, we consider the hard-core bosonic system and obtain the ground state properties.

In the absence of any rung couplings, the model shown in Eq.~(\ref{eq:ham}) decouples to two one-dimensional chains with dimerized NN interactions. As already mentioned before, physics of such dimerized chains have been discussed in Ref.~\cite{Mondal_topology} revealing the BO$_0$, BO$_\pi$ and the CDW phases. Note that in the spin-1/2 language, the individual legs can be described by the XXZ model with dimerized ZZ couplings. In this case one obtains similar physics compared to the hardcore boson model where the BO$_0$, BO$_\pi$ and the CDW phases can be identified as the trivial Ising paramagnet (IPM$_0$), topological Ising paramagnet (IPM$_\pi$) and the Ising N\'eel (IN) phases respectively~\cite{harsh_xxz}. In this study, however, we explore the scenario that emerges when the rung couplings (hopping and interaction) are turned on. For this purpose, we study the ground state properties of the system by utilizing the density matrix renormalization group (DMRG) method~\cite{white1992,schollowck_dmrg_rev} based on the matrix product states (MPS) ansatz~\cite{Verstraete_rev,schollowck_mps} with systems under open boundary condition. We consider the ladder of length up to $L=140$ rungs (which is equivalent to $280$ sites) at various fillings of hard-core bosons, i.e., $\rho = N/2L$, where $N$ denotes the number of bosons. Unless specified otherwise, we extrapolate all the quantities to the thermodynamic limit ($L\to\infty$) to avoid finite-size effects. We consider a maximum bond dimension up to $500$ to minimize truncation errors.

\section{Results}
In this section, we present our main findings in detail by considering dimerized NN interaction along the legs of the ladder. The dimerization is chosen in such a way that in the decoupled leg limit (i.e. $t_p=0$ and $V_p=0$), one of the legs of the ladder exhibits topological phase whereas the other remains in the trivial phase (see Ref.~\cite{Mondal_topology} for details). We call this configuration the staggered dimerization in the NN interaction. As already discussed in the previous section, the ground state of the system described by the model in Eq.~\ref{eq:ham} without the dimerized NN interaction is a gapped RMI phase~\cite{rmi_laflorencie}. In the following, we will discuss the rich interplay between the dimerized NN interaction along the legs, the rung hopping and the rung interaction. First, we present the ground state properties by varying the dimerization strength and keeping uniform hopping strengths throughout the ladder i.e. $t=t_p$. Then we discuss the role of rung hopping and rung interaction on the topological properties of the system for fixed dimerization strengths. In the end, we briefly discuss the case when uniform dimerization pattern in the NN interaction is assumed and show that the topological nature is not favorable.

\begin{figure}[t]
    \centering
    \includegraphics[width=0.8\linewidth]{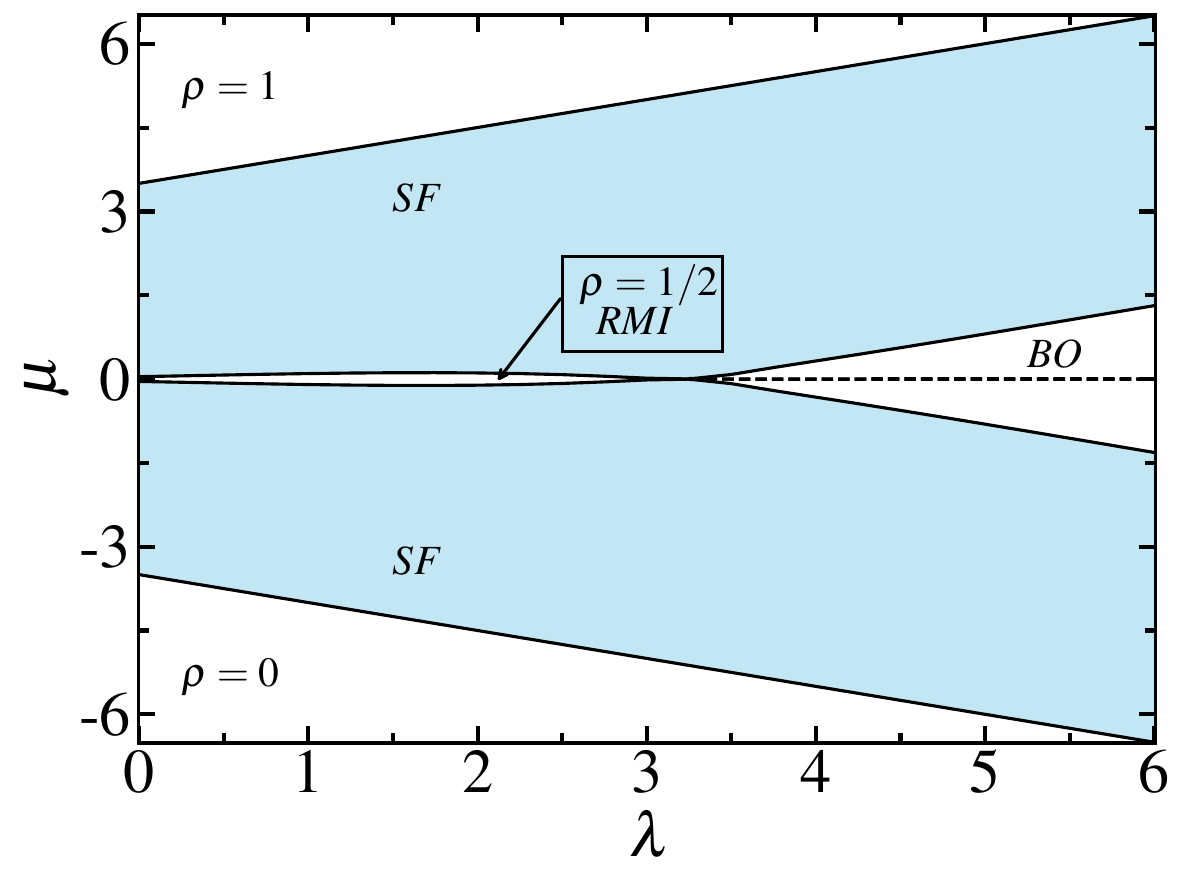}
    \caption{The phase diagram of the model shown in Eq.~\ref{eq:ham} in the $\lambda-\mu$ plane for the staggered dimerization case for fixed rung hopping $t_p=1$. The NN interactions along the legs are chosen as $V_1=V_2=1$. The shaded regions represent the gapless superfluid (SF) phase. The solid lines mark the boundary of the gapped phases at half filling (the white regions in the middle) and the dashed lines mark the pair of edge modes. The white regions for $\rho=0$ and $\rho=1$ denote the empty and full states, respectively.}
    \label{fig:effect_of_delta_staggered}
\end{figure}
\begin{figure}[b]
    \centering
    \includegraphics[width=0.8\linewidth]{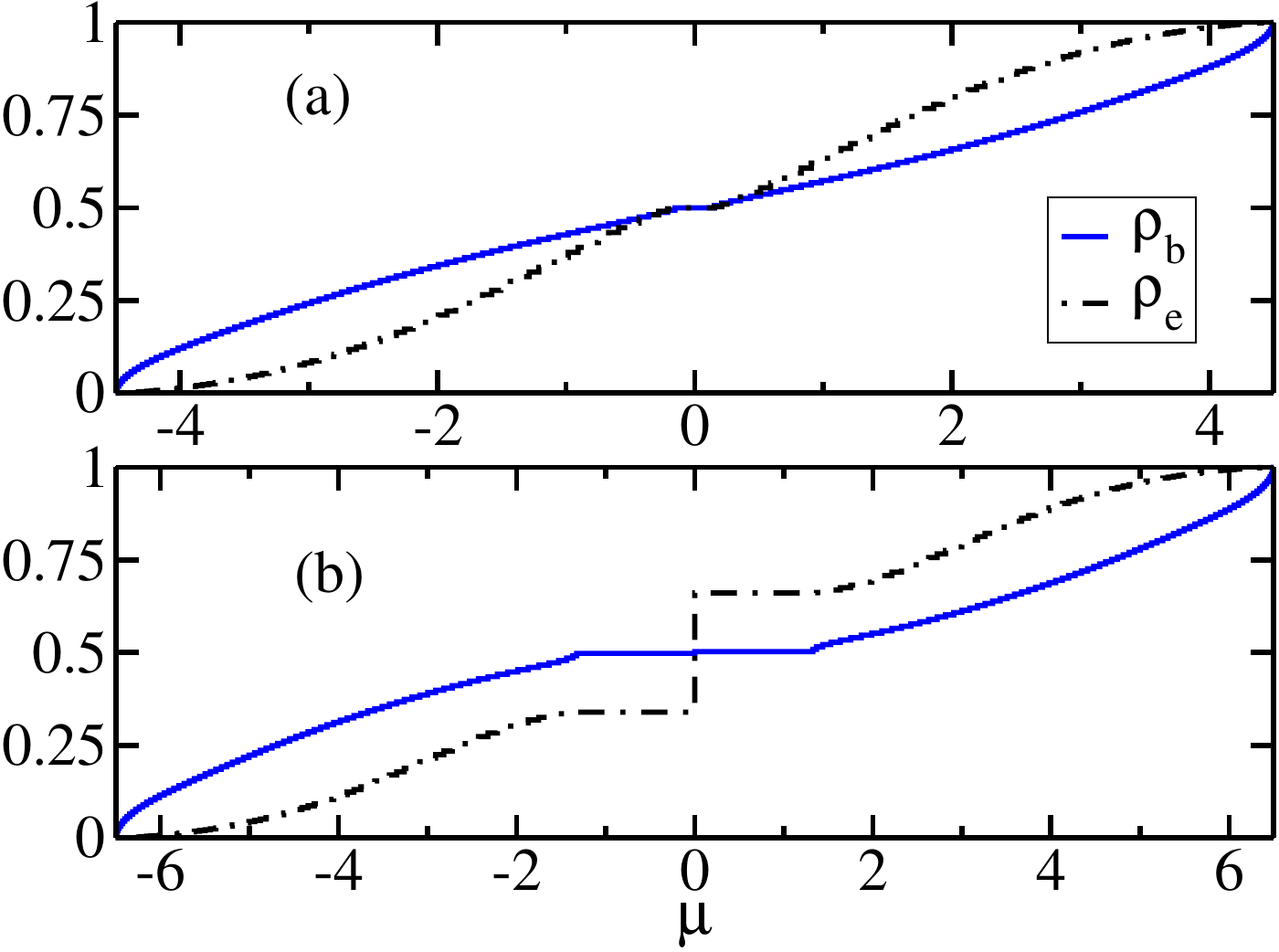}
    \caption{The figure depicts the bulk and the edge densities represented by $\rho_b$ (blue solid lines) and $\rho_e$ (dashed-dotted black lines) respectively as a function of $\mu$ for two different cuts through the phase diagram shown in Fig.~\ref{fig:effect_of_delta_staggered} at (a) $\lambda=2$ and (b) $\lambda=6$. This shows the nature of bulk phases (gapped or gapless) when edge states are being filled in different parameter regimes.}
    \label{fig:rho_mu_staggered_for_different_delta}
\end{figure}
\begin{figure}[t]
    \centering
    \includegraphics[width=0.8\linewidth]{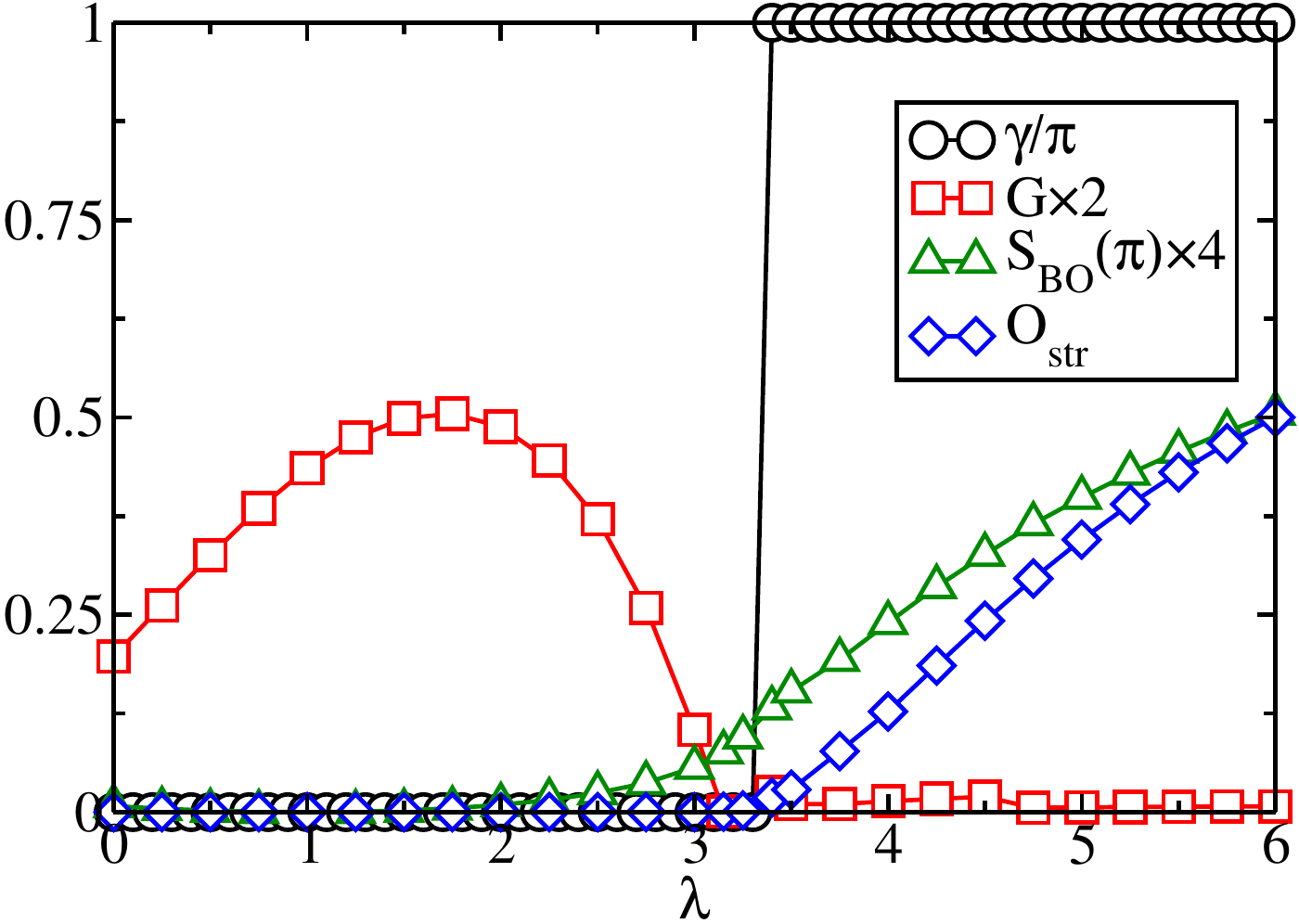}
    \caption{Shows the winding number ($\gamma/\pi$) (black circles) for $L=8$ rungs, extrapolated values of the single particle excitation gap $G$ (red squares), extrapolated values of the bond order structure factor ($S_{BO}(\pi)$) (green triangles) and the string order parameter $O_{str}$ (blue diamonds) for $L=140$ rungs as functions of $\lambda$ for $V_1=V_2=1$ and $t_p=1$. Here $G\times 2$ and $S_{BO}(\pi)\times 4$ have been plotted for better visibility.}
    \label{fig:wind_num_with_dimerization}
\end{figure}

\subsection{$t_p\neq 0$ and $V_p=0$}

\subsubsection{Ground state phase diagram}
In this subsection, we consider the system without the rung interaction i.e. $V_p=0$. We first obtain the ground state phase diagram of the model with uniform hopping strength throughout the ladder i.e. $t=t_p=1$ and by varying the dimerization strength $\lambda$. The phase diagram in this case is shown in $\lambda-\mu$ plane in Fig.~\ref{fig:effect_of_delta_staggered}. The white regions in the middle surrounded by black solid boundaries are the gapped phases at half filling ($\rho=1/2$) and the blue shaded regions on either side of the gapped phases are the gapless regions at densities away from half filling. The phase diagram depicts a gapped-gapped phase transition as a function of $\lambda$ through a gap-closing critical point at $\lambda \sim 3.3$. However, we obtain that the gapped phase for higher values of $\lambda$ exhibits zero energy edge states. 
The boundaries of the gapped phases are obtained from the chemical potentials defined as  
\begin{equation}
\label{eq:mus}
 \mu^+=(E_{N+1}-E_N) ~\text{and} ~\mu^-=(E_N-E_{N-1})
\end{equation} 
for different values of $\lambda$ and extrapolating them to thermodynamic limit ($L\to\infty$). Here, $E_N$ is the ground state energy of the system consisting of $N$ bosons.  The difference between the chemical potentials, i.e., $G=\mu^+-\mu^-$ denotes the gap in the system. While this difference is finite up to $\lambda\sim3.3$, it vanishes for $\lambda \gtrsim 3.3$ which can be seen as the dashed line in Fig.~\ref{fig:effect_of_delta_staggered}. This is an indication of the presence of zero energy edge states in the system. In this region, however, there exists a finite gap for two-particle excitation and accordingly, the gap is finite after $\lambda \sim 3.3$. The appearance of bulk gap and edge states can be clearly understood by analyzing the behavior of the bulk and edge densities defined as 
\begin{equation}
 \rho_b=\frac{1}{2L-4}\sum\limits_{i=2}^{L-1}[\langle n_{1,i}\rangle+\langle n_{2,i}\rangle]
\end{equation}and
\begin{equation}
\rho_e= \frac{1}{4}\big[\langle n_{1,1}\rangle + \langle n_{2,1}\rangle + \langle n_{1,L}\rangle+\langle n_{2,L}\rangle\big]
\end{equation}
respectively. In Fig.~\ref{fig:rho_mu_staggered_for_different_delta} we plot $\rho_b$ (blue solid curve) and $\rho_e$ (black dot-dashed curve) with respect to $\mu$  for  two different values of dimerization such as $\lambda=2$ (Fig.~\ref{fig:rho_mu_staggered_for_different_delta}(a)) and $\lambda=6$ (Fig.~\ref{fig:rho_mu_staggered_for_different_delta}(b)) which cut through the two gapped phases in Fig.~\ref{fig:effect_of_delta_staggered}. The gapped regions are clearly identified as the plateaus in $\rho_b$ near $\rho_b=0.5$ in both the figures. However, in Fig.~\ref{fig:rho_mu_staggered_for_different_delta}(b), we see a sharp jump in $\rho_e$ (dot-dashed curve) at $\mu=0$, indicating the presence of the edge states. The shoulder around the plateaus in $\rho_b$ are due to the gapless phases corresponding to the blue shaded regions in Fig.~\ref{fig:effect_of_delta_staggered} which are found to be the SF phases. We obtain that the gapped region before (after) the gap closing point is an RMI (BO) phase and the BO phase is topological in nature. We characterize these phases in detail in the following.

Before quantifying the gapped phase arising for $\lambda\lesssim 3.3$ we quantify the BO phase which is characterized by a finite oscillation in the bond kinetic energy $\langle B_i\rangle$ where $B_i=(a_i^\dagger a_{i+1} + \text{H.c.})$, resulting in a finite peak in the bond order structure factor defined as 
\begin{equation}
    S_{BO}(k)=\frac{1}{L^2}\sum_{i,j}e^{ik|i-j|}\langle B_iB_j \rangle
\end{equation}
on leg-1. We plot the finite size extrapolated values of $S_{BO}(\pi)$ (green triangles) in Fig.~\ref{fig:wind_num_with_dimerization} as a function of $\lambda$ which becomes finite for $\lambda\gtrsim 3.3$ confirming the transition to the BO phase. We also plot the single particle excitation gap $G$ (red squares) as a function of $\lambda$ along with $S_{BO}(\pi)$. The gap $G$ in the thermodynamic limit remains finite in the beginning but tends to vanish after the critical $\lambda \sim 3.3$ due to the presence of the zero energy edge states (see also  Fig.~\ref{fig:rho_mu_staggered_for_different_delta}(b)).

To concretely quantify the topological nature of the BO phase we compute the Berry phase ($\gamma$) which gives us the topological invariant of the system. The topological invariant in such an interacting system is calculated using the twisted boundary condition, i.e. by setting the hopping strength at the boundary in both the legs as $t\xrightarrow{}te^{i\theta}$ where, $\theta$ is the twist angle. When $\theta$ is varied from $0$ to $2\pi$ adiabatically, the ground state $|\psi(\theta)\rangle$ picks up a phase called the Berry phase and is calculated using the formula 
\begin{equation}
\gamma = \int_0^{2\pi}\langle\psi(\theta)|\partial_\theta|\psi(\theta)\rangle d\theta.
\label{eq:berry_phase}
\end{equation}
\begin{figure}[t]
    \centering
    \includegraphics[width=0.8\linewidth]{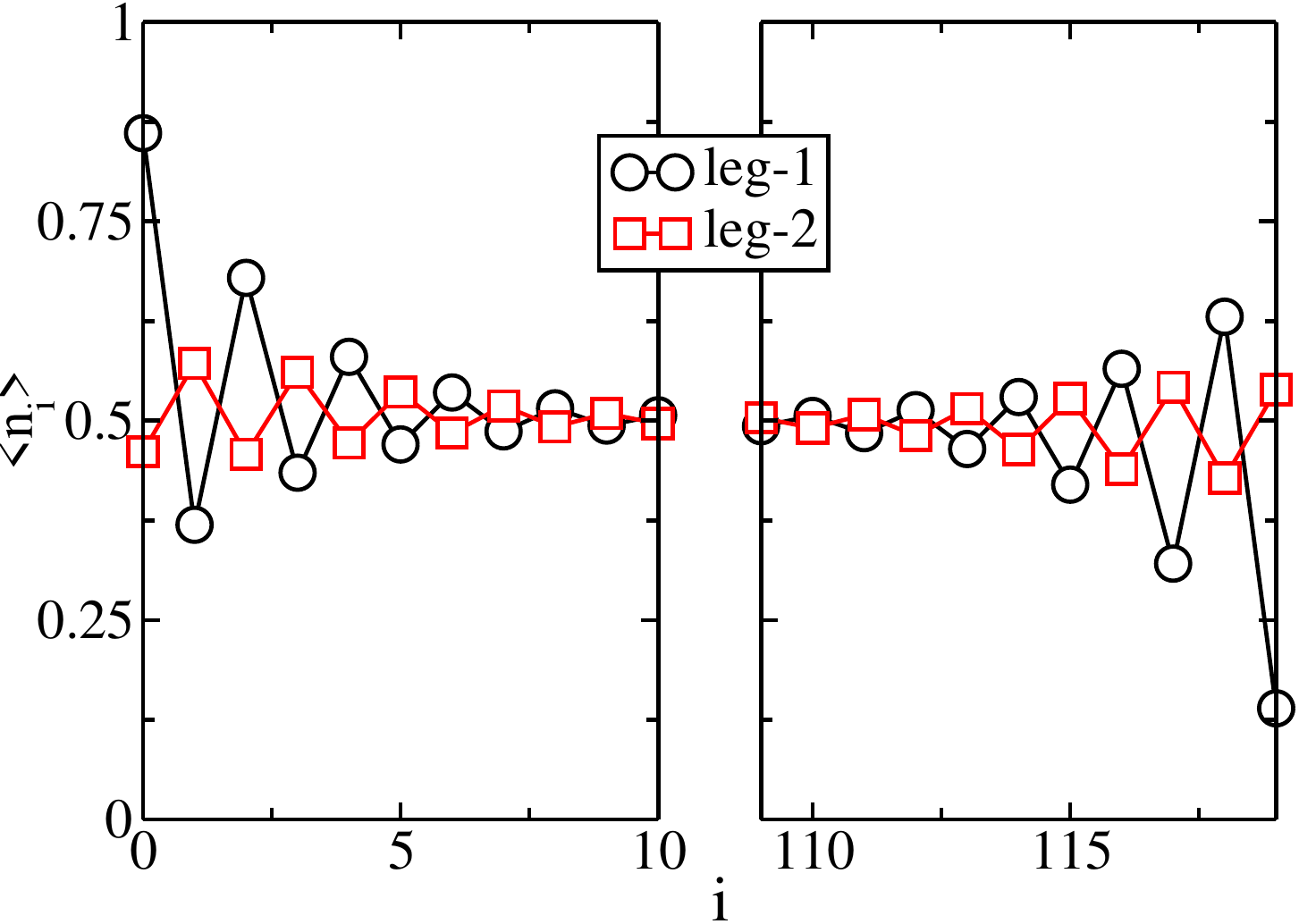}
    \caption{Onsite particle densities ($\langle n_i\rangle$) are plotted as a function of the site index $i$ along leg-1 (black line with circles) and leg-2 (red line with squares) for $\lambda=6$ with $V_1=V_2=1$ and $L=120$ rungs.}
    \label{fig:desnity_vs_site_index}
\end{figure}
If the bulk is topologically non-trivial, the acquired phase is $\pi$, while it is $0$ for a trivial bulk. In Fig.~\ref{fig:wind_num_with_dimerization} we plot $\gamma/\pi$ (black circles) as a function of $\lambda$ which represents the winding number. Here, $\gamma$ is computed using the exact diagonalization technique with a small system size of $L=8$ rungs ($16$ sites). A sharp change of $\gamma/\pi=0$ to $\gamma/\pi=1$  is the signature of the topological phase transition at the critical $\lambda\sim3.3$. 

Furthermore, topological phases exhibit non-local string correlations~\cite{den_nijs,Tasaki,Hida,string_ladder}, quantified by the string order parameter ($O_{\text{str}}$) defined as
\begin{equation}
    O_{\text{str}}(r) = -\langle Z_{\text{rung}}(i)e^{i\frac{\pi}{2}\sum_{j=i+1}^{k-1}Z_{\text{rung}}(j)}Z_{\text{rung}}(k)\rangle,
\label{eq:string_order_parameter}
\end{equation}
where $Z_{\text{rung}}(i)=Z_1(i)+Z_2(i)$ and $Z=1-2a^\dagger a$. $i$ denotes the $i^{th}$ rung of the ladder and $r=|i-j|$. To avoid edge rungs, we have chosen $i=2$ and $k=L-1$. We plot $O_\text{str}$ (blue diamonds) as a function of $\lambda$ in Fig.~\ref{fig:wind_num_with_dimerization} for a system consisting of $L=140$ rungs. Comparing the behavior of $\gamma$, $G$, $S_{BO}(\pi)$ and $O_\text{str}$, one can conclude that a topological phase transition occurs as a function of $\lambda$ at $\lambda\sim 3.3$ in this case.
\begin{figure}[t]
    \centering
    \includegraphics[width=0.8\linewidth]{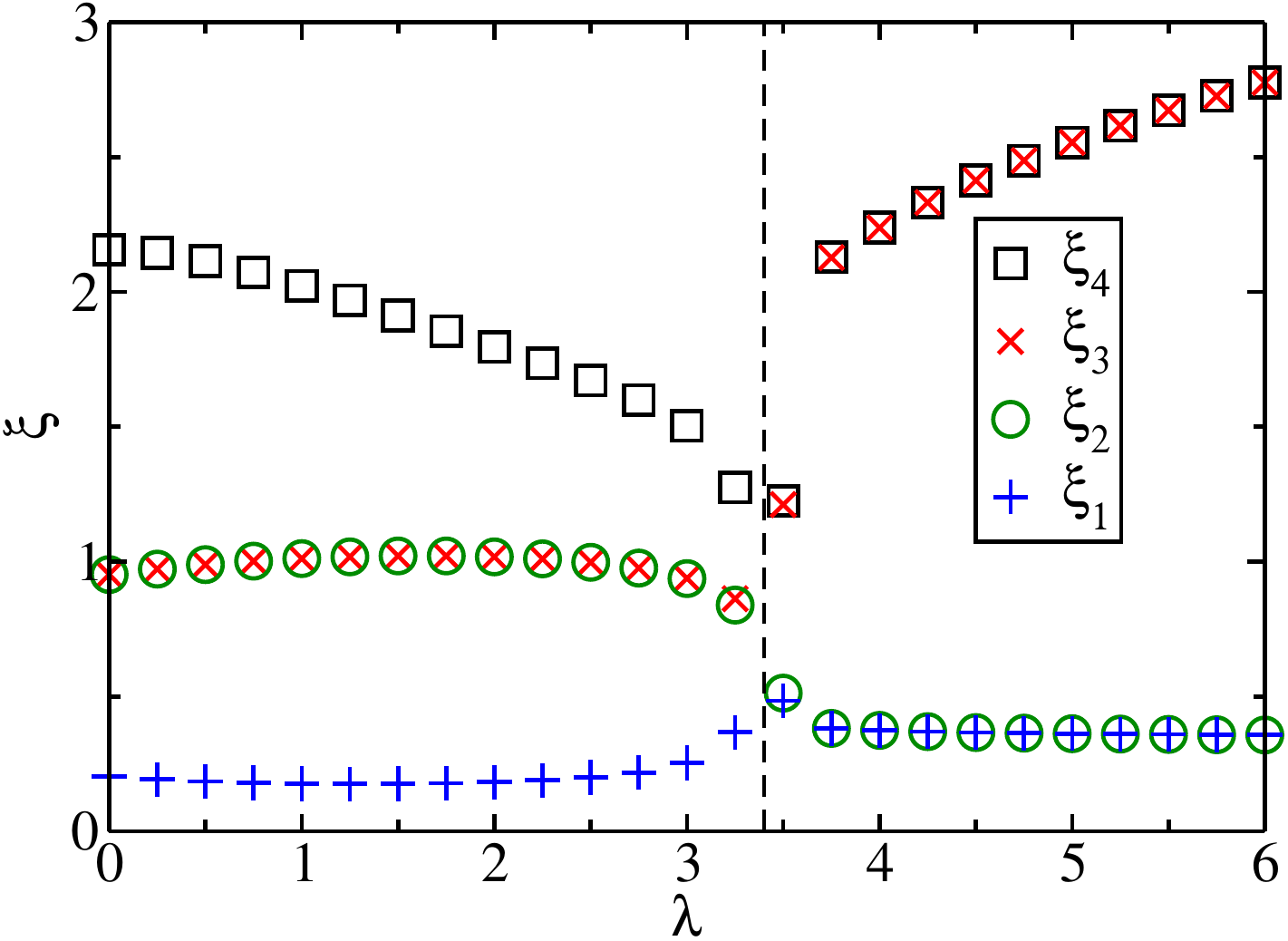}
    \caption{Shows the lower four values of the entanglement spectrum ($\xi_1,~\xi_2,~\xi_3,~\xi_4$) for a system of $L=140$ rungs ($280$ sites) as a function of $\lambda$ for $V_1=1$, $V_2=1$ and $t_p=1$. The dashed vertical line marks the critical point of transition from the RMI to the topological BO phase.}
    \label{fig:ent_spectrum_tp_1_vp_0}
\end{figure}
\begin{figure}[b]
    \centering
    \includegraphics[width=0.8\linewidth]{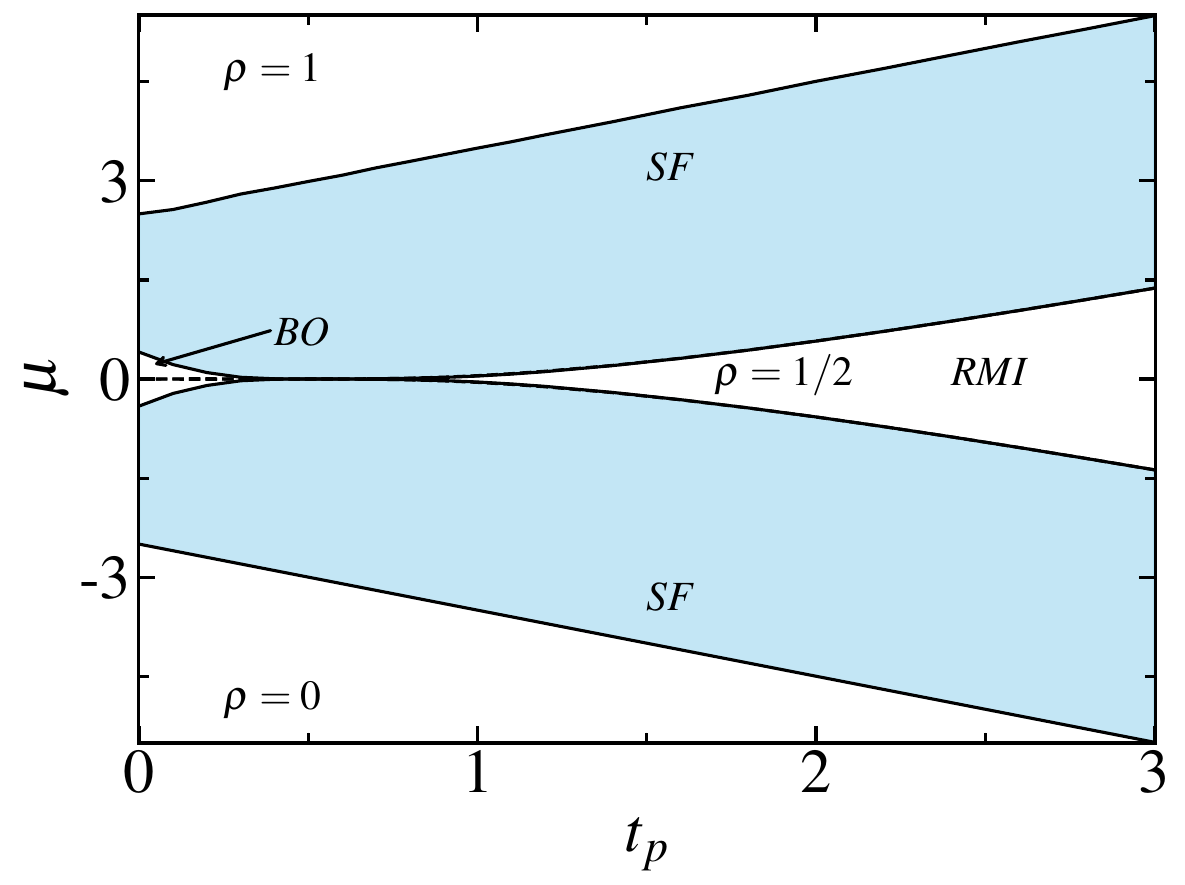}
    \caption{The phase diagram in the $t_p-\mu$ plane for the model given in Eq.~\ref{eq:ham} in the staggered dimerization configuration for $V_1=V_2=1$ and $\lambda=0$ in the absence of rung interaction. The central white regions correspond to the gapped phases at $\rho=1/2$. The white regions for $\rho=0$ and $\rho=1$ denote the empty and full states, respectively. }
    \label{fig:staggered_phase_diagram_v1_1_v2_0}
\end{figure}

We also compare the onsite particle densities $\langle n_i \rangle$ of both the legs in the topological phase at $\lambda=6$ which is shown in Fig.~\ref{fig:desnity_vs_site_index}. The nature of the edge states can be seen from the values of  $\langle n_i\rangle$ in leg-1 (black circles) which is uniform throughout the lattice except at the edges where $\langle n_i\rangle \sim  0.86$ (left edge) and $\langle n_i\rangle \sim 0.14$ (right edge). Such asymmetry in densities in both the edge sites is the characteristic of the polarised edge states which ideally tend to values one and zero on the left and right edge sites respectively in the limit of strong $\lambda$. However, the densities in leg-2 (red squares) oscillate around the value $\langle n_i\rangle=0.5$ throughout the lattice.

Another notable characteristic for identifying topological order in strongly correlated systems is the degeneracy in the entanglement spectrum of the ground state since the entanglement spectrum is associated with the energy spectrum of edge excitations~\cite{haldane,kawakami,pollman_entropy,sjia}. The entanglement spectrum is determined by logarithmically rescaling the singular values and it is obtained as
\begin{equation}
    \xi_i=-\text{ln}(\Lambda_i^2),
\label{eq:entanglement_spectrum}
\end{equation}
where $\Lambda_i$'s are the singular values corresponding to a particular bipartition in the MPS wavefunction of the ground state~\cite{Hayward2018, mondal_pumping}. We plot the lower four values of the entanglement spectrum ($\xi_1,~\xi_2,~\xi_3,~\xi_4$) for a bipartition at the central bond of the system as a function of $\lambda$ for a system with $L=140$ rungs for different dimerization strengths in Fig.~\ref{fig:ent_spectrum_tp_1_vp_0}. It is clear from the spectrum that there exists two-fold degeneracy in the lowest entanglement spectrum (i.e., $\xi_1=\xi_2$ and $\xi_3=\xi_4$) after $\lambda\sim3.3$ for which the system is topological in nature (compare with Fig.~\ref{fig:wind_num_with_dimerization}). Before the critical $\lambda$, such degeneracy is lifted due to the presence of the trivial phase.

\begin{figure}[t]
    \centering
    \includegraphics[width=0.8\linewidth]{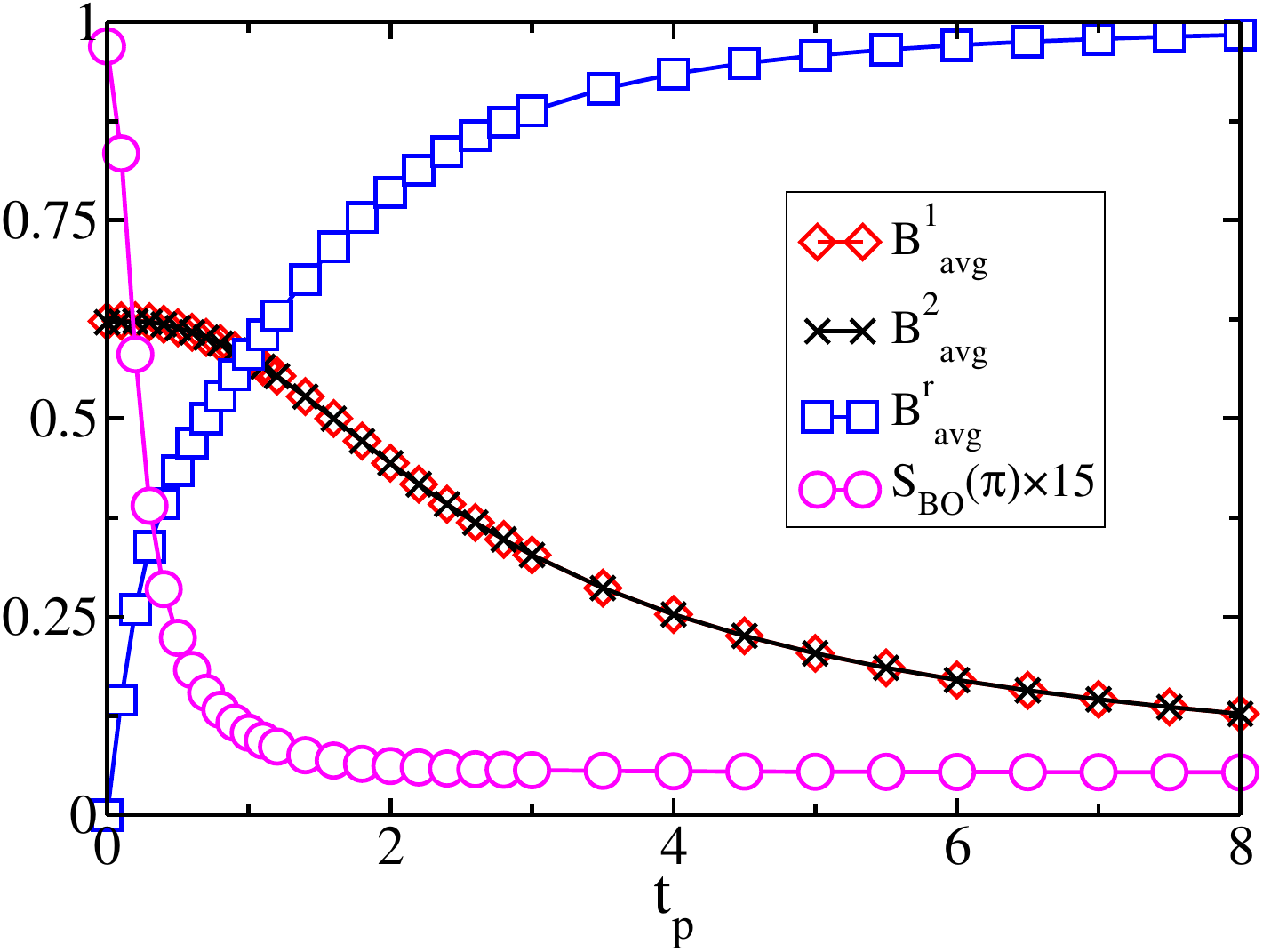}
    \caption{Shows the behavior of average bond energies $B_{avg}^{1}$ (red diamonds), $B_{avg}^{2}$ (black crosses) and $B_{avg}^r$ (blue squares) computed by averaging over all the respective bonds for $200$ sites ($L = 100$ rungs) at $\rho = 1/2$ as a function of $t_p$. Also shown is the scaled BO structure factor $S_{BO}(\pi)$ (magenta circles)  with varying $t_p$. Here the data for $S_{BO}(\pi)$ is magnified $15$ times for better visibility. For all the cases we set $\lambda=0$ and $V_1=V_2=1$.}
    \label{fig:bo_stfc_v1_1_v2_0}
\end{figure}
\begin{figure*}[t]
\centering
\includegraphics[width=0.8\linewidth]{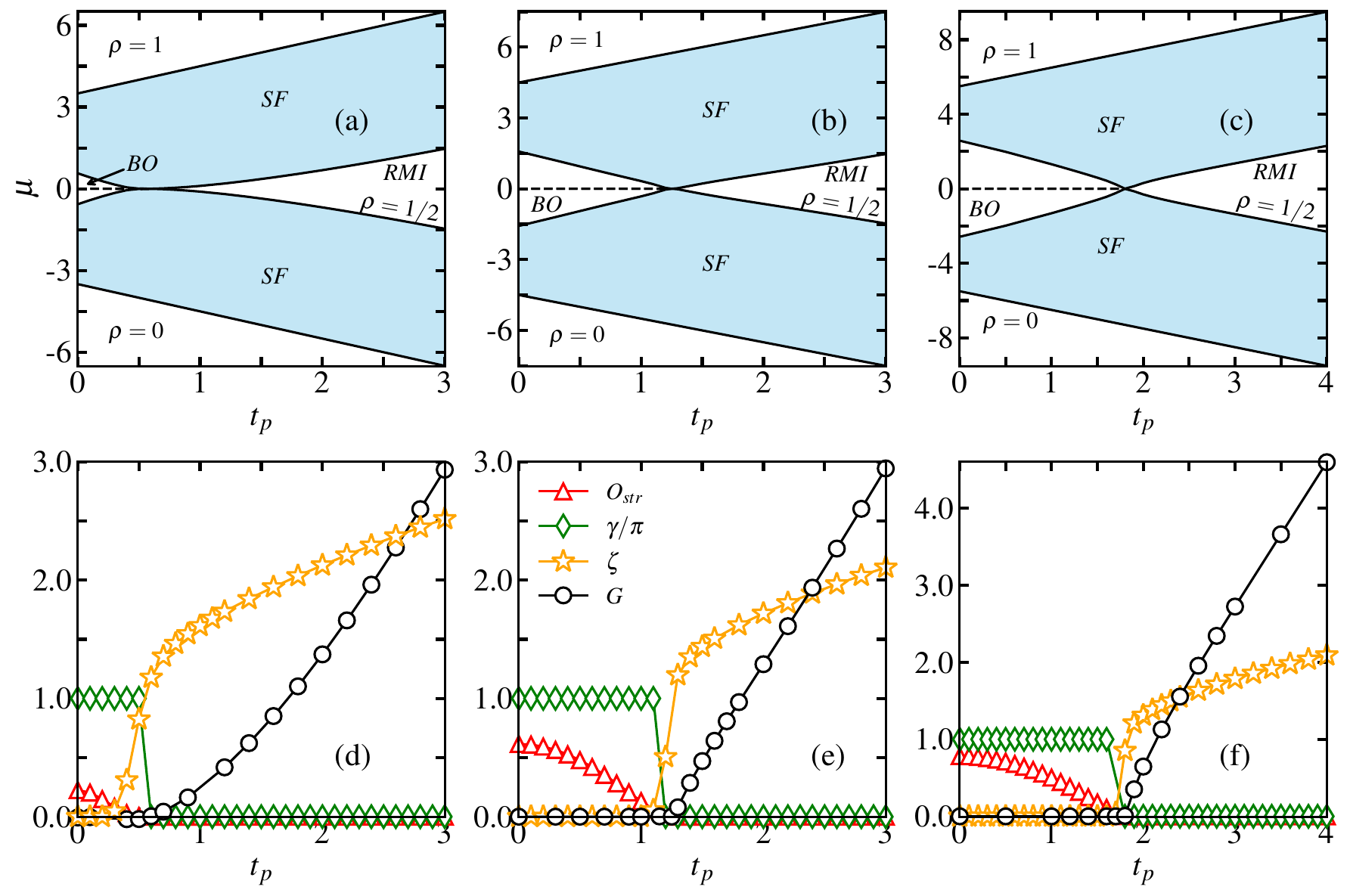}
\caption{(Top panel) Phase diagrams of model  shown in Eq.~\ref{eq:ham} in the $t_p$-$\mu$ plane for  staggered dimerization with $V_p=0$ and dimerization strength: (a) $\lambda = 2$, (b) $\lambda = 4$, and (c) $\lambda = 6$. The shaded regions in each figure denote the gapless superfluid (SF) phase. The black solid lines mark the boundary of the gapped phases and the black dashed lines indicate the pair of degenerate edge modes. While the white regions at $\rho=1/2$ are the gapped phases, the regions for $\rho=0$ and $\rho=1$ denote the empty and full states, respectively. (Bottom panel) Shows the extrapolated values of $G$ (black circles),  $\gamma/\pi$ (green diamonds) computed using the exact diagonalization method for a system consisting of $L=8$ rungs ($16$ sites), $\zeta$ (orange stars) and $O_\text{str}$ (red triangles) obtained with a system of $L=140$ rungs ($280$ sites) as functions of $t_p$ for the dimerization strength (d) $\lambda=2$, (e) $\lambda=4$, and (f) $\lambda=6$. }
\label{fig:staggered_phase_with_tp}
\end{figure*}
\begin{figure}[b]
\centering
\includegraphics[width=0.8\linewidth]{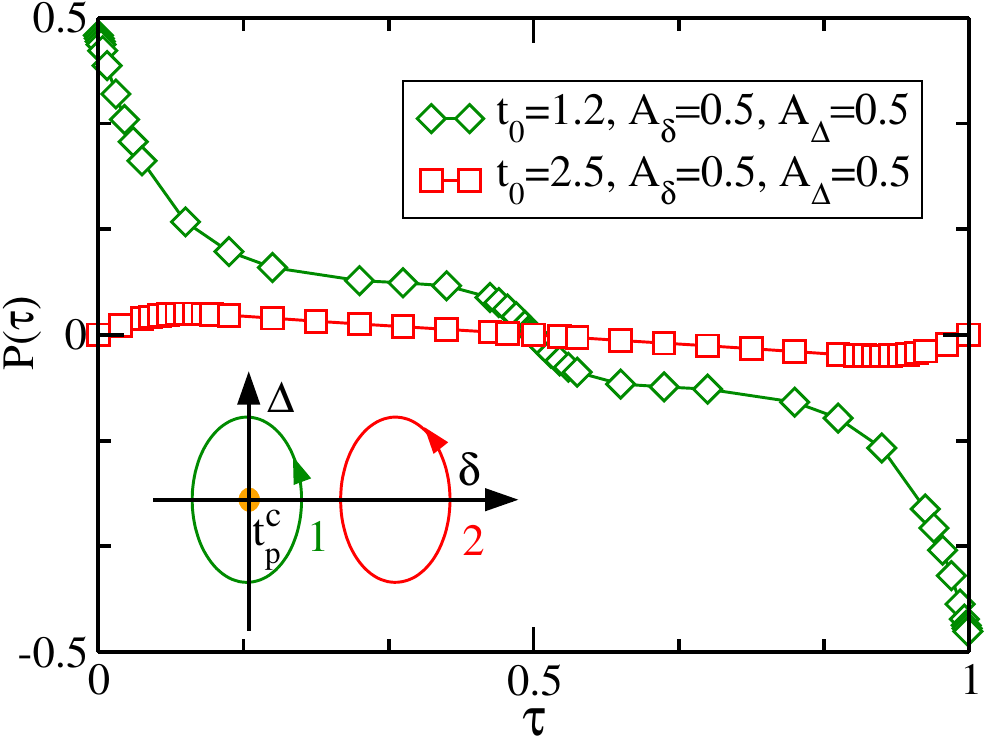}
\caption{Shows the polarization ($P(\tau)$) as a function of $\tau$ for a system consisting of $L=100$ rungs ($200$ sites) for two distinct pumping cycles, associated with different values $t_0$, as illustrated in the inset. In the plot, green diamonds represent the change in polarization for cycle $1$, while red squares correspond to cycle $2$. Here the dimerization is set to be $\lambda=4$. As the critical rung-hopping $t_{p}^c\sim1.2$ for the topological transition  lies within cycle $1$, we achieve robust pumping whereas no charge is being pumped during cycle $2$.}
\label{fig:polarization_with_tp}
\end{figure}

We now try to understand the nature of the gapped phase in the regime before the gap closing point $\lambda\sim3.3$ shown in Fig.~\ref{fig:effect_of_delta_staggered}. This can be understood by starting from the limit of $\lambda=0$ for which the system exhibits a small gap for $t_p=1$ as can be seen in Fig.~\ref{fig:effect_of_delta_staggered}. In order to have a better insight about the phase we obtain the phase diagram in the $t_p-\mu$ plane which is shown in Fig.~\ref{fig:staggered_phase_diagram_v1_1_v2_0}. For this case we have assumed $V_1=1$ and $V_2=1$ and fix $\lambda=0$. In this case, when $t_p=0$, the two chains are completely decoupled from each other, and the leg-1 (leg-2) chain exhibits a trivial (topological) BO phase ~\cite{Mondal_topology}. As a result of this the system is initially gapped when $t_p=0$ and exhibits a pair of edge states corresponding to the topological leg (leg-2). However, with the increase in $t_p$, the gap slowly decreases and tends to vanish near $t_p\sim 0.6$ and reopens again and in this second gapped phase ($t_p>0.6$), the mid-gap states do not appear. While the first gapped phase with the mid-gap states is found to exhibit topological character (not shown) we obtain that the gapped phase for $t_p \gtrsim 0.6$ is an RMI phase by comparing the average NN bond energies and the BO structure factor. We compute the average NN bond energy by using the formula 
\begin{eqnarray}
    B_{avg}^\alpha=\frac{1}{L}\sum_i\langle B_i^\alpha\rangle ~~; B_{avg}^r=\frac{1}{L}\sum_i\langle B_i^r\rangle
\end{eqnarray}
along the legs $\alpha=1,~2$ and rungs respectively. 
In Fig.~\ref{fig:bo_stfc_v1_1_v2_0}, we plot $B_{avg}^1$ (red diamonds), $B_{avg}^2$ (black crosses), $B_{avg}^r$ (blue squares) along with the extrapolated values of $S_{BO}(\pi)$ (magenta circles) as functions of $t_p$. It can be seen that in the decoupled leg limit the BO structure factor $S_{BO}(\pi)$ is large due to the bond ordering in the individual legs as expected. The features of bond ordering is also seen from the finite and almost equal values of $B_{avg}^1$ and $B_{avg}^2$. However, as $t_p$ increases, all the three quantities namely $S_{BO}(\pi)$, $B_{avg}^1$ and $B_{avg}^2$ tend to decrease indicating a decrease in the bond ordering along the legs. At the same time we obtain that the average bond energy along the rungs i.e.  $B_{avg}^r$ increases as a function of $t_p$ which was initially zero in the decoupled leg limit. These behavior together confirms that the gapped phase for $t_p \gtrsim 0.6$ is an RMI phase for $\lambda=0$. This clarifies that the phase at $\lambda=0$ in the phase diagram Fig.~\ref{fig:effect_of_delta_staggered} for $t_p=1$ is an RMI phase which survives up to $\lambda\sim3.3$ before entering into the topological BO phase.

The above analysis reveals that there occurs a phase transition from an RMI phase to a topological BO phase if the dimerization strength $\lambda$ is increased from zero when the hopping strength in the ladder is uniform i.e. $t=t_p$. This also suggests that when the dimerization $\lambda=0$, the system which was topological in the decoupled leg limit becomes a trivial RMI phase with an increase in $t_p$. These findings raise the question on the dependence of the topological BO to RMI phase transition as a function of $t_p$ on the dimerization strength. To address this question, we obtain the phase diagram in the $\mu-t_p$ plane for three different values of dimerization such as $\lambda=2,~4$ and $6$ which are shown in Fig.~\ref{fig:staggered_phase_with_tp}(a), (b) and (c) respectively. These figures exhibit the transition from the gapped BO phase to the RMI phase in each case.  We characterize the topological nature of the BO phase and the transition to the RMI phase by plotting  $G$ (black circles), $O_{str}$ (red triangles), and $\gamma/\pi$ (green diamonds) together as functions of $t_p$ in Fig.~\ref{fig:staggered_phase_with_tp}(d-f) for the parameter values considered in Fig.~\ref{fig:staggered_phase_with_tp}(a-c) respectively. Additionally, we plot a quantity $\zeta=\xi_1-\xi_2+\xi_3-\xi_4$ (orange stars) which tracks the degeneracy in the entanglement spectrum. If the spectrum is two-fold degenerate (i.e., $\xi_1=\xi_2$ and $\xi_3=\xi_4$), $\zeta$ should vanish and become finite otherwise. By comparing these quantities we obtain that the BO phase in each case is topological in nature and the transitions to the RMI phase occur at the critical rung-hopping strengths $t_p^c\sim0.5$, $\sim1.2$, and $\sim1.8$ for $\lambda=2,~4$ and $6$ respectively. We observe that the $t_p^c$ for the transition from the topological BO phase to the RMI phase shifts towards higher values with an increase in $\lambda$. Most importantly, we can see that a topological phase can be established on a ladder with uniform leg and rung hopping strengths (i.e. $t=t_p=1$) by varying the dimerization strength $\lambda$ associated to the interactions. Note that one of the legs of the ladder is always topological for any finite value of $\lambda$ for the configuration of dimerized interaction favoring the topological BO phase (i.e. weak-strong-weak-strong-weak configuration).

\subsubsection{Thouless charge pumping}
In this part of the paper, we show the dynamical signature of the topological phases obtained in the preceding section. In this context, we explore the Thouless charge pumping~\cite{Thouless1983} or topological charge pumping (TCP) which involves a quantized transport of particles through the adiabatic evolution of a system through periodic modulation of the parameters. During the pumping cycle, the system goes adiabatically from a topological to a trivial state and then comes back to a topological state after completing a single cycle and the pumping path winds around the topological phase transition point. The chosen system parameter for pumping corresponds to the one that governs the topological phase transition, and the amount of charge pumped in a cycle is quantized and directly linked to the topological invariant of the system. 
Recently the TCP has emerged as a promising platform for exploring the signatures of topological phases in non-interacting and interacting systems both from the theoretical as well as experimental perspective~\cite{rice_mele,monika_review,Lohse2016,Takahashi2016pumping,Asboth2016_rm,bound_pump1,bound_pump3, ArguelloLuengo2024stabilizationof,Kuno2017,bertok_pump,mondal_phonon,Hayward2018,spin_pumping,pumping_quasicrystals,Taddia2017,pumping_1d,hubbarad_thouless_pump,qubit_pumping}. Here, we aim to complement the aforementioned topological phase transition through the TCP. The TCP in general is achieved by writing the appropriate Rice-Mele model corresponding to the system under consideration which allows the adiabatic evolution of the system by introducing an additional symmetry-breaking term to keep the bulk gap open throughout the pumping cycle - a necessary condition for pumping.

\begin{figure*}[t]
\centering
\includegraphics[width=0.8\linewidth]{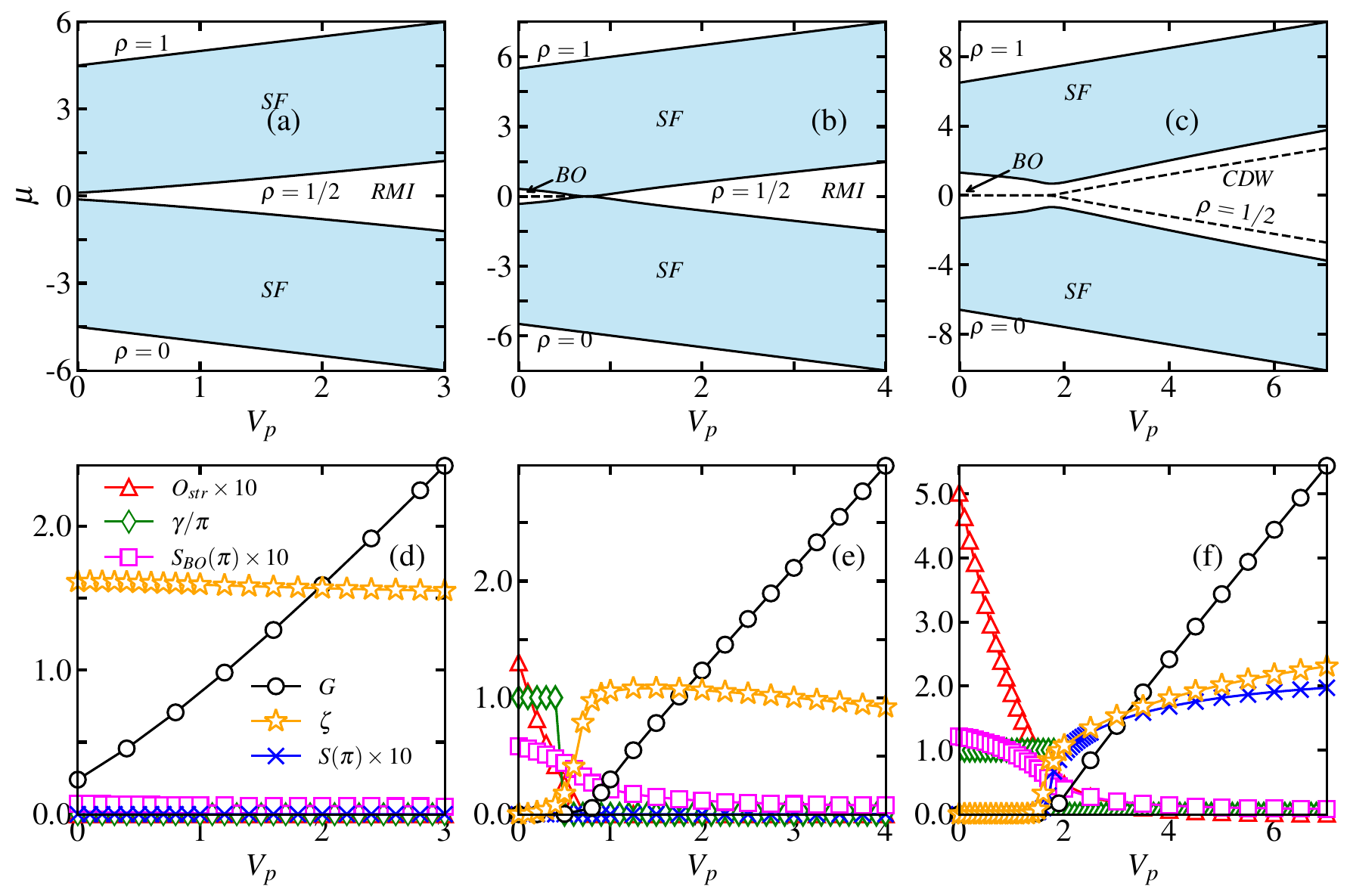}
\caption{(Top panel) Shows the phase diagram of model shown in Eq.~\ref{eq:ham} in the $V_p$-$\mu$ plane for the  staggered dimerization case with parameters $t_p=1$, $V_1=V_2=1$ and dimerization strength: (a) $\lambda = 2$, (b) $\lambda = 4$ and (c) $\lambda = 6$. The white regions at $\rho=1/2$ are the gapped phases. The shaded region denotes the gapless SF phase. The black dashed lines denote the pair of degenerate edge states in (b) and (c). The white regions for $\rho=0$ and $\rho=1$ denote the empty and full states, respectively. (d-f) Show that variation of $O_{str}$ (red triangles),  $\gamma/\pi$ (green diamonds), $S_{BO}(\pi)$ (magenta squares), $G$ (black circles), $\zeta$ (orange stars) and $S(\pi)$ (blue crosses)  as a functions $V_p$. Here $O_{str}$, $S_{BO}(\pi)$ and  $S(\pi)$ are magnified by $10$ times for better visibility. }
\label{fig:staggered_phase_with_vp}
\end{figure*}

Here, we present the pumping protocol for the two-leg ladder system based on the Hamiltonian (Eq.~\ref{eq:ham}), incorporating an additional symmetry-breaking parameter $\Delta$ for the specific dimerization strength $\lambda=4$. The corresponding model Hamiltonian is expressed by
\begin{eqnarray}
 H_{1,p}(\tau)&=-&t\sum_{\alpha\in(1,2)}\sum\limits_{i}({a}^\dagger_{\alpha,i}{a}_{\alpha,i+1}+\text{H.c.}) \nonumber\\&+&\sum_{\alpha\in(1,2)}\sum\limits_{i}V_{\alpha,i} \left(n_{\alpha,i}-\frac{1}{2}\right)\left(n_{\alpha,i+1}-\frac{1}{2}\right)\nonumber\\&-& (t_0 + \delta(\tau))\sum\limits_{i}(a^\dagger_{1,i}a_{2,i}+\text{H.c.})\nonumber\\&+&\Delta(\tau)\sum\limits_i((-1)^in_{1,i}+(-1)^{i+1}n_{2,i}).
 \label{eq:pumping_hamiltonian_with_tp]}
\end{eqnarray}
Here, $\tau$ represents the adiabatic pumping parameter. $\delta(\tau) = A_\delta$cos$(2\pi\tau)$ and $\Delta(\tau) = A_\Delta$sin$(2\pi\tau)$ are functions of $\tau$ representing periodic modulation in the rung hopping and staggered onsite potential, respectively. The origin of the pumping cycle is $t_p=t_0$ in the $\delta-\Delta$ plane. To study the TCP, we have chosen two sets of parameters corresponding to two pumping cycles i.e. (a) $t_0=1.2$, $A_\delta=0.5$ and $A_\Delta=0.5$ (green continuous line in the inset of Fig.~\ref{fig:polarization_with_tp} and (b) $t_0=2.5$, $A_\delta=0.5$ and $A_\Delta=0.5$ (red continuous line in the inset of Fig.~\ref{fig:polarization_with_tp}) and calculate the polarization given by
\begin{eqnarray}
 P(\tau) = \frac{1}{2L}\sum\limits_{i=1}^{2L}\bra{\psi(\tau)}(i-(2L-1)/2)\ket{\psi(\tau)},
 \label{eq:polarization}
\end{eqnarray}
for a system size of $L=100$ rungs, where $\ket{\psi(\tau)}$ is the ground state wave function of the Hamiltonian given in Eq.~\ref{eq:pumping_hamiltonian_with_tp]}. The amount of charge pumped $Q$ in a cycle is given by
\begin{equation}
Q = \int_0^{1}d\tau\partial_\tau P(\tau).
\label{eq:total_charge}
\end{equation}
Note that the pumping can only occur when the pumping path encloses the gap closing critical point $t_{p}^c$ (orange circle in the inset of Fig.~\ref{fig:polarization_with_tp}). Since, for $\lambda=4$ the critical $t_{p}^c\sim 1.2$, it lies on the $\delta$ axis for the cycle $1$ ($t_0-A_\delta<t_{p}^c< t_0+A_\delta$) only. Consequently, the polarization changes from $0.5$ to $-0.5$ for cycle $1$ (green diamonds in Fig.~\ref{fig:polarization_with_tp}), resulting in a total charge transfer of $|Q|=1$, while no charge is pumped for cycle $2$ (red squares in Fig.~\ref{fig:polarization_with_tp}). This analysis concretely establishes that there is a topological phase transition happening between the topological BO phase and the trivial RMI phase through a gap-closing critical point. Now we turn our focus to understand the effect of rung interaction.

\subsection{$t_p\neq 0$ , $V_p\neq 0$}
After analyzing the effect of the rung coupling $t_p$ alone, in this part of the paper we will focus on studying the combined effect of $t_p$ and $V_p$ (i.e. the rung interaction) on the quantum phases.
To investigate the combined effect of $t_p$ and $V_p$, we set $t_p=1$ and examine the impact of $V_p$ for the values of $\lambda=2,~4$ and $6$. It is important to note that, in the scenario under consideration, the system is in the trivial RMI phase for $\lambda=2$ and in the topological BO phase for both $\lambda=4$ and $6$.

\subsubsection{Phase diagram}
The ground state phase diagrams in the $V_p-\mu$ plane for $t=t_p=1$ are depicted in Fig. ~\ref{fig:staggered_phase_with_vp}(a), (b) and (c) for $\lambda=2,~4$ and $6$, respectively. It is evident from the diagrams that the topological BO phases exist at $V_p=0$ for $\lambda=4$ and $6$ only and the RMI phase is observed for $\lambda=2$ at half-filling. We obtain that for $\lambda=2$, the system remains in the RMI phase with an increase in $V_p$ (Fig.~\ref{fig:staggered_phase_with_vp}(a)). However, for $\lambda=4$ and $6$ the topological BO phase is unstable against $V_p$. While for $\lambda=4$ we obtain a BO-RMI phase transition as a function of $V_p$ through a gap closing point (Fig.~\ref{fig:staggered_phase_with_vp}(b)), for $\lambda=6$, the topological BO phase undergoes a transition to a CDW phase. After entering the CDW phase, the zero energy edge modes become energetic and remain within the gap which can be seen from the behavior of gap in Fig.~\ref{fig:staggered_phase_with_vp}(c). 
Note that the BO-RMI transition for $\lambda=4$ involves a gap-closing point, whereas no gap-closing occurs for $\lambda=6$. Rather, in the latter case the gap reaches a minimum and reopens again after $V_p\sim1.7$, indicating a gapped-gapped transition.

The above mentioned phases have been identified by comparing the already defined observables such as $\gamma/\pi$ (green diamonds), $O_{str}$ (red triangles), $G$ (black circles), and $\zeta$ (orange stars)  with respect to $V_p$ in Fig.~\ref{fig:staggered_phase_with_vp} (bottom panel). For $\lambda=2$ (Fig.~\ref{fig:staggered_phase_with_vp}(d)), no topological phase is observed with increasing $V_p$, as evidenced by a zero winding number and a vanishing string order parameter along with a finite excitation gap and finite $\zeta$. At the same time for $\lambda=4$ (Fig.~\ref{fig:staggered_phase_with_vp}(e)) and $6$ (Fig.~\ref{fig:staggered_phase_with_vp}(f)), the topological signatures vanish after the critical $V_p\sim0.4$ and $V_p\sim1.7$ respectively. We identify the gapped phase for higher values of $V_p\gtrsim1.7$ for $\lambda=6$ as a CDW phase characterized by long-range density-density correlation and hence finite density structure factor defined as 
\begin{equation}
    S(k)=\frac{1}{L^2}\sum_{i,j}e^{ik|i-j|}\langle n_in_j\rangle
\end{equation}
on leg-1. We plot $S(\pi)$ as a function of $V_p$ in 
Fig. ~\ref{fig:staggered_phase_with_vp}(d-f) (blue crosses) which clearly shows that for $\lambda=6$, $S(\pi)$ becomes finite after the critical point $V_p\sim1.7$, indicating the onset of a CDW phase and for other cases the structure factor vanishes. 
Regarding the topological aspects, it is noteworthy that a pair of degenerate zero-energy modes is observed at $V_p=0$ for $\lambda=4$ and $6$ (Fig.~\ref{fig:staggered_phase_with_vp}(b) and (c)). This degeneracy persists up to the critical points for each case of dimerization which can be seen from Fig.~\ref{fig:staggered_phase_with_vp}.
\begin{figure}[b]
\centering
\includegraphics[width=0.9\linewidth]{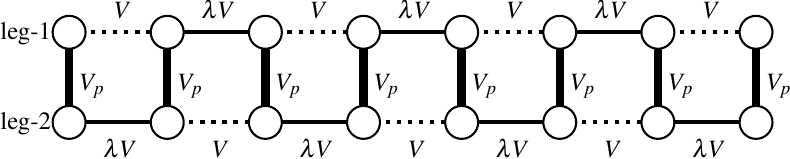}
\caption{The schematic represents the equivalent one-dimensional chain of dimerized NN interaction of type $V_p-\lambda V-V_p-\lambda V\ldots$ along the continuous line.}
\label{fig:cdw_schematic}
\end{figure}
\begin{figure}[b]
\centering
\includegraphics[width=0.8\linewidth]{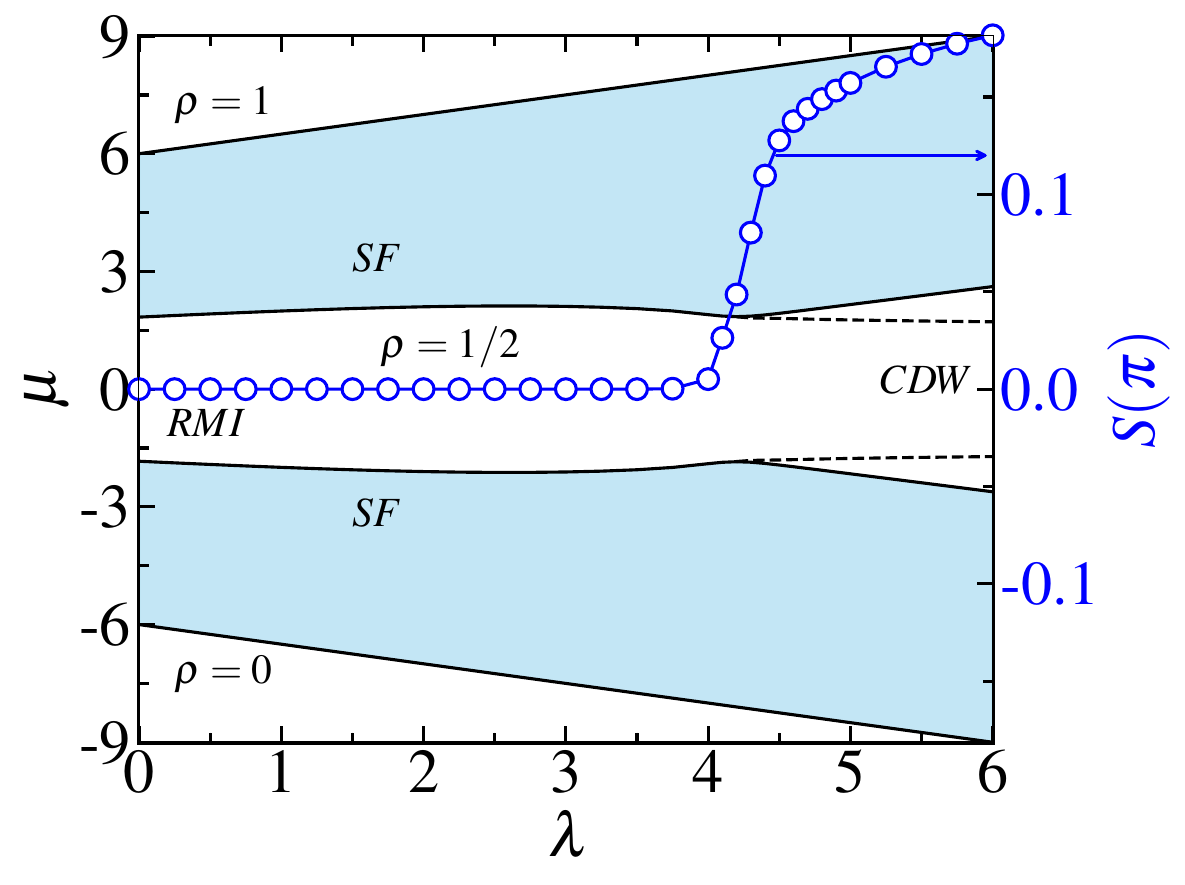}
\caption{The phase diagram in $\lambda-\mu$ plane with fixed $t_p=1$ and $V_p=5$. The central white region corresponds to the  gapped phases such as the RMI and the CDW phases at $\rho=1/2$. The white regions for $\rho=0$ and $\rho=1$ denote the empty and full states, respectively. The blue line with circles denotes the values of CDW structure factor $S(\pi)$. The dashed lines represent the mid gap states.}
\label{fig:staggered_phase_diagram_vp_5_lam}
\end{figure}
\begin{figure}[b]
    \centering
    \includegraphics[width=0.8\linewidth]{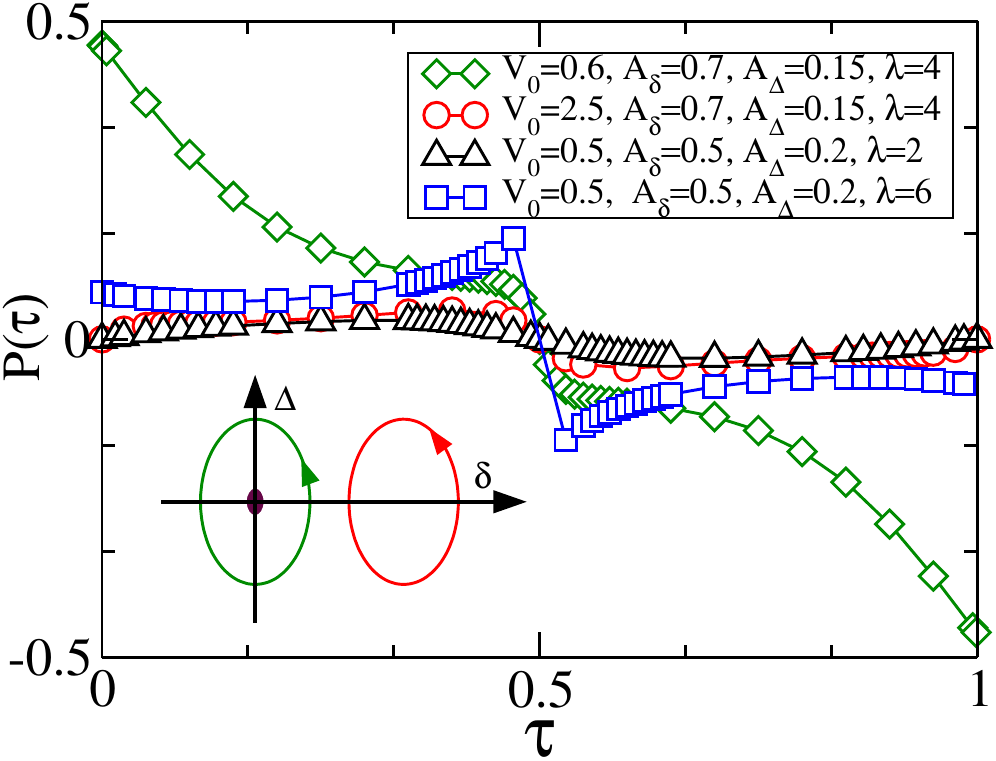}
    \caption{Shows the polarization $P(\tau)$ as a function of $\tau$ for a system with $L=140$ rungs.  The inset showcases two distinct pumping cycles corresponding to two different $V_0$ values for $\lambda=4$. Green diamonds represent the polarization change for cycle $1$, while red squares denote cycle $2$. As the critical $V_{p}\sim0.5$ lies within cycle $1$, only this cycle demonstrates robust pumping, with no charge pumped during cycle $2$. Additionally, black triangles and blue squares indicate the polarization evolution for $\lambda=2$ and $6$, respectively. For $\lambda=2$, the system always remains in a trivial phase, resulting in no charge pumping. Similarly, for $\lambda=6$, pumping ceases due to the absence of a gap closing point within the pumping cycle.}
    \label{fig:polarization_with_vp}
\end{figure}

These findings can be understood through energy minimization considerations. For $\lambda=4$, there is already a strong dimerization along the legs for $V_p=0$. However, with increasing $V_p$, the energy cost of having two particles on a single rung becomes increasingly unfavorable due to the repulsive rung interactions. Consequently, particles tend to dimerize along the rungs to minimize energy rather than along the bonds of the legs, leading to the RMI phase. As a result, there is a transition from the BO phase to the RMI phase for $\lambda=4$. Due to the same reason, the RMI phase for $\lambda=2$ becomes more and more stable with an increase in $V_p$. However, stronger dimerization like $\lambda=6$ favors a topological BO phase over the RMI phase when $V_p$ is small. Surprisingly, when $V_p$ becomes sufficiently strong, the bond ordering along the legs gets suppressed and a CDW phase is favored instead even at larger values of dimerization strength. The reason for this can be attributed to the following. When $V_p$ is very strong compared to $V_1=V_2=V=1$ and $\lambda$ is large, the interactions along the bonds associated to $V$ are negligible as compared to that along the bonds associated to $\lambda V$ and along the rungs. Thus, the system behaves like a 1D chain with dimerized interaction strengths $V_p-\lambda V-V_p-\lambda V\ldots$ (see Fig.~\ref{fig:cdw_schematic}), and the system exhibits a CDW structure with one particle and one hole residing in a rung followed by one hole and one particle in the alternate rung~\cite{Mondal_topology}.

To obtain a clearer picture on this, in Fig.~\ref{fig:staggered_phase_diagram_vp_5_lam} we plot the phase diagram for a sufficiently strong $V_p=5$ in the $\lambda-\mu$ plane. The figure clearly depicts that the gap does not close for any values of $\lambda$, and a gapped-gapped (RMI-CDW) phase transition occurs at $\lambda\sim 4$. We also plot the CDW structure factor $S(\pi)$ as a function of $V_p$ in Fig.~\ref{fig:staggered_phase_diagram_vp_5_lam} which shows that $S(\pi)=0$ in the RMI phase and becomes finite in the CDW phase. From this analysis, we can see that the RMI phase for $t_p=1$ is favored by the increase in $V_p$ for small dimerization along the legs. However, the RMI character gets destroyed and a CDW phase is favored when the dimerization is large.

\subsubsection{Thouless charge pumping}
We now obtain the signature of the topological phases through Thouless charge pumping which has been introduced in the previous section.  The corresponding  Hamiltonian is given by
\begin{eqnarray}
 H_{2,p}(\tau)&=-&t\sum\limits_{\alpha\in(1,2)}\sum\limits_{i}({a}^\dagger_{\alpha,i}{a}_{\alpha,i+1}+\text{H.c.}) \nonumber\\&+&\sum\limits_{\alpha\in(1,2)}\sum\limits_{i}V_{\alpha,i} \left(n_{\alpha,i}-\frac{1}{2}\right)\left(n_{\alpha,i+1}-\frac{1}{2}\right)\nonumber\\&-& t_p\sum\limits_{i}(a^\dagger_{1,i}a_{2,i}+\text{H.c.})\nonumber\\&+&(V_0+\delta(\tau))\sum\limits_i\left(n_{1,i}-\frac{1}{2}\right)\left(n_{2,i}-\frac{1}{2}\right)\nonumber\\&+&\Delta(\tau)\sum\limits_i((-1)^in_{1,i}+(-1)^{i+1}n_{2,i})
 \label{eq:pumping_hamiltonian_with_vp]}
\end{eqnarray}
which involves varying parameters $\delta(\tau)=A_\delta$cos$(2\pi\tau)$ and $\Delta(\tau)=A_\Delta$sin$(2\pi\tau)$ periodically, where $\tau$ represents the adiabatic pumping parameter. The pumping cycle originates at $V_p=V_0$ in the $\delta-\Delta$ plane. According to the results shown in Fig.~\ref{fig:staggered_phase_with_vp}, a robust charge pumping is expected for $\lambda=4$, whereas for other cases the pumping should break down. To demonstrate this we choose four sets of parameter values and compute the polarization as per Eq.~\ref{eq:polarization} and plot them as a function of $\tau$ in Fig.~\ref{fig:polarization_with_vp}. Two of them correspond to $\lambda=4$ to illustrate both robust pumping and its breakdown, and another two for $\lambda=2$ and $6$ to demonstrate only the breakdown of pumping. For $\lambda=4$, with $V_0=0.6$ and $A_\delta=0.7$, robust pumping occurs as the critical $V_{p}$ falls within the pumping cycle, resulting in a polarization change from $0.5$ to $-0.5$ (green diamonds), hence in a transfer of total charge $|Q|=1$ per cycle. Conversely, no charge is pumped when the gap closing point lies outside the cycle (red circles). A clear breakdown of pumping  (blue squares) is observed for $\lambda=6$, indicating the absence of a gap-closing point within the cycle despite the transition point lying within the pumping path. This is due to fact that the BO-CDW phase transition in this case breaks the symmetries of the topological BO phase. Similarly, we observe that no charge is pumped for $\lambda=2$ (black triangles). This analysis clearly supports the topological phase transition that occurs for $\lambda=4$.

So far, we have demonstrated that the system of hardcore bosons on a two-leg ladder with staggered NN dimerization exhibits topological phases and phase transitions to trivial phases. However, we examined that if the dimerization is considered to be uniform in both the legs (i.e. both the legs support topological phases in the decoupled leg limit) the onset of rung couplings destroy the topological properties of the system. We will briefly discuss this in the following for completeness. We will show that unlike the staggered dimerization case, here the topology is weak and we observe no well-defined topological phase transition as a function of the rung coupling. We will first discuss the situation when the rung interaction is zero and then we will discuss the physics by varying the rung interaction.

\subsection{Uniform dimerization}
\begin{figure}[t]
\centering
\includegraphics[width=0.9\linewidth]{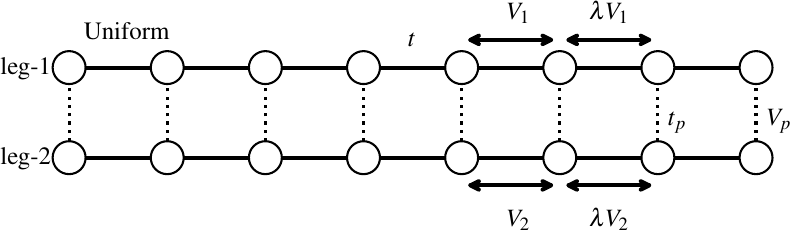}
\caption{Schematic diagram illustrating a two-leg ladder with uniform dimerization pattern along the legs.}
\label{fig:dimerizaed_ladder_uniform}
\end{figure}

The Hamiltonian corresponding to the uniform dimerization configuration in the NN interaction is given by
\begin{eqnarray}
 H&=-&t\sum\limits_{\alpha\in(1,2)}\sum\limits_{i}({a}^\dagger_{\alpha,i}{a}_{\alpha,i+1}+\text{H.c.}) \nonumber\\&+&\sum\limits_{\alpha\in(1,2)}\sum\limits_{i\in\text{odd}}V_{\alpha} \left(n_{\alpha,i}-\frac{1}{2}\right)\left(n_{\alpha,i+1}-\frac{1}{2}\right)\nonumber\\&+&\sum\limits_{\alpha\in(1,2)}\sum\limits_{i\in\text{even}}\lambda V_{\alpha} \left(n_{\alpha,i}-\frac{1}{2}\right)\left(n_{\alpha,i+1}-\frac{1}{2}\right)\nonumber\\&-& t_p\sum\limits_{i}(a^\dagger_{1,i}a_{2,i}+\text{H.c.})\nonumber\\&+&V_p\sum\limits_{i}\left(n_{1,i}-\frac{1}{2}\right)\left(n_{2,i}-\frac{1}{2}\right)
 \label{eq:ham2}
\end{eqnarray}
where the parameters have their usual meaning as defined earlier and the schematic corresponding to this configuration in Fig.~\ref{fig:dimerizaed_ladder_uniform}. In the following, we will first discuss the case of $V_p=0$ and then study the effect of finite $V_p$.
\subsubsection{$t_p\neq 0$, $V_p=0$}
\begin{figure}[b]
\centering
\includegraphics[width=0.8\linewidth]{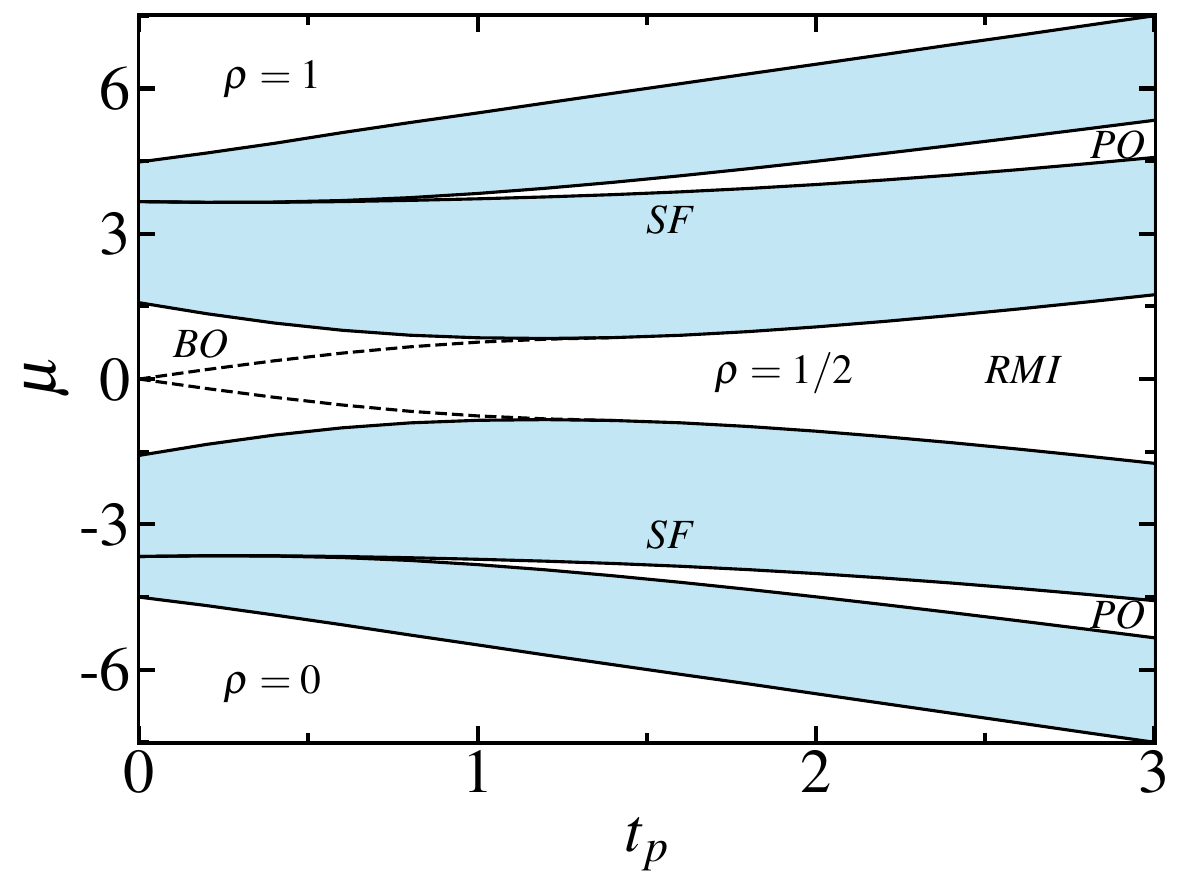}
\caption{Phase diagram of the model shown in Eq.~\ref{eq:ham2} in the $t_p$-$\mu$ plane with $V_p=0$, $V_1 = V_2 = 1$ and $\lambda = 4$ when uniform dimerization along the legs is considered. The white region in the middle denotes the transition from the gapped BO phase to the gapped RMI phase at $\rho=1/2$. The blue shaded regions represent the gapless superfluid (SF) phase. The white regions for $\rho=0$ and $\rho=1$ denote the empty and full states, respectively.  The white regions at $\rho=1/4$ and $3/4$ are the gapped plaquette order (PO) phases. }
\label{fig:uniform_phase_diagram}
\end{figure}

The phase diagram of the model shown in Eq.~\ref{eq:ham2} for uniform dimerization is depicted in Fig.~\ref{fig:uniform_phase_diagram} in the $t_p-\mu$ plane with fixed $V_1 = V_2= 1$ and $\lambda = 4$. Here we obtain that in the absence of $t_p$, the system exhibits a gapped phase at $\rho=1/2$ with four degenerate zero energy edge modes lying inside the bulk gap. This is expected since both the decoupled legs exhibit topological phases. When $t_p$ increases, the bulk gap shrinks but never closes, and after a certain value, the gap increases again, indicating a crossover between two gapped phases. On the other hand, the zero energy edge modes split by moving away from the zero energy limit and eventually merge into the bulk when $t_p$ increases (dashed black lines). Thus, the system can be considered as a weak topological insulator and here we do not observe a gap-closing topological phase transition as a function of $t_p$. Further, by analyzing the onsite particle distribution and the bond energy along all possible bonds of the lattice, we identify the first gapped phase to be a BO phase and the second gapped phase to be an RMI phase (not shown). Additionally, we obtain two other gapped phases at $\rho=1/4$ and $\rho=3/4$, which are dual to each other. We identify these phases to be the plaquette order (PO) phases where a single particle (hole) is localized in each plaquette for the $\rho=1/4$ ($=3/4$) case. At incommensurate densities, however, the system exhibits the behavior of a SF phase (shaded blue area).

The above analysis reveals that connecting two topologically non-trivial chains via finite rung hopping induces energetic edge modes, rendering the system to be weakly topological one. Consequently, there occurs a crossover from the BO to the RMI phase as a function of $t_p$ at half-filling. For further analysis, we will now explore the effect of rung interaction ($V_p$) on the phase diagram depicted in Fig.~\ref{fig:uniform_phase_diagram} for two different values of $t_p$, specifically within the BO and the RMI regions.

\subsubsection{Effect of rung interaction}
\begin{figure}[t]
\centering
\includegraphics[width=0.8\linewidth]{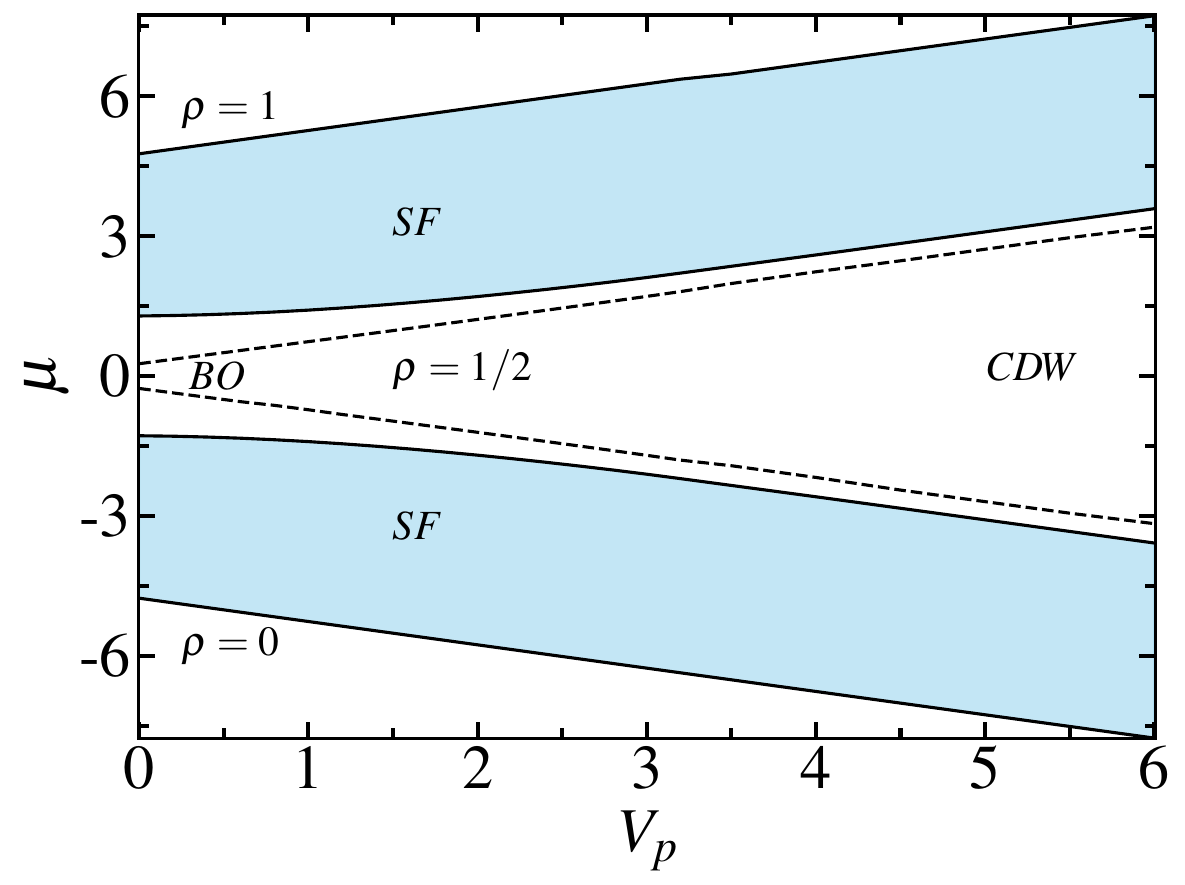}
\caption{Phase diagram of the model shown in Eq.~\ref{eq:ham2} in the $V_p$-$\mu$ plane with fixed $t_p=0.25$, $V_1 = V_2 = 1$ and $\lambda = 4$ when uniform dimerization along the legs is considered. The white region in the middle shows the gapped phases (BO and CDW) at $\rho=1/2$. The blue shaded regions represent the gapless superfluid (SF) phase. The white regions for $\rho=0$ and $\rho=1$ denote the empty and full states, respectively. The dashed lines represent the mid gap states.   }
\label{fig:uniform_dimer_phase_diagram_with_v}
\end{figure}

\begin{figure}[t]
\centering
\includegraphics[width=0.8\linewidth]{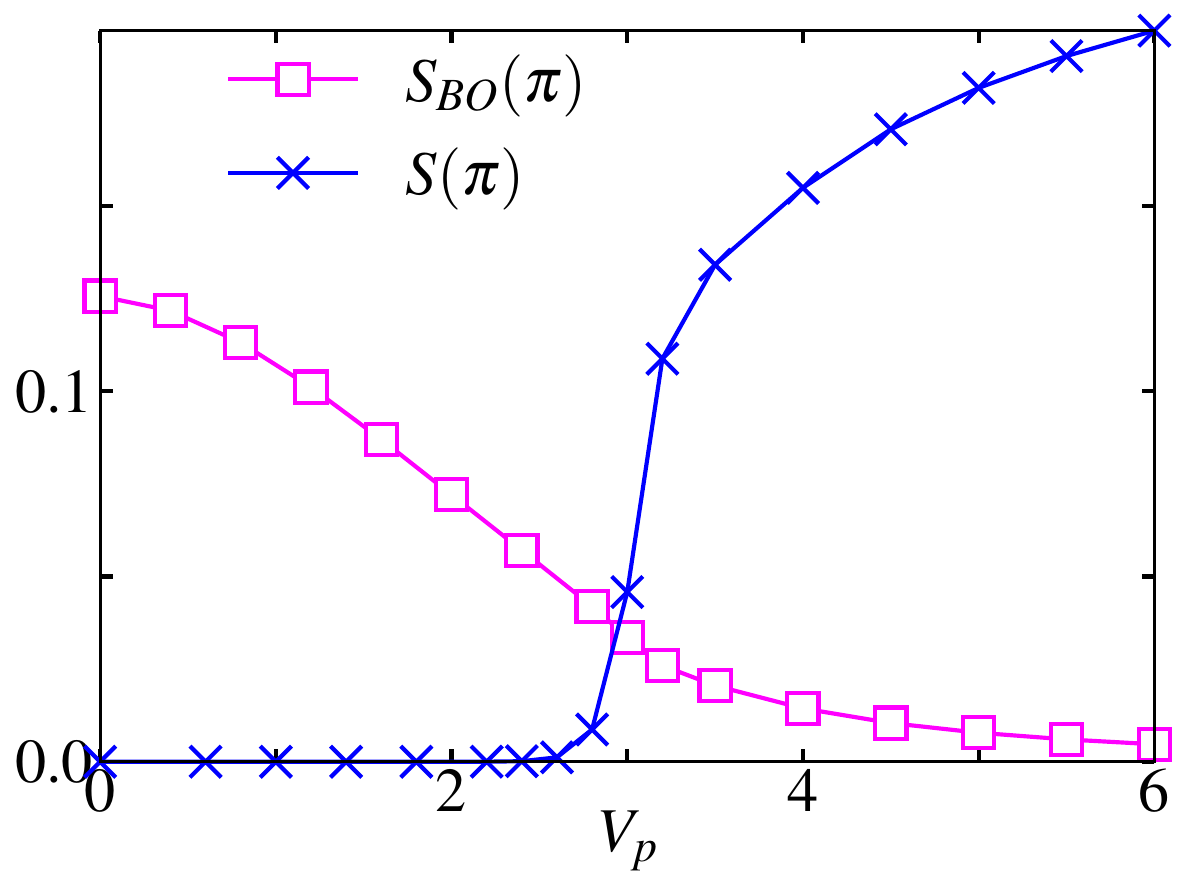}
\caption{The extrapolated values of BO structure factor $S_{BO}(\pi)$ (magenta squares) and the CDW structure factor $S(\pi)$ (blue crosses) with varying rung interaction $V_p$ for the cut at $t_p=0.25$ in the phase diagram  shown in Fig.~\ref{fig:uniform_phase_diagram}.}
\label{fig:bo_cdw_stfc_uniform_with_v}
\end{figure}

To analyze the effect of rung interaction ($V_p$) on the bond order phase of Fig.~\ref{fig:uniform_phase_diagram}, we fix the value of rung hopping at $t_p=0.25$ and plot the phase diagram in $V_p-\mu$ plane as shown in Fig. ~\ref{fig:uniform_dimer_phase_diagram_with_v}. The figure shows that there is a single gapped phase at $\rho=1/2$ except for the full and empty states. On the other hand, the system exhibits an SF phase at incommensurate densities. Moreover, we find that the system undergoes a transition from the BO phase to a CDW phase at $\rho=1/2$ and the edge states present at $V_p=0$ remain within the bulk gap with an increase in $V_p$. In order to clearly identify these phases and to mark the critical point of transition, we plot the BO structure factor $S_{BO}(\pi)$ (magenta line with squares) and CDW structure factor $S(\pi)$ (blue line with crosses) in Fig.~\ref{fig:bo_cdw_stfc_uniform_with_v}. While $S_{BO}(\pi)$ remains finite for $V_p\lesssim 3.0$, $S(\pi)$ vanishes in this regime, indicating a BO phase. However, for $V_p\gtrsim 3.0$, $S_{BO}(\pi)$ vanishes, and $S(\pi)$ remains finite, indicating the appearance of a CDW phase. On the other hand, the RMI phase upon increasing $V_p$ remains RMI (not shown). 

This analysis suggests that the topological phase transition is not favored in the case of uniform dimerization.

\section{Conclusions}\label{conc}
In summary, we numerically studied the ground state properties of hardcore bosons on a two-leg ladder with dimerized nearest-neighbor interaction along the legs using the DMRG method. We considered a dimerization pattern such that one of the legs of the ladder is topological and the other leg is trivial in nature. Our analyses disclosed a topological phase transition from a trivial RMI to a topological BO phase as a function of the dimerization strength for a fixed rung hopping when the rung interaction is zero. For fixed dimerization strengths, however, the system is shown to exhibit a gap-closing BO-RMI phase transition as a function of the rung hopping. We further showed that depending on whether the dimerization is weak or strong, the system either exhibits a BO-RMI phase transition through a gap-closing point or a BO-CDW phase transition with no gap-closing as a function of the rung interaction for uniform hopping throughout the ladder. Finally, we considered a system with uniform dimerization pattern, i.e., when both the legs of the ladder are topological in nature, and found no signatures of a topological phase transition.

Our studies revealed an interaction-driven topological phase transition in a two-leg ladder system which is intermediate to one and two-dimensional lattices. In the present scenario the topological phases and phase transitions are driven purely by interaction and the non-interacting counterpart is a trivial phase. More insights on these phase transitions can be obtained by using an appropriate Bosonization analysis~\cite{four_spin_furukawa,frustrated_xxz_furukawa,spt_spin_ladder,harsh_xxz}. Such studies can in principle be extended to two-dimensional lattices where topological phases can be achieved only due to interactions. It will also be informative to study the effect of onsite interaction among the bosons by relaxing the hardcore constraint. Another direction can be to understand the effect of attractive NN interactions on such systems. Finally, we want to note that the emerging physics reported here can be simulated using a set-up of Rydberg tweezer arrays~\cite{browyes}. The Rydberg atom platform offers a highly flexible experimental system for constructing arbitrary lattice geometries~\cite{broywes_geometry}. The hopping structure can be adjusted by carefully tuning the dipolar interaction between Rydberg atoms. Additionally, by suitably selecting Rydberg states, the van der Waals interaction can be made either isotropic or anisotropic~\cite{browaeys_ising} which can be used to achieve the dimerized interaction pattern considered here.

\section{Acknowledgements}

We thank Suman Mondal for useful discussions. T.M. acknowledges support from Science and Engineering Research Board (SERB), Government of India, through projects No. MTR/2022/000382 and No. STR/2022/000023.

\bibliography{ladder.bib}

\begin{thebibliography}{80}%
\makeatletter
\providecommand \@ifxundefined [1]{%
 \@ifx{#1\undefined}
}%
\providecommand \@ifnum [1]{%
 \ifnum #1\expandafter \@firstoftwo
 \else \expandafter \@secondoftwo
 \fi
}%
\providecommand \@ifx [1]{%
 \ifx #1\expandafter \@firstoftwo
 \else \expandafter \@secondoftwo
 \fi
}%
\providecommand \natexlab [1]{#1}%
\providecommand \enquote  [1]{``#1''}%
\providecommand \bibnamefont  [1]{#1}%
\providecommand \bibfnamefont [1]{#1}%
\providecommand \citenamefont [1]{#1}%
\providecommand \href@noop [0]{\@secondoftwo}%
\providecommand \href [0]{\begingroup \@sanitize@url \@href}%
\providecommand \@href[1]{\@@startlink{#1}\@@href}%
\providecommand \@@href[1]{\endgroup#1\@@endlink}%
\providecommand \@sanitize@url [0]{\catcode `\\12\catcode `\$12\catcode `\&12\catcode `\#12\catcode `\^12\catcode `\_12\catcode `\%12\relax}%
\providecommand \@@startlink[1]{}%
\providecommand \@@endlink[0]{}%
\providecommand \url  [0]{\begingroup\@sanitize@url \@url }%
\providecommand \@url [1]{\endgroup\@href {#1}{\urlprefix }}%
\providecommand \urlprefix  [0]{URL }%
\providecommand \Eprint [0]{\href }%
\providecommand \doibase [0]{http://dx.doi.org/}%
\providecommand \selectlanguage [0]{\@gobble}%
\providecommand \bibinfo  [0]{\@secondoftwo}%
\providecommand \bibfield  [0]{\@secondoftwo}%
\providecommand \translation [1]{[#1]}%
\providecommand \BibitemOpen [0]{}%
\providecommand \bibitemStop [0]{}%
\providecommand \bibitemNoStop [0]{.\EOS\space}%
\providecommand \EOS [0]{\spacefactor3000\relax}%
\providecommand \BibitemShut  [1]{\csname bibitem#1\endcsname}%
\let\auto@bib@innerbib\@empty
\bibitem [{\citenamefont {Klitzing}\ \emph {et~al.}(1980)\citenamefont {Klitzing}, \citenamefont {Dorda},\ and\ \citenamefont {Pepper}}]{Klitzing}%
  \BibitemOpen
  \bibfield  {author} {\bibinfo {author} {\bibfnamefont {K.~v.}\ \bibnamefont {Klitzing}}, \bibinfo {author} {\bibfnamefont {G.}~\bibnamefont {Dorda}}, \ and\ \bibinfo {author} {\bibfnamefont {M.}~\bibnamefont {Pepper}},\ }\bibfield  {title} {\enquote {\bibinfo {title} {New method for high-accuracy determination of the fine-structure constant based on quantized hall resistance},}\ }\href {\doibase 10.1103/PhysRevLett.45.494} {\bibfield  {journal} {\bibinfo  {journal} {Phys. Rev. Lett.}\ }\textbf {\bibinfo {volume} {45}},\ \bibinfo {pages} {494--497} (\bibinfo {year} {1980})}\BibitemShut {NoStop}%
\bibitem [{\citenamefont {Thouless}\ \emph {et~al.}(1982)\citenamefont {Thouless}, \citenamefont {Kohmoto}, \citenamefont {Nightingale},\ and\ \citenamefont {den Nijs}}]{qhe_thouless}%
  \BibitemOpen
  \bibfield  {author} {\bibinfo {author} {\bibfnamefont {D.~J.}\ \bibnamefont {Thouless}}, \bibinfo {author} {\bibfnamefont {M.}~\bibnamefont {Kohmoto}}, \bibinfo {author} {\bibfnamefont {M.~P.}\ \bibnamefont {Nightingale}}, \ and\ \bibinfo {author} {\bibfnamefont {M.}~\bibnamefont {den Nijs}},\ }\bibfield  {title} {\enquote {\bibinfo {title} {Quantized hall conductance in a two-dimensional periodic potential},}\ }\href {\doibase 10.1103/PhysRevLett.49.405} {\bibfield  {journal} {\bibinfo  {journal} {Phys. Rev. Lett.}\ }\textbf {\bibinfo {volume} {49}},\ \bibinfo {pages} {405--408} (\bibinfo {year} {1982})}\BibitemShut {NoStop}%
\bibitem [{\citenamefont {Hasan}\ and\ \citenamefont {Kane}(2010)}]{Hassanreview}%
  \BibitemOpen
  \bibfield  {author} {\bibinfo {author} {\bibfnamefont {M.~Z.}\ \bibnamefont {Hasan}}\ and\ \bibinfo {author} {\bibfnamefont {C.~L.}\ \bibnamefont {Kane}},\ }\bibfield  {title} {\enquote {\bibinfo {title} {Colloquium: Topological insulators},}\ }\href {\doibase 10.1103/RevModPhys.82.3045} {\bibfield  {journal} {\bibinfo  {journal} {Rev. Mod. Phys.}\ }\textbf {\bibinfo {volume} {82}},\ \bibinfo {pages} {3045--3067} (\bibinfo {year} {2010})}\BibitemShut {NoStop}%
\bibitem [{\citenamefont {von Klitzing}(2017)}]{vonKlitzing2017}%
  \BibitemOpen
  \bibfield  {author} {\bibinfo {author} {\bibfnamefont {Klaus}\ \bibnamefont {von Klitzing}},\ }\bibfield  {title} {\enquote {\bibinfo {title} {Metrology in 2019},}\ }\href {\doibase 10.1038/nphys4029} {\bibfield  {journal} {\bibinfo  {journal} {Nature Physics}\ }\textbf {\bibinfo {volume} {13}},\ \bibinfo {pages} {198--198} (\bibinfo {year} {2017})}\BibitemShut {NoStop}%
\bibitem [{\citenamefont {Senthil}(2015)}]{senthil_review}%
  \BibitemOpen
  \bibfield  {author} {\bibinfo {author} {\bibfnamefont {T.}~\bibnamefont {Senthil}},\ }\bibfield  {title} {\enquote {\bibinfo {title} {Symmetry-protected topological phases of quantum matter},}\ }\href {\doibase 10.1146/annurev-conmatphys-031214-014740} {\bibfield  {journal} {\bibinfo  {journal} {Annual Review of Condensed Matter Physics}\ }\textbf {\bibinfo {volume} {6}},\ \bibinfo {pages} {299--324} (\bibinfo {year} {2015})}\BibitemShut {NoStop}%
\bibitem [{\citenamefont {Fidkowski}\ and\ \citenamefont {Kitaev}(2011)}]{Fidkowski}%
  \BibitemOpen
  \bibfield  {author} {\bibinfo {author} {\bibfnamefont {Lukasz}\ \bibnamefont {Fidkowski}}\ and\ \bibinfo {author} {\bibfnamefont {Alexei}\ \bibnamefont {Kitaev}},\ }\bibfield  {title} {\enquote {\bibinfo {title} {Topological phases of fermions in one dimension},}\ }\href {\doibase 10.1103/PhysRevB.83.075103} {\bibfield  {journal} {\bibinfo  {journal} {Phys. Rev. B}\ }\textbf {\bibinfo {volume} {83}},\ \bibinfo {pages} {075103} (\bibinfo {year} {2011})}\BibitemShut {NoStop}%
\bibitem [{\citenamefont {Pollmann}\ \emph {et~al.}(2010)\citenamefont {Pollmann}, \citenamefont {Turner}, \citenamefont {Berg},\ and\ \citenamefont {Oshikawa}}]{Oshikawa}%
  \BibitemOpen
  \bibfield  {author} {\bibinfo {author} {\bibfnamefont {Frank}\ \bibnamefont {Pollmann}}, \bibinfo {author} {\bibfnamefont {Ari~M.}\ \bibnamefont {Turner}}, \bibinfo {author} {\bibfnamefont {Erez}\ \bibnamefont {Berg}}, \ and\ \bibinfo {author} {\bibfnamefont {Masaki}\ \bibnamefont {Oshikawa}},\ }\bibfield  {title} {\enquote {\bibinfo {title} {Entanglement spectrum of a topological phase in one dimension},}\ }\href {\doibase 10.1103/PhysRevB.81.064439} {\bibfield  {journal} {\bibinfo  {journal} {Phys. Rev. B}\ }\textbf {\bibinfo {volume} {81}},\ \bibinfo {pages} {064439} (\bibinfo {year} {2010})}\BibitemShut {NoStop}%
\bibitem [{\citenamefont {Gu}\ and\ \citenamefont {Wen}(2009)}]{Xiao-Gang}%
  \BibitemOpen
  \bibfield  {author} {\bibinfo {author} {\bibfnamefont {Zheng-Cheng}\ \bibnamefont {Gu}}\ and\ \bibinfo {author} {\bibfnamefont {Xiao-Gang}\ \bibnamefont {Wen}},\ }\bibfield  {title} {\enquote {\bibinfo {title} {Tensor-entanglement-filtering renormalization approach and symmetry-protected topological order},}\ }\href {\doibase 10.1103/PhysRevB.80.155131} {\bibfield  {journal} {\bibinfo  {journal} {Phys. Rev. B}\ }\textbf {\bibinfo {volume} {80}},\ \bibinfo {pages} {155131} (\bibinfo {year} {2009})}\BibitemShut {NoStop}%
\bibitem [{\citenamefont {Chen}\ \emph {et~al.}(2012)\citenamefont {Chen}, \citenamefont {Gu}, \citenamefont {Liu},\ and\ \citenamefont {Wen}}]{sptinteract}%
  \BibitemOpen
  \bibfield  {author} {\bibinfo {author} {\bibfnamefont {Xie}\ \bibnamefont {Chen}}, \bibinfo {author} {\bibfnamefont {Zheng-Cheng}\ \bibnamefont {Gu}}, \bibinfo {author} {\bibfnamefont {Zheng-Xin}\ \bibnamefont {Liu}}, \ and\ \bibinfo {author} {\bibfnamefont {Xiao-Gang}\ \bibnamefont {Wen}},\ }\bibfield  {title} {\enquote {\bibinfo {title} {Symmetry-protected topological orders in interacting bosonic systems},}\ }\href {\doibase 10.1126/science.1227224} {\bibfield  {journal} {\bibinfo  {journal} {Science}\ }\textbf {\bibinfo {volume} {338}},\ \bibinfo {pages} {1604--1606} (\bibinfo {year} {2012})}\BibitemShut {NoStop}%
\bibitem [{\citenamefont {Pachos}\ and\ \citenamefont {Simon}(2014)}]{Pachos_2014}%
  \BibitemOpen
  \bibfield  {author} {\bibinfo {author} {\bibfnamefont {Jiannis~K}\ \bibnamefont {Pachos}}\ and\ \bibinfo {author} {\bibfnamefont {Steven~H}\ \bibnamefont {Simon}},\ }\bibfield  {title} {\enquote {\bibinfo {title} {Focus on topological quantum computation},}\ }\href {\doibase 10.1088/1367-2630/16/6/065003} {\bibfield  {journal} {\bibinfo  {journal} {New Journal of Physics}\ }\textbf {\bibinfo {volume} {16}},\ \bibinfo {pages} {065003} (\bibinfo {year} {2014})}\BibitemShut {NoStop}%
\bibitem [{\citenamefont {Asb{\'o}th}\ \emph {et~al.}(2016{\natexlab{a}})\citenamefont {Asb{\'o}th}, \citenamefont {Oroszl{\'a}ny},\ and\ \citenamefont {P{\'a}lyi}}]{Asboth2016_ssh}%
  \BibitemOpen
  \bibfield  {author} {\bibinfo {author} {\bibfnamefont {J{\'a}nos~K.}\ \bibnamefont {Asb{\'o}th}}, \bibinfo {author} {\bibfnamefont {L{\'a}szl{\'o}}\ \bibnamefont {Oroszl{\'a}ny}}, \ and\ \bibinfo {author} {\bibfnamefont {Andr{\'a}s}\ \bibnamefont {P{\'a}lyi}},\ }\enquote {\bibinfo {title} {The su-schrieffer-heeger (ssh) model},}\ in\ \href {\doibase 10.1007/978-3-319-25607-8_1} {\emph {\bibinfo {booktitle} {A Short Course on Topological Insulators: Band Structure and Edge States in One and Two Dimensions}}}\ (\bibinfo  {publisher} {Springer International Publishing},\ \bibinfo {address} {Cham},\ \bibinfo {year} {2016})\ pp.\ \bibinfo {pages} {1--22}\BibitemShut {NoStop}%
\bibitem [{\citenamefont {Su}\ \emph {et~al.}(1979)\citenamefont {Su}, \citenamefont {Schrieffer},\ and\ \citenamefont {Heeger}}]{ssh_model}%
  \BibitemOpen
  \bibfield  {author} {\bibinfo {author} {\bibfnamefont {W.~P.}\ \bibnamefont {Su}}, \bibinfo {author} {\bibfnamefont {J.~R.}\ \bibnamefont {Schrieffer}}, \ and\ \bibinfo {author} {\bibfnamefont {A.~J.}\ \bibnamefont {Heeger}},\ }\bibfield  {title} {\enquote {\bibinfo {title} {Solitons in polyacetylene},}\ }\href {\doibase 10.1103/PhysRevLett.42.1698} {\bibfield  {journal} {\bibinfo  {journal} {Phys. Rev. Lett.}\ }\textbf {\bibinfo {volume} {42}},\ \bibinfo {pages} {1698--1701} (\bibinfo {year} {1979})}\BibitemShut {NoStop}%
\bibitem [{\citenamefont {Zak}(1989)}]{Zak1989}%
  \BibitemOpen
  \bibfield  {author} {\bibinfo {author} {\bibfnamefont {J.}~\bibnamefont {Zak}},\ }\bibfield  {title} {\enquote {\bibinfo {title} {Berry's phase for energy bands in solids},}\ }\href {\doibase 10.1103/PhysRevLett.62.2747} {\bibfield  {journal} {\bibinfo  {journal} {Phys. Rev. Lett.}\ }\textbf {\bibinfo {volume} {62}},\ \bibinfo {pages} {2747--2750} (\bibinfo {year} {1989})}\BibitemShut {NoStop}%
\bibitem [{\citenamefont {Atala}\ \emph {et~al.}(2013)\citenamefont {Atala}, \citenamefont {Aidelsburger}, \citenamefont {Barreiro}, \citenamefont {Abanin}, \citenamefont {Kitagawa}, \citenamefont {Demler},\ and\ \citenamefont {Bloch}}]{Atala2013}%
  \BibitemOpen
  \bibfield  {author} {\bibinfo {author} {\bibfnamefont {Marcos}\ \bibnamefont {Atala}}, \bibinfo {author} {\bibfnamefont {Monika}\ \bibnamefont {Aidelsburger}}, \bibinfo {author} {\bibfnamefont {Julio~T.}\ \bibnamefont {Barreiro}}, \bibinfo {author} {\bibfnamefont {Dmitry}\ \bibnamefont {Abanin}}, \bibinfo {author} {\bibfnamefont {Takuya}\ \bibnamefont {Kitagawa}}, \bibinfo {author} {\bibfnamefont {Eugene}\ \bibnamefont {Demler}}, \ and\ \bibinfo {author} {\bibfnamefont {Immanuel}\ \bibnamefont {Bloch}},\ }\bibfield  {title} {\enquote {\bibinfo {title} {Direct measurement of the zak phase in topological bloch bands},}\ }\href {\doibase https://doi.org/10.1038/nphys2790} {\bibfield  {journal} {\bibinfo  {journal} {Nature Physics}\ }\textbf {\bibinfo {volume} {9}},\ \bibinfo {pages} {795} (\bibinfo {year} {2013})}\BibitemShut {NoStop}%
\bibitem [{\citenamefont {Nakajima}\ \emph {et~al.}(2016)\citenamefont {Nakajima}, \citenamefont {Tomita}, \citenamefont {Taie}, \citenamefont {Ichinose}, \citenamefont {Ozawa}, \citenamefont {Wang}, \citenamefont {Troyer},\ and\ \citenamefont {Takahashi}}]{Takahashi2016pumping}%
  \BibitemOpen
  \bibfield  {author} {\bibinfo {author} {\bibfnamefont {Shuta}\ \bibnamefont {Nakajima}}, \bibinfo {author} {\bibfnamefont {Takafumi}\ \bibnamefont {Tomita}}, \bibinfo {author} {\bibfnamefont {Shintaro}\ \bibnamefont {Taie}}, \bibinfo {author} {\bibfnamefont {Tomohiro}\ \bibnamefont {Ichinose}}, \bibinfo {author} {\bibfnamefont {Hideki}\ \bibnamefont {Ozawa}}, \bibinfo {author} {\bibfnamefont {Lei}\ \bibnamefont {Wang}}, \bibinfo {author} {\bibfnamefont {Matthias}\ \bibnamefont {Troyer}}, \ and\ \bibinfo {author} {\bibfnamefont {Yoshiro}\ \bibnamefont {Takahashi}},\ }\bibfield  {title} {\enquote {\bibinfo {title} {Topological thouless pumping of ultracold fermions},}\ }\href {\doibase http://dx.doi.org/10.1038/nphys3622 10.1038/nphys3622} {\bibfield  {journal} {\bibinfo  {journal} {Nature Physics}\ }\textbf {\bibinfo {volume} {12}},\ \bibinfo {pages} {296} (\bibinfo {year} {2016})}\BibitemShut {NoStop}%
\bibitem [{\citenamefont {Lohse}\ \emph {et~al.}(2016)\citenamefont {Lohse}, \citenamefont {Schweizer}, \citenamefont {Zilberberg}, \citenamefont {Aidelsburger},\ and\ \citenamefont {Bloch}}]{Lohse2016}%
  \BibitemOpen
  \bibfield  {author} {\bibinfo {author} {\bibfnamefont {M.}~\bibnamefont {Lohse}}, \bibinfo {author} {\bibfnamefont {C.}~\bibnamefont {Schweizer}}, \bibinfo {author} {\bibfnamefont {O.}~\bibnamefont {Zilberberg}}, \bibinfo {author} {\bibfnamefont {M.}~\bibnamefont {Aidelsburger}}, \ and\ \bibinfo {author} {\bibfnamefont {I.}~\bibnamefont {Bloch}},\ }\bibfield  {title} {\enquote {\bibinfo {title} {A thouless quantum pump with ultracold bosonic atoms in an optical superlattice},}\ }\href {\doibase 10.1038/nphys3584} {\bibfield  {journal} {\bibinfo  {journal} {Nature Physics}\ }\textbf {\bibinfo {volume} {12}},\ \bibinfo {pages} {350--354} (\bibinfo {year} {2016})}\BibitemShut {NoStop}%
\bibitem [{\citenamefont {Mukherjee}\ \emph {et~al.}(2017)\citenamefont {Mukherjee}, \citenamefont {Spracklen}, \citenamefont {Valiente}, \citenamefont {Andersson}, \citenamefont {{\"O}hberg}, \citenamefont {Goldman},\ and\ \citenamefont {Thomson}}]{Mukherjee2017}%
  \BibitemOpen
  \bibfield  {author} {\bibinfo {author} {\bibfnamefont {Sebabrata}\ \bibnamefont {Mukherjee}}, \bibinfo {author} {\bibfnamefont {Alexander}\ \bibnamefont {Spracklen}}, \bibinfo {author} {\bibfnamefont {Manuel}\ \bibnamefont {Valiente}}, \bibinfo {author} {\bibfnamefont {Erika}\ \bibnamefont {Andersson}}, \bibinfo {author} {\bibfnamefont {Patrik}\ \bibnamefont {{\"O}hberg}}, \bibinfo {author} {\bibfnamefont {Nathan}\ \bibnamefont {Goldman}}, \ and\ \bibinfo {author} {\bibfnamefont {Robert~R.}\ \bibnamefont {Thomson}},\ }\bibfield  {title} {\enquote {\bibinfo {title} {Experimental observation of anomalous topological edge modes in a slowly driven photonic lattice},}\ }\href {\doibase 10.1038/ncomms13918} {\bibfield  {journal} {\bibinfo  {journal} {Nature Communications}\ }\textbf {\bibinfo {volume} {8}},\ \bibinfo {pages} {13918} (\bibinfo {year} {2017})}\BibitemShut {NoStop}%
\bibitem [{\citenamefont {Lu}\ \emph {et~al.}(2014)\citenamefont {Lu}, \citenamefont {Joannopoulos},\ and\ \citenamefont {Solja{\v{c}}i{\'{c}}}}]{Lu2014}%
  \BibitemOpen
  \bibfield  {author} {\bibinfo {author} {\bibfnamefont {Ling}\ \bibnamefont {Lu}}, \bibinfo {author} {\bibfnamefont {John~D.}\ \bibnamefont {Joannopoulos}}, \ and\ \bibinfo {author} {\bibfnamefont {Marin}\ \bibnamefont {Solja{\v{c}}i{\'{c}}}},\ }\bibfield  {title} {\enquote {\bibinfo {title} {Topological photonics},}\ }\href {\doibase 10.1038/nphoton.2014.248} {\bibfield  {journal} {\bibinfo  {journal} {Nature Photonics}\ }\textbf {\bibinfo {volume} {8}},\ \bibinfo {pages} {821--829} (\bibinfo {year} {2014})}\BibitemShut {NoStop}%
\bibitem [{\citenamefont {Meier}\ \emph {et~al.}(2016)\citenamefont {Meier}, \citenamefont {An},\ and\ \citenamefont {Gadway}}]{ssh_expt_2}%
  \BibitemOpen
  \bibfield  {author} {\bibinfo {author} {\bibfnamefont {Eric~J.}\ \bibnamefont {Meier}}, \bibinfo {author} {\bibfnamefont {Fangzhao~Alex}\ \bibnamefont {An}}, \ and\ \bibinfo {author} {\bibfnamefont {Bryce}\ \bibnamefont {Gadway}},\ }\bibfield  {title} {\enquote {\bibinfo {title} {Observation of the topological soliton state in the su--schrieffer--heeger model},}\ }\href {\doibase 10.1038/ncomms13986} {\bibfield  {journal} {\bibinfo  {journal} {Nature Communications}\ }\textbf {\bibinfo {volume} {7}},\ \bibinfo {pages} {13986} (\bibinfo {year} {2016})}\BibitemShut {NoStop}%
\bibitem [{\citenamefont {Kitagawa}\ \emph {et~al.}(2012)\citenamefont {Kitagawa}, \citenamefont {Broome}, \citenamefont {Fedrizzi}, \citenamefont {Rudner}, \citenamefont {Berg}, \citenamefont {Kassal}, \citenamefont {Aspuru-Guzik}, \citenamefont {Demler},\ and\ \citenamefont {White}}]{Kitagawa2012}%
  \BibitemOpen
  \bibfield  {author} {\bibinfo {author} {\bibfnamefont {Takuya}\ \bibnamefont {Kitagawa}}, \bibinfo {author} {\bibfnamefont {Matthew~A.}\ \bibnamefont {Broome}}, \bibinfo {author} {\bibfnamefont {Alessandro}\ \bibnamefont {Fedrizzi}}, \bibinfo {author} {\bibfnamefont {Mark~S.}\ \bibnamefont {Rudner}}, \bibinfo {author} {\bibfnamefont {Erez}\ \bibnamefont {Berg}}, \bibinfo {author} {\bibfnamefont {Ivan}\ \bibnamefont {Kassal}}, \bibinfo {author} {\bibfnamefont {Al{\'a}n}\ \bibnamefont {Aspuru-Guzik}}, \bibinfo {author} {\bibfnamefont {Eugene}\ \bibnamefont {Demler}}, \ and\ \bibinfo {author} {\bibfnamefont {Andrew~G.}\ \bibnamefont {White}},\ }\bibfield  {title} {\enquote {\bibinfo {title} {Observation of topologically protected bound states in photonic quantum walks},}\ }\href {\doibase 10.1038/ncomms1872} {\bibfield  {journal} {\bibinfo  {journal} {Nature Communications}\ }\textbf {\bibinfo {volume} {3}},\ \bibinfo {pages} {882} (\bibinfo {year} {2012})}\BibitemShut {NoStop}%
\bibitem [{\citenamefont {Leder}\ \emph {et~al.}(2016)\citenamefont {Leder}, \citenamefont {Grossert}, \citenamefont {Sitta}, \citenamefont {Genske}, \citenamefont {Rosch},\ and\ \citenamefont {Weitz}}]{Leder2016}%
  \BibitemOpen
  \bibfield  {author} {\bibinfo {author} {\bibfnamefont {Martin}\ \bibnamefont {Leder}}, \bibinfo {author} {\bibfnamefont {Christopher}\ \bibnamefont {Grossert}}, \bibinfo {author} {\bibfnamefont {Lukas}\ \bibnamefont {Sitta}}, \bibinfo {author} {\bibfnamefont {Maximilian}\ \bibnamefont {Genske}}, \bibinfo {author} {\bibfnamefont {Achim}\ \bibnamefont {Rosch}}, \ and\ \bibinfo {author} {\bibfnamefont {Martin}\ \bibnamefont {Weitz}},\ }\bibfield  {title} {\enquote {\bibinfo {title} {Real-space imaging of a topologically protected edge state with ultracold atoms in an amplitude-chirped optical lattice},}\ }\href {\doibase 10.1038/ncomms13112} {\bibfield  {journal} {\bibinfo  {journal} {Nature Communications}\ }\textbf {\bibinfo {volume} {7}},\ \bibinfo {pages} {13112} (\bibinfo {year} {2016})}\BibitemShut {NoStop}%
\bibitem [{\citenamefont {Jangjan}\ and\ \citenamefont {Hosseini}(2020)}]{floquet_ssh_ladder}%
  \BibitemOpen
  \bibfield  {author} {\bibinfo {author} {\bibfnamefont {Milad}\ \bibnamefont {Jangjan}}\ and\ \bibinfo {author} {\bibfnamefont {Mir~Vahid}\ \bibnamefont {Hosseini}},\ }\bibfield  {title} {\enquote {\bibinfo {title} {Floquet engineering of topological metal states and hybridization of edge states with bulk states in dimerized two-leg ladders},}\ }\href {\doibase 10.1038/s41598-020-71196-3} {\bibfield  {journal} {\bibinfo  {journal} {Scientific Reports}\ }\textbf {\bibinfo {volume} {10}},\ \bibinfo {pages} {14256} (\bibinfo {year} {2020})}\BibitemShut {NoStop}%
\bibitem [{\citenamefont {Rachel}(2018)}]{rachel_review}%
  \BibitemOpen
  \bibfield  {author} {\bibinfo {author} {\bibfnamefont {Stephan}\ \bibnamefont {Rachel}},\ }\bibfield  {title} {\enquote {\bibinfo {title} {Interacting topological insulators: a review},}\ }\href {\doibase 10.1088/1361-6633/aad6a6} {\bibfield  {journal} {\bibinfo  {journal} {Reports on Progress in Physics}\ }\textbf {\bibinfo {volume} {81}},\ \bibinfo {pages} {116501} (\bibinfo {year} {2018})}\BibitemShut {NoStop}%
\bibitem [{\citenamefont {Grusdt}\ \emph {et~al.}(2013)\citenamefont {Grusdt}, \citenamefont {H\"oning},\ and\ \citenamefont {Fleischhauer}}]{fleishauer_prl}%
  \BibitemOpen
  \bibfield  {author} {\bibinfo {author} {\bibfnamefont {Fabian}\ \bibnamefont {Grusdt}}, \bibinfo {author} {\bibfnamefont {Michael}\ \bibnamefont {H\"oning}}, \ and\ \bibinfo {author} {\bibfnamefont {Michael}\ \bibnamefont {Fleischhauer}},\ }\bibfield  {title} {\enquote {\bibinfo {title} {Topological edge states in the one-dimensional superlattice bose-hubbard model},}\ }\href {\doibase 10.1103/PhysRevLett.110.260405} {\bibfield  {journal} {\bibinfo  {journal} {Phys. Rev. Lett.}\ }\textbf {\bibinfo {volume} {110}},\ \bibinfo {pages} {260405} (\bibinfo {year} {2013})}\BibitemShut {NoStop}%
\bibitem [{\citenamefont {Greschner}\ \emph {et~al.}(2020)\citenamefont {Greschner}, \citenamefont {Mondal},\ and\ \citenamefont {Mishra}}]{mondal_pumping}%
  \BibitemOpen
  \bibfield  {author} {\bibinfo {author} {\bibfnamefont {Sebastian}\ \bibnamefont {Greschner}}, \bibinfo {author} {\bibfnamefont {Suman}\ \bibnamefont {Mondal}}, \ and\ \bibinfo {author} {\bibfnamefont {Tapan}\ \bibnamefont {Mishra}},\ }\bibfield  {title} {\enquote {\bibinfo {title} {Topological charge pumping of bound bosonic pairs},}\ }\href {\doibase 10.1103/PhysRevA.101.053630} {\bibfield  {journal} {\bibinfo  {journal} {Phys. Rev. A}\ }\textbf {\bibinfo {volume} {101}},\ \bibinfo {pages} {053630} (\bibinfo {year} {2020})}\BibitemShut {NoStop}%
\bibitem [{\citenamefont {Di~Liberto}\ \emph {et~al.}(2016)\citenamefont {Di~Liberto}, \citenamefont {Recati}, \citenamefont {Carusotto},\ and\ \citenamefont {Menotti}}]{DiLiberto2016}%
  \BibitemOpen
  \bibfield  {author} {\bibinfo {author} {\bibfnamefont {M.}~\bibnamefont {Di~Liberto}}, \bibinfo {author} {\bibfnamefont {A.}~\bibnamefont {Recati}}, \bibinfo {author} {\bibfnamefont {I.}~\bibnamefont {Carusotto}}, \ and\ \bibinfo {author} {\bibfnamefont {C.}~\bibnamefont {Menotti}},\ }\bibfield  {title} {\enquote {\bibinfo {title} {Two-body physics in the su-schrieffer-heeger model},}\ }\href {\doibase 10.1103/PhysRevA.94.062704} {\bibfield  {journal} {\bibinfo  {journal} {Phys. Rev. A}\ }\textbf {\bibinfo {volume} {94}},\ \bibinfo {pages} {062704} (\bibinfo {year} {2016})}\BibitemShut {NoStop}%
\bibitem [{\citenamefont {Di~Liberto}\ \emph {et~al.}(2017)\citenamefont {Di~Liberto}, \citenamefont {Recati}, \citenamefont {Carusotto},\ and\ \citenamefont {Menotti}}]{DiLiberto2017}%
  \BibitemOpen
  \bibfield  {author} {\bibinfo {author} {\bibfnamefont {M.}~\bibnamefont {Di~Liberto}}, \bibinfo {author} {\bibfnamefont {A.}~\bibnamefont {Recati}}, \bibinfo {author} {\bibfnamefont {I.}~\bibnamefont {Carusotto}}, \ and\ \bibinfo {author} {\bibfnamefont {C.}~\bibnamefont {Menotti}},\ }\bibfield  {title} {\enquote {\bibinfo {title} {Two-body bound and edge states in the extended ssh bose-hubbard model},}\ }\href {\doibase 10.1140/epjst/e2016-60388-y} {\bibfield  {journal} {\bibinfo  {journal} {The European Physical Journal Special Topics}\ }\textbf {\bibinfo {volume} {226}},\ \bibinfo {pages} {2751--2762} (\bibinfo {year} {2017})}\BibitemShut {NoStop}%
\bibitem [{\citenamefont {Zhou}\ \emph {et~al.}(2023)\citenamefont {Zhou}, \citenamefont {Pan},\ and\ \citenamefont {Jia}}]{sjia}%
  \BibitemOpen
  \bibfield  {author} {\bibinfo {author} {\bibfnamefont {Xiaofan}\ \bibnamefont {Zhou}}, \bibinfo {author} {\bibfnamefont {Jian-Song}\ \bibnamefont {Pan}}, \ and\ \bibinfo {author} {\bibfnamefont {Suotang}\ \bibnamefont {Jia}},\ }\bibfield  {title} {\enquote {\bibinfo {title} {Exploring interacting topological insulator in the extended su-schrieffer-heeger model},}\ }\href {\doibase 10.1103/PhysRevB.107.054105} {\bibfield  {journal} {\bibinfo  {journal} {Phys. Rev. B}\ }\textbf {\bibinfo {volume} {107}},\ \bibinfo {pages} {054105} (\bibinfo {year} {2023})}\BibitemShut {NoStop}%
\bibitem [{\citenamefont {Li}\ \emph {et~al.}(2013)\citenamefont {Li}, \citenamefont {Zhao},\ and\ \citenamefont {Vincent~Liu}}]{wu_vincet_liu}%
  \BibitemOpen
  \bibfield  {author} {\bibinfo {author} {\bibfnamefont {Xiaopeng}\ \bibnamefont {Li}}, \bibinfo {author} {\bibfnamefont {Erhai}\ \bibnamefont {Zhao}}, \ and\ \bibinfo {author} {\bibfnamefont {W.}~\bibnamefont {Vincent~Liu}},\ }\bibfield  {title} {\enquote {\bibinfo {title} {Topological states in a ladder-like optical lattice containing ultracold atoms in higher orbital bands},}\ }\href {\doibase 10.1038/ncomms2523} {\bibfield  {journal} {\bibinfo  {journal} {Nature Communications}\ }\textbf {\bibinfo {volume} {4}},\ \bibinfo {pages} {1523} (\bibinfo {year} {2013})}\BibitemShut {NoStop}%
\bibitem [{\citenamefont {Taddia}\ \emph {et~al.}(2017)\citenamefont {Taddia}, \citenamefont {Cornfeld}, \citenamefont {Rossini}, \citenamefont {Mazza}, \citenamefont {Sela},\ and\ \citenamefont {Fazio}}]{Taddia2017}%
  \BibitemOpen
  \bibfield  {author} {\bibinfo {author} {\bibfnamefont {Luca}\ \bibnamefont {Taddia}}, \bibinfo {author} {\bibfnamefont {Eyal}\ \bibnamefont {Cornfeld}}, \bibinfo {author} {\bibfnamefont {Davide}\ \bibnamefont {Rossini}}, \bibinfo {author} {\bibfnamefont {Leonardo}\ \bibnamefont {Mazza}}, \bibinfo {author} {\bibfnamefont {Eran}\ \bibnamefont {Sela}}, \ and\ \bibinfo {author} {\bibfnamefont {Rosario}\ \bibnamefont {Fazio}},\ }\bibfield  {title} {\enquote {\bibinfo {title} {Topological fractional pumping with alkaline-earth-like atoms in synthetic lattices},}\ }\href {\doibase 10.1103/PhysRevLett.118.230402} {\bibfield  {journal} {\bibinfo  {journal} {Phys. Rev. Lett.}\ }\textbf {\bibinfo {volume} {118}},\ \bibinfo {pages} {230402} (\bibinfo {year} {2017})}\BibitemShut {NoStop}%
\bibitem [{\citenamefont {Fraxanet}\ \emph {et~al.}(2022)\citenamefont {Fraxanet}, \citenamefont {Gonz\'alez-Cuadra}, \citenamefont {Pfau}, \citenamefont {Lewenstein}, \citenamefont {Langen},\ and\ \citenamefont {Barbiero}}]{fraxanet}%
  \BibitemOpen
  \bibfield  {author} {\bibinfo {author} {\bibfnamefont {Joana}\ \bibnamefont {Fraxanet}}, \bibinfo {author} {\bibfnamefont {Daniel}\ \bibnamefont {Gonz\'alez-Cuadra}}, \bibinfo {author} {\bibfnamefont {Tilman}\ \bibnamefont {Pfau}}, \bibinfo {author} {\bibfnamefont {Maciej}\ \bibnamefont {Lewenstein}}, \bibinfo {author} {\bibfnamefont {Tim}\ \bibnamefont {Langen}}, \ and\ \bibinfo {author} {\bibfnamefont {Luca}\ \bibnamefont {Barbiero}},\ }\bibfield  {title} {\enquote {\bibinfo {title} {Topological quantum critical points in the extended bose-hubbard model},}\ }\href {\doibase 10.1103/PhysRevLett.128.043402} {\bibfield  {journal} {\bibinfo  {journal} {Phys. Rev. Lett.}\ }\textbf {\bibinfo {volume} {128}},\ \bibinfo {pages} {043402} (\bibinfo {year} {2022})}\BibitemShut {NoStop}%
\bibitem [{\citenamefont {Juli\`a-Farr\'e}\ \emph {et~al.}(2022)\citenamefont {Juli\`a-Farr\'e}, \citenamefont {Gonz\'alez-Cuadra}, \citenamefont {Patscheider}, \citenamefont {Mark}, \citenamefont {Ferlaino}, \citenamefont {Lewenstein}, \citenamefont {Barbiero},\ and\ \citenamefont {Dauphin}}]{juliafarre}%
  \BibitemOpen
  \bibfield  {author} {\bibinfo {author} {\bibfnamefont {Sergi}\ \bibnamefont {Juli\`a-Farr\'e}}, \bibinfo {author} {\bibfnamefont {Daniel}\ \bibnamefont {Gonz\'alez-Cuadra}}, \bibinfo {author} {\bibfnamefont {Alexander}\ \bibnamefont {Patscheider}}, \bibinfo {author} {\bibfnamefont {Manfred~J.}\ \bibnamefont {Mark}}, \bibinfo {author} {\bibfnamefont {Francesca}\ \bibnamefont {Ferlaino}}, \bibinfo {author} {\bibfnamefont {Maciej}\ \bibnamefont {Lewenstein}}, \bibinfo {author} {\bibfnamefont {Luca}\ \bibnamefont {Barbiero}}, \ and\ \bibinfo {author} {\bibfnamefont {Alexandre}\ \bibnamefont {Dauphin}},\ }\bibfield  {title} {\enquote {\bibinfo {title} {Revealing the topological nature of the bond order wave in a strongly correlated quantum system},}\ }\href {\doibase 10.1103/PhysRevResearch.4.L032005} {\bibfield  {journal} {\bibinfo  {journal} {Phys. Rev. Res.}\ }\textbf {\bibinfo {volume} {4}},\ \bibinfo {pages} {L032005} (\bibinfo {year} {2022})}\BibitemShut {NoStop}%
\bibitem [{\citenamefont {Mondal}\ \emph {et~al.}(2021)\citenamefont {Mondal}, \citenamefont {Greschner}, \citenamefont {Santos},\ and\ \citenamefont {Mishra}}]{mondal_sshhubbard}%
  \BibitemOpen
  \bibfield  {author} {\bibinfo {author} {\bibfnamefont {Suman}\ \bibnamefont {Mondal}}, \bibinfo {author} {\bibfnamefont {Sebastian}\ \bibnamefont {Greschner}}, \bibinfo {author} {\bibfnamefont {Luis}\ \bibnamefont {Santos}}, \ and\ \bibinfo {author} {\bibfnamefont {Tapan}\ \bibnamefont {Mishra}},\ }\bibfield  {title} {\enquote {\bibinfo {title} {Topological inheritance in two-component hubbard models with single-component su-schrieffer-heeger dimerization},}\ }\href {\doibase 10.1103/PhysRevA.104.013315} {\bibfield  {journal} {\bibinfo  {journal} {Phys. Rev. A}\ }\textbf {\bibinfo {volume} {104}},\ \bibinfo {pages} {013315} (\bibinfo {year} {2021})}\BibitemShut {NoStop}%
\bibitem [{\citenamefont {Nersesyan}(2020)}]{Nersesyan}%
  \BibitemOpen
  \bibfield  {author} {\bibinfo {author} {\bibfnamefont {A.~A.}\ \bibnamefont {Nersesyan}},\ }\bibfield  {title} {\enquote {\bibinfo {title} {Phase diagram of an interacting staggered su-schrieffer-heeger two-chain ladder close to a quantum critical point},}\ }\href {\doibase 10.1103/PhysRevB.102.045108} {\bibfield  {journal} {\bibinfo  {journal} {Phys. Rev. B}\ }\textbf {\bibinfo {volume} {102}},\ \bibinfo {pages} {045108} (\bibinfo {year} {2020})}\BibitemShut {NoStop}%
\bibitem [{\citenamefont {Zhang}\ and\ \citenamefont {Zhou}(2017)}]{ssh_ladder_glide_symmetry}%
  \BibitemOpen
  \bibfield  {author} {\bibinfo {author} {\bibfnamefont {Shao-Liang}\ \bibnamefont {Zhang}}\ and\ \bibinfo {author} {\bibfnamefont {Qi}~\bibnamefont {Zhou}},\ }\bibfield  {title} {\enquote {\bibinfo {title} {Two-leg su-schrieffer-heeger chain with glide reflection symmetry},}\ }\href {\doibase 10.1103/PhysRevA.95.061601} {\bibfield  {journal} {\bibinfo  {journal} {Phys. Rev. A}\ }\textbf {\bibinfo {volume} {95}},\ \bibinfo {pages} {061601} (\bibinfo {year} {2017})}\BibitemShut {NoStop}%
\bibitem [{\citenamefont {Padhan}\ \emph {et~al.}(2024)\citenamefont {Padhan}, \citenamefont {Mondal}, \citenamefont {Vishveshwara},\ and\ \citenamefont {Mishra}}]{padhan}%
  \BibitemOpen
  \bibfield  {author} {\bibinfo {author} {\bibfnamefont {Ashirbad}\ \bibnamefont {Padhan}}, \bibinfo {author} {\bibfnamefont {Suman}\ \bibnamefont {Mondal}}, \bibinfo {author} {\bibfnamefont {Smitha}\ \bibnamefont {Vishveshwara}}, \ and\ \bibinfo {author} {\bibfnamefont {Tapan}\ \bibnamefont {Mishra}},\ }\bibfield  {title} {\enquote {\bibinfo {title} {Interacting bosons on a su-schrieffer-heeger ladder: Topological phases and thouless pumping},}\ }\href {\doibase 10.1103/PhysRevB.109.085120} {\bibfield  {journal} {\bibinfo  {journal} {Phys. Rev. B}\ }\textbf {\bibinfo {volume} {109}},\ \bibinfo {pages} {085120} (\bibinfo {year} {2024})}\BibitemShut {NoStop}%
\bibitem [{\citenamefont {Mondal}\ \emph {et~al.}(2023)\citenamefont {Mondal}, \citenamefont {Agarwala}, \citenamefont {Mishra},\ and\ \citenamefont {Prakash}}]{suman_adhip}%
  \BibitemOpen
  \bibfield  {author} {\bibinfo {author} {\bibfnamefont {Suman}\ \bibnamefont {Mondal}}, \bibinfo {author} {\bibfnamefont {Adhip}\ \bibnamefont {Agarwala}}, \bibinfo {author} {\bibfnamefont {Tapan}\ \bibnamefont {Mishra}}, \ and\ \bibinfo {author} {\bibfnamefont {Abhishodh}\ \bibnamefont {Prakash}},\ }\bibfield  {title} {\enquote {\bibinfo {title} {Symmetry-enriched criticality in a coupled spin ladder},}\ }\href {\doibase 10.1103/PhysRevB.108.245135} {\bibfield  {journal} {\bibinfo  {journal} {Phys. Rev. B}\ }\textbf {\bibinfo {volume} {108}},\ \bibinfo {pages} {245135} (\bibinfo {year} {2023})}\BibitemShut {NoStop}%
\bibitem [{\citenamefont {Feng}\ \emph {et~al.}(2022)\citenamefont {Feng}, \citenamefont {Xing}, \citenamefont {Poletti}, \citenamefont {Scalettar},\ and\ \citenamefont {Batrouni}}]{batrouni}%
  \BibitemOpen
  \bibfield  {author} {\bibinfo {author} {\bibfnamefont {Chunhan}\ \bibnamefont {Feng}}, \bibinfo {author} {\bibfnamefont {Bo}~\bibnamefont {Xing}}, \bibinfo {author} {\bibfnamefont {Dario}\ \bibnamefont {Poletti}}, \bibinfo {author} {\bibfnamefont {Richard}\ \bibnamefont {Scalettar}}, \ and\ \bibinfo {author} {\bibfnamefont {George}\ \bibnamefont {Batrouni}},\ }\bibfield  {title} {\enquote {\bibinfo {title} {Phase diagram of the su-schrieffer-heeger-hubbard model on a square lattice},}\ }\href {\doibase 10.1103/PhysRevB.106.L081114} {\bibfield  {journal} {\bibinfo  {journal} {Phys. Rev. B}\ }\textbf {\bibinfo {volume} {106}},\ \bibinfo {pages} {L081114} (\bibinfo {year} {2022})}\BibitemShut {NoStop}%
\bibitem [{\citenamefont {de~Léséleuc}\ \emph {et~al.}(2019)\citenamefont {de~Léséleuc}, \citenamefont {Lienhard}, \citenamefont {Scholl}, \citenamefont {Barredo}, \citenamefont {Weber}, \citenamefont {Lang}, \citenamefont {Büchler}, \citenamefont {Lahaye},\ and\ \citenamefont {Browaeys}}]{browyes}%
  \BibitemOpen
  \bibfield  {author} {\bibinfo {author} {\bibfnamefont {Sylvain}\ \bibnamefont {de~Léséleuc}}, \bibinfo {author} {\bibfnamefont {Vincent}\ \bibnamefont {Lienhard}}, \bibinfo {author} {\bibfnamefont {Pascal}\ \bibnamefont {Scholl}}, \bibinfo {author} {\bibfnamefont {Daniel}\ \bibnamefont {Barredo}}, \bibinfo {author} {\bibfnamefont {Sebastian}\ \bibnamefont {Weber}}, \bibinfo {author} {\bibfnamefont {Nicolai}\ \bibnamefont {Lang}}, \bibinfo {author} {\bibfnamefont {Hans~Peter}\ \bibnamefont {Büchler}}, \bibinfo {author} {\bibfnamefont {Thierry}\ \bibnamefont {Lahaye}}, \ and\ \bibinfo {author} {\bibfnamefont {Antoine}\ \bibnamefont {Browaeys}},\ }\bibfield  {title} {\enquote {\bibinfo {title} {Observation of a symmetry-protected topological phase of interacting bosons with rydberg atoms},}\ }\href {\doibase 10.1126/science.aav9105} {\bibfield  {journal} {\bibinfo  {journal} {Science}\ }\textbf {\bibinfo {volume} {365}},\ \bibinfo {pages} {775--780} (\bibinfo {year} {2019})}\BibitemShut {NoStop}%
\bibitem [{\citenamefont {Sompet}\ \emph {et~al.}(2022)\citenamefont {Sompet}, \citenamefont {Hirthe}, \citenamefont {Bourgund}, \citenamefont {Chalopin}, \citenamefont {Bibo}, \citenamefont {Koepsell}, \citenamefont {Bojović}, \citenamefont {Verresen}, \citenamefont {Pollmann}, \citenamefont {Salomon}, \citenamefont {Gross}, \citenamefont {Hilker},\ and\ \citenamefont {Bloch}}]{bloch_haldane}%
  \BibitemOpen
  \bibfield  {author} {\bibinfo {author} {\bibfnamefont {Pimonpan}\ \bibnamefont {Sompet}}, \bibinfo {author} {\bibfnamefont {Sarah}\ \bibnamefont {Hirthe}}, \bibinfo {author} {\bibfnamefont {Dominik}\ \bibnamefont {Bourgund}}, \bibinfo {author} {\bibfnamefont {Thomas}\ \bibnamefont {Chalopin}}, \bibinfo {author} {\bibfnamefont {Julian}\ \bibnamefont {Bibo}}, \bibinfo {author} {\bibfnamefont {Joannis}\ \bibnamefont {Koepsell}}, \bibinfo {author} {\bibfnamefont {Petar}\ \bibnamefont {Bojović}}, \bibinfo {author} {\bibfnamefont {Ruben}\ \bibnamefont {Verresen}}, \bibinfo {author} {\bibfnamefont {Frank}\ \bibnamefont {Pollmann}}, \bibinfo {author} {\bibfnamefont {Guillaume}\ \bibnamefont {Salomon}}, \bibinfo {author} {\bibfnamefont {Christian}\ \bibnamefont {Gross}}, \bibinfo {author} {\bibfnamefont {Timon~A.}\ \bibnamefont {Hilker}}, \ and\ \bibinfo {author} {\bibfnamefont {Immanuel}\ \bibnamefont {Bloch}},\ }\bibfield  {title} {\enquote {\bibinfo {title} {Realizing the symmetry-protected haldane phase in
  fermi–hubbard ladders},}\ }\href {\doibase 10.1038/s41586-022-04688-z} {\bibfield  {journal} {\bibinfo  {journal} {Nature}\ }\textbf {\bibinfo {volume} {606}},\ \bibinfo {pages} {484–488} (\bibinfo {year} {2022})}\BibitemShut {NoStop}%
\bibitem [{\citenamefont {Thatcher}\ \emph {et~al.}(2022)\citenamefont {Thatcher}, \citenamefont {Fairfield}, \citenamefont {Merlo-Ramírez},\ and\ \citenamefont {Merlo}}]{ssh_expt_1}%
  \BibitemOpen
  \bibfield  {author} {\bibinfo {author} {\bibfnamefont {Luke}\ \bibnamefont {Thatcher}}, \bibinfo {author} {\bibfnamefont {Parker}\ \bibnamefont {Fairfield}}, \bibinfo {author} {\bibfnamefont {Lázaro}\ \bibnamefont {Merlo-Ramírez}}, \ and\ \bibinfo {author} {\bibfnamefont {Juan~M}\ \bibnamefont {Merlo}},\ }\bibfield  {title} {\enquote {\bibinfo {title} {Experimental observation of topological phase transitions in a mechanical 1d-ssh model},}\ }\href {\doibase 10.1088/1402-4896/ac4ed2} {\bibfield  {journal} {\bibinfo  {journal} {Physica Scripta}\ }\textbf {\bibinfo {volume} {97}},\ \bibinfo {pages} {035702} (\bibinfo {year} {2022})}\BibitemShut {NoStop}%
\bibitem [{\citenamefont {Xie}\ \emph {et~al.}(2019)\citenamefont {Xie}, \citenamefont {Gou}, \citenamefont {Xiao}, \citenamefont {Gadway},\ and\ \citenamefont {Yan}}]{ssh4_expt}%
  \BibitemOpen
  \bibfield  {author} {\bibinfo {author} {\bibfnamefont {Dizhou}\ \bibnamefont {Xie}}, \bibinfo {author} {\bibfnamefont {Wei}\ \bibnamefont {Gou}}, \bibinfo {author} {\bibfnamefont {Teng}\ \bibnamefont {Xiao}}, \bibinfo {author} {\bibfnamefont {Bryce}\ \bibnamefont {Gadway}}, \ and\ \bibinfo {author} {\bibfnamefont {Bo}~\bibnamefont {Yan}},\ }\bibfield  {title} {\enquote {\bibinfo {title} {Topological characterizations of an extended su--schrieffer--heeger model},}\ }\href {\doibase 10.1038/s41534-019-0159-6} {\bibfield  {journal} {\bibinfo  {journal} {npj Quantum Information}\ }\textbf {\bibinfo {volume} {5}},\ \bibinfo {pages} {55} (\bibinfo {year} {2019})}\BibitemShut {NoStop}%
\bibitem [{\citenamefont {Li}\ \emph {et~al.}(2023)\citenamefont {Li}, \citenamefont {Wang}, \citenamefont {Zhao}, \citenamefont {Du}, \citenamefont {Zhang}, \citenamefont {Hu}, \citenamefont {Mei}, \citenamefont {Xiao}, \citenamefont {Ma},\ and\ \citenamefont {Jia}}]{suotang_jia_expt}%
  \BibitemOpen
  \bibfield  {author} {\bibinfo {author} {\bibfnamefont {Yuqing}\ \bibnamefont {Li}}, \bibinfo {author} {\bibfnamefont {Yunfei}\ \bibnamefont {Wang}}, \bibinfo {author} {\bibfnamefont {Hongxing}\ \bibnamefont {Zhao}}, \bibinfo {author} {\bibfnamefont {Huiying}\ \bibnamefont {Du}}, \bibinfo {author} {\bibfnamefont {Jiahui}\ \bibnamefont {Zhang}}, \bibinfo {author} {\bibfnamefont {Ying}\ \bibnamefont {Hu}}, \bibinfo {author} {\bibfnamefont {Feng}\ \bibnamefont {Mei}}, \bibinfo {author} {\bibfnamefont {Liantuan}\ \bibnamefont {Xiao}}, \bibinfo {author} {\bibfnamefont {Jie}\ \bibnamefont {Ma}}, \ and\ \bibinfo {author} {\bibfnamefont {Suotang}\ \bibnamefont {Jia}},\ }\bibfield  {title} {\enquote {\bibinfo {title} {Interaction-induced breakdown of chiral dynamics in the su-schrieffer-heeger model},}\ }\href {\doibase 10.1103/PhysRevResearch.5.L032035} {\bibfield  {journal} {\bibinfo  {journal} {Phys. Rev. Res.}\ }\textbf {\bibinfo {volume} {5}},\ \bibinfo {pages} {L032035} (\bibinfo {year} {2023})}\BibitemShut
  {NoStop}%
\bibitem [{\citenamefont {Mondal}\ \emph {et~al.}(2022{\natexlab{a}})\citenamefont {Mondal}, \citenamefont {Padhan},\ and\ \citenamefont {Mishra}}]{Mondal_topology}%
  \BibitemOpen
  \bibfield  {author} {\bibinfo {author} {\bibfnamefont {Suman}\ \bibnamefont {Mondal}}, \bibinfo {author} {\bibfnamefont {Ashirbad}\ \bibnamefont {Padhan}}, \ and\ \bibinfo {author} {\bibfnamefont {Tapan}\ \bibnamefont {Mishra}},\ }\bibfield  {title} {\enquote {\bibinfo {title} {Realizing a symmetry protected topological phase through dimerized interactions},}\ }\href {\doibase 10.1103/PhysRevB.106.L201106} {\bibfield  {journal} {\bibinfo  {journal} {Phys. Rev. B}\ }\textbf {\bibinfo {volume} {106}},\ \bibinfo {pages} {L201106} (\bibinfo {year} {2022}{\natexlab{a}})}\BibitemShut {NoStop}%
\bibitem [{\citenamefont {Het\'enyi}(2020)}]{Hetenyi_tvvp}%
  \BibitemOpen
  \bibfield  {author} {\bibinfo {author} {\bibfnamefont {Bal\'azs}\ \bibnamefont {Het\'enyi}},\ }\bibfield  {title} {\enquote {\bibinfo {title} {Interaction-driven polarization shift in the $t\text{\ensuremath{-}}v\text{\ensuremath{-}}{V}^{\ensuremath{'}}$ lattice fermion model at half filling: Emergent haldane phase},}\ }\href {\doibase 10.1103/PhysRevResearch.2.023277} {\bibfield  {journal} {\bibinfo  {journal} {Phys. Rev. Res.}\ }\textbf {\bibinfo {volume} {2}},\ \bibinfo {pages} {023277} (\bibinfo {year} {2020})}\BibitemShut {NoStop}%
\bibitem [{\citenamefont {Cr\'epin}\ \emph {et~al.}(2011)\citenamefont {Cr\'epin}, \citenamefont {Laflorencie}, \citenamefont {Roux},\ and\ \citenamefont {Simon}}]{rmi_laflorencie}%
  \BibitemOpen
  \bibfield  {author} {\bibinfo {author} {\bibfnamefont {Fran\ifmmode \mbox{\c{c}}\else~\c{c}\fi{}ois}\ \bibnamefont {Cr\'epin}}, \bibinfo {author} {\bibfnamefont {Nicolas}\ \bibnamefont {Laflorencie}}, \bibinfo {author} {\bibfnamefont {Guillaume}\ \bibnamefont {Roux}}, \ and\ \bibinfo {author} {\bibfnamefont {Pascal}\ \bibnamefont {Simon}},\ }\bibfield  {title} {\enquote {\bibinfo {title} {Phase diagram of hard-core bosons on clean and disordered two-leg ladders: Mott insulator--luttinger liquid--bose glass},}\ }\href {\doibase 10.1103/PhysRevB.84.054517} {\bibfield  {journal} {\bibinfo  {journal} {Phys. Rev. B}\ }\textbf {\bibinfo {volume} {84}},\ \bibinfo {pages} {054517} (\bibinfo {year} {2011})}\BibitemShut {NoStop}%
\bibitem [{\citenamefont {Citro}\ and\ \citenamefont {Aidelsburger}(2023)}]{monika_review}%
  \BibitemOpen
  \bibfield  {author} {\bibinfo {author} {\bibfnamefont {Roberta}\ \bibnamefont {Citro}}\ and\ \bibinfo {author} {\bibfnamefont {Monika}\ \bibnamefont {Aidelsburger}},\ }\bibfield  {title} {\enquote {\bibinfo {title} {Thouless pumping and topology},}\ }\href {\doibase 10.1038/s42254-022-00545-0} {\bibfield  {journal} {\bibinfo  {journal} {Nature Reviews Physics}\ }\textbf {\bibinfo {volume} {5}},\ \bibinfo {pages} {87--101} (\bibinfo {year} {2023})}\BibitemShut {NoStop}%
\bibitem [{\citenamefont {Thouless}(1983)}]{Thouless1983}%
  \BibitemOpen
  \bibfield  {author} {\bibinfo {author} {\bibfnamefont {D.~J.}\ \bibnamefont {Thouless}},\ }\bibfield  {title} {\enquote {\bibinfo {title} {Quantization of particle transport},}\ }\href {\doibase 10.1103/PhysRevB.27.6083} {\bibfield  {journal} {\bibinfo  {journal} {Phys. Rev. B}\ }\textbf {\bibinfo {volume} {27}},\ \bibinfo {pages} {6083--6087} (\bibinfo {year} {1983})}\BibitemShut {NoStop}%
\bibitem [{\citenamefont {Holstein}\ and\ \citenamefont {Primakoff}(1940)}]{hp_transformation}%
  \BibitemOpen
  \bibfield  {author} {\bibinfo {author} {\bibfnamefont {T.}~\bibnamefont {Holstein}}\ and\ \bibinfo {author} {\bibfnamefont {H.}~\bibnamefont {Primakoff}},\ }\bibfield  {title} {\enquote {\bibinfo {title} {Field dependence of the intrinsic domain magnetization of a ferromagnet},}\ }\href {\doibase 10.1103/PhysRev.58.1098} {\bibfield  {journal} {\bibinfo  {journal} {Phys. Rev.}\ }\textbf {\bibinfo {volume} {58}},\ \bibinfo {pages} {1098--1113} (\bibinfo {year} {1940})}\BibitemShut {NoStop}%
\bibitem [{\citenamefont {Nigam}\ \emph {et~al.}(2024)\citenamefont {Nigam}, \citenamefont {Padhan}, \citenamefont {Sen}, \citenamefont {Mishra},\ and\ \citenamefont {Bhattacharjee}}]{harsh_xxz}%
  \BibitemOpen
  \bibfield  {author} {\bibinfo {author} {\bibfnamefont {Harsh}\ \bibnamefont {Nigam}}, \bibinfo {author} {\bibfnamefont {Ashirbad}\ \bibnamefont {Padhan}}, \bibinfo {author} {\bibfnamefont {Diptiman}\ \bibnamefont {Sen}}, \bibinfo {author} {\bibfnamefont {Tapan}\ \bibnamefont {Mishra}}, \ and\ \bibinfo {author} {\bibfnamefont {Subhro}\ \bibnamefont {Bhattacharjee}},\ }\href@noop {} {\enquote {\bibinfo {title} {Phases and phase transitions in a dimerized spin-$\mathbf{\frac{1}{2}}$ xxz chain},}\ } (\bibinfo {year} {2024}),\ \Eprint {http://arxiv.org/abs/2408.14474} {arXiv:2408.14474 [cond-mat.str-el]} \BibitemShut {NoStop}%
\bibitem [{\citenamefont {White}(1992)}]{white1992}%
  \BibitemOpen
  \bibfield  {author} {\bibinfo {author} {\bibfnamefont {Steven~R.}\ \bibnamefont {White}},\ }\bibfield  {title} {\enquote {\bibinfo {title} {Density matrix formulation for quantum renormalization groups},}\ }\href {\doibase 10.1103/PhysRevLett.69.2863} {\bibfield  {journal} {\bibinfo  {journal} {Phys. Rev. Lett.}\ }\textbf {\bibinfo {volume} {69}},\ \bibinfo {pages} {2863--2866} (\bibinfo {year} {1992})}\BibitemShut {NoStop}%
\bibitem [{\citenamefont {Schollw\"ock}(2005)}]{schollowck_dmrg_rev}%
  \BibitemOpen
  \bibfield  {author} {\bibinfo {author} {\bibfnamefont {U.}~\bibnamefont {Schollw\"ock}},\ }\bibfield  {title} {\enquote {\bibinfo {title} {The density-matrix renormalization group},}\ }\href {\doibase 10.1103/RevModPhys.77.259} {\bibfield  {journal} {\bibinfo  {journal} {Rev. Mod. Phys.}\ }\textbf {\bibinfo {volume} {77}},\ \bibinfo {pages} {259--315} (\bibinfo {year} {2005})}\BibitemShut {NoStop}%
\bibitem [{\citenamefont {Cirac}\ \emph {et~al.}(2021)\citenamefont {Cirac}, \citenamefont {P\'erez-Garc\'{\i}a}, \citenamefont {Schuch},\ and\ \citenamefont {Verstraete}}]{Verstraete_rev}%
  \BibitemOpen
  \bibfield  {author} {\bibinfo {author} {\bibfnamefont {J.~Ignacio}\ \bibnamefont {Cirac}}, \bibinfo {author} {\bibfnamefont {David}\ \bibnamefont {P\'erez-Garc\'{\i}a}}, \bibinfo {author} {\bibfnamefont {Norbert}\ \bibnamefont {Schuch}}, \ and\ \bibinfo {author} {\bibfnamefont {Frank}\ \bibnamefont {Verstraete}},\ }\bibfield  {title} {\enquote {\bibinfo {title} {Matrix product states and projected entangled pair states: Concepts, symmetries, theorems},}\ }\href {\doibase 10.1103/RevModPhys.93.045003} {\bibfield  {journal} {\bibinfo  {journal} {Rev. Mod. Phys.}\ }\textbf {\bibinfo {volume} {93}},\ \bibinfo {pages} {045003} (\bibinfo {year} {2021})}\BibitemShut {NoStop}%
\bibitem [{\citenamefont {Schollwöck}(2011)}]{schollowck_mps}%
  \BibitemOpen
  \bibfield  {author} {\bibinfo {author} {\bibfnamefont {Ulrich}\ \bibnamefont {Schollwöck}},\ }\bibfield  {title} {\enquote {\bibinfo {title} {The density-matrix renormalization group in the age of matrix product states},}\ }\href {\doibase https://doi.org/10.1016/j.aop.2010.09.012} {\bibfield  {journal} {\bibinfo  {journal} {Annals of Physics}\ }\textbf {\bibinfo {volume} {326}},\ \bibinfo {pages} {96--192} (\bibinfo {year} {2011})},\ \bibinfo {note} {january 2011 Special Issue}\BibitemShut {NoStop}%
\bibitem [{\citenamefont {den Nijs}\ and\ \citenamefont {Rommelse}(1989)}]{den_nijs}%
  \BibitemOpen
  \bibfield  {author} {\bibinfo {author} {\bibfnamefont {Marcel}\ \bibnamefont {den Nijs}}\ and\ \bibinfo {author} {\bibfnamefont {Koos}\ \bibnamefont {Rommelse}},\ }\bibfield  {title} {\enquote {\bibinfo {title} {Preroughening transitions in crystal surfaces and valence-bond phases in quantum spin chains},}\ }\href {\doibase 10.1103/PhysRevB.40.4709} {\bibfield  {journal} {\bibinfo  {journal} {Phys. Rev. B}\ }\textbf {\bibinfo {volume} {40}},\ \bibinfo {pages} {4709--4734} (\bibinfo {year} {1989})}\BibitemShut {NoStop}%
\bibitem [{\citenamefont {Tasaki}(1991)}]{Tasaki}%
  \BibitemOpen
  \bibfield  {author} {\bibinfo {author} {\bibfnamefont {Hal}\ \bibnamefont {Tasaki}},\ }\bibfield  {title} {\enquote {\bibinfo {title} {Quantum liquid in antiferromagnetic chains: A stochastic geometric approach to the haldane gap},}\ }\href {\doibase 10.1103/PhysRevLett.66.798} {\bibfield  {journal} {\bibinfo  {journal} {Phys. Rev. Lett.}\ }\textbf {\bibinfo {volume} {66}},\ \bibinfo {pages} {798--801} (\bibinfo {year} {1991})}\BibitemShut {NoStop}%
\bibitem [{\citenamefont {Hida}(1992)}]{Hida}%
  \BibitemOpen
  \bibfield  {author} {\bibinfo {author} {\bibfnamefont {Kazuo}\ \bibnamefont {Hida}},\ }\bibfield  {title} {\enquote {\bibinfo {title} {Crossover between the haldane-gap phase and the dimer phase in the spin-1/2 alternating heisenberg chain},}\ }\href {\doibase 10.1103/PhysRevB.45.2207} {\bibfield  {journal} {\bibinfo  {journal} {Phys. Rev. B}\ }\textbf {\bibinfo {volume} {45}},\ \bibinfo {pages} {2207--2212} (\bibinfo {year} {1992})}\BibitemShut {NoStop}%
\bibitem [{\citenamefont {Gibson}\ \emph {et~al.}(2011)\citenamefont {Gibson}, \citenamefont {Meyer},\ and\ \citenamefont {Chitov}}]{string_ladder}%
  \BibitemOpen
  \bibfield  {author} {\bibinfo {author} {\bibfnamefont {Sandra~J.}\ \bibnamefont {Gibson}}, \bibinfo {author} {\bibfnamefont {R.}~\bibnamefont {Meyer}}, \ and\ \bibinfo {author} {\bibfnamefont {Gennady~Y.}\ \bibnamefont {Chitov}},\ }\bibfield  {title} {\enquote {\bibinfo {title} {Numerical study of critical properties and hidden orders in dimerized spin ladders},}\ }\href {\doibase 10.1103/PhysRevB.83.104423} {\bibfield  {journal} {\bibinfo  {journal} {Phys. Rev. B}\ }\textbf {\bibinfo {volume} {83}},\ \bibinfo {pages} {104423} (\bibinfo {year} {2011})}\BibitemShut {NoStop}%
\bibitem [{\citenamefont {Li}\ and\ \citenamefont {Haldane}(2008)}]{haldane}%
  \BibitemOpen
  \bibfield  {author} {\bibinfo {author} {\bibfnamefont {Hui}\ \bibnamefont {Li}}\ and\ \bibinfo {author} {\bibfnamefont {F.~D.~M.}\ \bibnamefont {Haldane}},\ }\bibfield  {title} {\enquote {\bibinfo {title} {Entanglement spectrum as a generalization of entanglement entropy: Identification of topological order in non-abelian fractional quantum hall effect states},}\ }\href {\doibase 10.1103/PhysRevLett.101.010504} {\bibfield  {journal} {\bibinfo  {journal} {Phys. Rev. Lett.}\ }\textbf {\bibinfo {volume} {101}},\ \bibinfo {pages} {010504} (\bibinfo {year} {2008})}\BibitemShut {NoStop}%
\bibitem [{\citenamefont {Yoshida}\ \emph {et~al.}(2014)\citenamefont {Yoshida}, \citenamefont {Peters}, \citenamefont {Fujimoto},\ and\ \citenamefont {Kawakami}}]{kawakami}%
  \BibitemOpen
  \bibfield  {author} {\bibinfo {author} {\bibfnamefont {Tsuneya}\ \bibnamefont {Yoshida}}, \bibinfo {author} {\bibfnamefont {Robert}\ \bibnamefont {Peters}}, \bibinfo {author} {\bibfnamefont {Satoshi}\ \bibnamefont {Fujimoto}}, \ and\ \bibinfo {author} {\bibfnamefont {Norio}\ \bibnamefont {Kawakami}},\ }\bibfield  {title} {\enquote {\bibinfo {title} {Characterization of a topological mott insulator in one dimension},}\ }\href {\doibase 10.1103/PhysRevLett.112.196404} {\bibfield  {journal} {\bibinfo  {journal} {Phys. Rev. Lett.}\ }\textbf {\bibinfo {volume} {112}},\ \bibinfo {pages} {196404} (\bibinfo {year} {2014})}\BibitemShut {NoStop}%
\bibitem [{\citenamefont {Turner}\ \emph {et~al.}(2011)\citenamefont {Turner}, \citenamefont {Pollmann},\ and\ \citenamefont {Berg}}]{pollman_entropy}%
  \BibitemOpen
  \bibfield  {author} {\bibinfo {author} {\bibfnamefont {Ari~M.}\ \bibnamefont {Turner}}, \bibinfo {author} {\bibfnamefont {Frank}\ \bibnamefont {Pollmann}}, \ and\ \bibinfo {author} {\bibfnamefont {Erez}\ \bibnamefont {Berg}},\ }\bibfield  {title} {\enquote {\bibinfo {title} {Topological phases of one-dimensional fermions: An entanglement point of view},}\ }\href {\doibase 10.1103/PhysRevB.83.075102} {\bibfield  {journal} {\bibinfo  {journal} {Phys. Rev. B}\ }\textbf {\bibinfo {volume} {83}},\ \bibinfo {pages} {075102} (\bibinfo {year} {2011})}\BibitemShut {NoStop}%
\bibitem [{\citenamefont {Hayward}\ \emph {et~al.}(2018)\citenamefont {Hayward}, \citenamefont {Schweizer}, \citenamefont {Lohse}, \citenamefont {Aidelsburger},\ and\ \citenamefont {Heidrich-Meisner}}]{Hayward2018}%
  \BibitemOpen
  \bibfield  {author} {\bibinfo {author} {\bibfnamefont {A.}~\bibnamefont {Hayward}}, \bibinfo {author} {\bibfnamefont {C.}~\bibnamefont {Schweizer}}, \bibinfo {author} {\bibfnamefont {M.}~\bibnamefont {Lohse}}, \bibinfo {author} {\bibfnamefont {M.}~\bibnamefont {Aidelsburger}}, \ and\ \bibinfo {author} {\bibfnamefont {F.}~\bibnamefont {Heidrich-Meisner}},\ }\bibfield  {title} {\enquote {\bibinfo {title} {Topological charge pumping in the interacting bosonic rice-mele model},}\ }\href {\doibase 10.1103/PhysRevB.98.245148} {\bibfield  {journal} {\bibinfo  {journal} {Phys. Rev. B}\ }\textbf {\bibinfo {volume} {98}},\ \bibinfo {pages} {245148} (\bibinfo {year} {2018})}\BibitemShut {NoStop}%
\bibitem [{\citenamefont {Rice}\ and\ \citenamefont {Mele}(1982)}]{rice_mele}%
  \BibitemOpen
  \bibfield  {author} {\bibinfo {author} {\bibfnamefont {M.~J.}\ \bibnamefont {Rice}}\ and\ \bibinfo {author} {\bibfnamefont {E.~J.}\ \bibnamefont {Mele}},\ }\bibfield  {title} {\enquote {\bibinfo {title} {Elementary excitations of a linearly conjugated diatomic polymer},}\ }\href {\doibase 10.1103/PhysRevLett.49.1455} {\bibfield  {journal} {\bibinfo  {journal} {Phys. Rev. Lett.}\ }\textbf {\bibinfo {volume} {49}},\ \bibinfo {pages} {1455--1459} (\bibinfo {year} {1982})}\BibitemShut {NoStop}%
\bibitem [{\citenamefont {Asb{\'o}th}\ \emph {et~al.}(2016{\natexlab{b}})\citenamefont {Asb{\'o}th}, \citenamefont {Oroszl{\'a}ny},\ and\ \citenamefont {P{\'a}lyi}}]{Asboth2016_rm}%
  \BibitemOpen
  \bibfield  {author} {\bibinfo {author} {\bibfnamefont {J{\'a}nos~K.}\ \bibnamefont {Asb{\'o}th}}, \bibinfo {author} {\bibfnamefont {L{\'a}szl{\'o}}\ \bibnamefont {Oroszl{\'a}ny}}, \ and\ \bibinfo {author} {\bibfnamefont {Andr{\'a}s}\ \bibnamefont {P{\'a}lyi}},\ }\enquote {\bibinfo {title} {Adiabatic charge pumping, rice-mele model},}\ in\ \href {\doibase 10.1007/978-3-319-25607-8_4} {\emph {\bibinfo {booktitle} {A Short Course on Topological Insulators: Band Structure and Edge States in One and Two Dimensions}}}\ (\bibinfo  {publisher} {Springer International Publishing},\ \bibinfo {address} {Cham},\ \bibinfo {year} {2016})\ pp.\ \bibinfo {pages} {55--68}\BibitemShut {NoStop}%
\bibitem [{\citenamefont {Lin}\ \emph {et~al.}(2020)\citenamefont {Lin}, \citenamefont {Ke},\ and\ \citenamefont {Lee}}]{bound_pump1}%
  \BibitemOpen
  \bibfield  {author} {\bibinfo {author} {\bibfnamefont {Ling}\ \bibnamefont {Lin}}, \bibinfo {author} {\bibfnamefont {Yongguan}\ \bibnamefont {Ke}}, \ and\ \bibinfo {author} {\bibfnamefont {Chaohong}\ \bibnamefont {Lee}},\ }\bibfield  {title} {\enquote {\bibinfo {title} {Interaction-induced topological bound states and thouless pumping in a one-dimensional optical lattice},}\ }\href {\doibase 10.1103/PhysRevA.101.023620} {\bibfield  {journal} {\bibinfo  {journal} {Phys. Rev. A}\ }\textbf {\bibinfo {volume} {101}},\ \bibinfo {pages} {023620} (\bibinfo {year} {2020})}\BibitemShut {NoStop}%
\bibitem [{\citenamefont {Ke}\ \emph {et~al.}(2017)\citenamefont {Ke}, \citenamefont {Qin}, \citenamefont {Kivshar},\ and\ \citenamefont {Lee}}]{bound_pump3}%
  \BibitemOpen
  \bibfield  {author} {\bibinfo {author} {\bibfnamefont {Yongguan}\ \bibnamefont {Ke}}, \bibinfo {author} {\bibfnamefont {Xizhou}\ \bibnamefont {Qin}}, \bibinfo {author} {\bibfnamefont {Yuri~S.}\ \bibnamefont {Kivshar}}, \ and\ \bibinfo {author} {\bibfnamefont {Chaohong}\ \bibnamefont {Lee}},\ }\bibfield  {title} {\enquote {\bibinfo {title} {Multiparticle wannier states and thouless pumping of interacting bosons},}\ }\href {\doibase 10.1103/PhysRevA.95.063630} {\bibfield  {journal} {\bibinfo  {journal} {Phys. Rev. A}\ }\textbf {\bibinfo {volume} {95}},\ \bibinfo {pages} {063630} (\bibinfo {year} {2017})}\BibitemShut {NoStop}%
\bibitem [{\citenamefont {Arg{\"{u}}ello-Luengo}\ \emph {et~al.}(2024)\citenamefont {Arg{\"{u}}ello-Luengo}, \citenamefont {Mark}, \citenamefont {Ferlaino}, \citenamefont {Lewenstein}, \citenamefont {Barbiero},\ and\ \citenamefont {Juli{\`{a}}-Farr{\'{e}}}}]{ArguelloLuengo2024stabilizationof}%
  \BibitemOpen
  \bibfield  {author} {\bibinfo {author} {\bibfnamefont {Javier}\ \bibnamefont {Arg{\"{u}}ello-Luengo}}, \bibinfo {author} {\bibfnamefont {Manfred~J.}\ \bibnamefont {Mark}}, \bibinfo {author} {\bibfnamefont {Francesca}\ \bibnamefont {Ferlaino}}, \bibinfo {author} {\bibfnamefont {Maciej}\ \bibnamefont {Lewenstein}}, \bibinfo {author} {\bibfnamefont {Luca}\ \bibnamefont {Barbiero}}, \ and\ \bibinfo {author} {\bibfnamefont {Sergi}\ \bibnamefont {Juli{\`{a}}-Farr{\'{e}}}},\ }\bibfield  {title} {\enquote {\bibinfo {title} {Stabilization of {H}ubbard-{T}houless pumps through nonlocal fermionic repulsion},}\ }\href {\doibase 10.22331/q-2024-03-14-1285} {\bibfield  {journal} {\bibinfo  {journal} {{Quantum}}\ }\textbf {\bibinfo {volume} {8}},\ \bibinfo {pages} {1285} (\bibinfo {year} {2024})}\BibitemShut {NoStop}%
\bibitem [{\citenamefont {Kuno}\ \emph {et~al.}(2017)\citenamefont {Kuno}, \citenamefont {Shimizu},\ and\ \citenamefont {Ichinose}}]{Kuno2017}%
  \BibitemOpen
  \bibfield  {author} {\bibinfo {author} {\bibfnamefont {Yoshihito}\ \bibnamefont {Kuno}}, \bibinfo {author} {\bibfnamefont {Keita}\ \bibnamefont {Shimizu}}, \ and\ \bibinfo {author} {\bibfnamefont {Ikuo}\ \bibnamefont {Ichinose}},\ }\bibfield  {title} {\enquote {\bibinfo {title} {Various topological mott insulators and topological bulk charge pumping in strongly-interacting boson system in one-dimensional superlattice},}\ }\href@noop {} {\bibfield  {journal} {\bibinfo  {journal} {New Journal of Physics}\ }\textbf {\bibinfo {volume} {19}},\ \bibinfo {pages} {123025} (\bibinfo {year} {2017})}\BibitemShut {NoStop}%
\bibitem [{\citenamefont {Bertok}\ \emph {et~al.}(2022)\citenamefont {Bertok}, \citenamefont {Heidrich-Meisner},\ and\ \citenamefont {Aligia}}]{bertok_pump}%
  \BibitemOpen
  \bibfield  {author} {\bibinfo {author} {\bibfnamefont {E.}~\bibnamefont {Bertok}}, \bibinfo {author} {\bibfnamefont {F.}~\bibnamefont {Heidrich-Meisner}}, \ and\ \bibinfo {author} {\bibfnamefont {A.~A.}\ \bibnamefont {Aligia}},\ }\bibfield  {title} {\enquote {\bibinfo {title} {Splitting of topological charge pumping in an interacting two-component fermionic rice-mele hubbard model},}\ }\href {\doibase 10.1103/PhysRevB.106.045141} {\bibfield  {journal} {\bibinfo  {journal} {Phys. Rev. B}\ }\textbf {\bibinfo {volume} {106}},\ \bibinfo {pages} {045141} (\bibinfo {year} {2022})}\BibitemShut {NoStop}%
\bibitem [{\citenamefont {Mondal}\ \emph {et~al.}(2022{\natexlab{b}})\citenamefont {Mondal}, \citenamefont {Bertok},\ and\ \citenamefont {Heidrich-Meisner}}]{mondal_phonon}%
  \BibitemOpen
  \bibfield  {author} {\bibinfo {author} {\bibfnamefont {Suman}\ \bibnamefont {Mondal}}, \bibinfo {author} {\bibfnamefont {Eric}\ \bibnamefont {Bertok}}, \ and\ \bibinfo {author} {\bibfnamefont {Fabian}\ \bibnamefont {Heidrich-Meisner}},\ }\bibfield  {title} {\enquote {\bibinfo {title} {Phonon-induced breakdown of thouless pumping in the rice-mele-holstein model},}\ }\href {\doibase 10.1103/PhysRevB.106.235118} {\bibfield  {journal} {\bibinfo  {journal} {Phys. Rev. B}\ }\textbf {\bibinfo {volume} {106}},\ \bibinfo {pages} {235118} (\bibinfo {year} {2022}{\natexlab{b}})}\BibitemShut {NoStop}%
\bibitem [{\citenamefont {Schweizer}\ \emph {et~al.}(2016)\citenamefont {Schweizer}, \citenamefont {Lohse}, \citenamefont {Citro},\ and\ \citenamefont {Bloch}}]{spin_pumping}%
  \BibitemOpen
  \bibfield  {author} {\bibinfo {author} {\bibfnamefont {C.}~\bibnamefont {Schweizer}}, \bibinfo {author} {\bibfnamefont {M.}~\bibnamefont {Lohse}}, \bibinfo {author} {\bibfnamefont {R.}~\bibnamefont {Citro}}, \ and\ \bibinfo {author} {\bibfnamefont {I.}~\bibnamefont {Bloch}},\ }\bibfield  {title} {\enquote {\bibinfo {title} {Spin pumping and measurement of spin currents in optical superlattices},}\ }\href {\doibase 10.1103/PhysRevLett.117.170405} {\bibfield  {journal} {\bibinfo  {journal} {Phys. Rev. Lett.}\ }\textbf {\bibinfo {volume} {117}},\ \bibinfo {pages} {170405} (\bibinfo {year} {2016})}\BibitemShut {NoStop}%
\bibitem [{\citenamefont {Kraus}\ \emph {et~al.}(2012)\citenamefont {Kraus}, \citenamefont {Lahini}, \citenamefont {Ringel}, \citenamefont {Verbin},\ and\ \citenamefont {Zilberberg}}]{pumping_quasicrystals}%
  \BibitemOpen
  \bibfield  {author} {\bibinfo {author} {\bibfnamefont {Yaacov~E.}\ \bibnamefont {Kraus}}, \bibinfo {author} {\bibfnamefont {Yoav}\ \bibnamefont {Lahini}}, \bibinfo {author} {\bibfnamefont {Zohar}\ \bibnamefont {Ringel}}, \bibinfo {author} {\bibfnamefont {Mor}\ \bibnamefont {Verbin}}, \ and\ \bibinfo {author} {\bibfnamefont {Oded}\ \bibnamefont {Zilberberg}},\ }\bibfield  {title} {\enquote {\bibinfo {title} {Topological states and adiabatic pumping in quasicrystals},}\ }\href {\doibase 10.1103/PhysRevLett.109.106402} {\bibfield  {journal} {\bibinfo  {journal} {Phys. Rev. Lett.}\ }\textbf {\bibinfo {volume} {109}},\ \bibinfo {pages} {106402} (\bibinfo {year} {2012})}\BibitemShut {NoStop}%
\bibitem [{\citenamefont {Wang}\ \emph {et~al.}(2013)\citenamefont {Wang}, \citenamefont {Troyer},\ and\ \citenamefont {Dai}}]{pumping_1d}%
  \BibitemOpen
  \bibfield  {author} {\bibinfo {author} {\bibfnamefont {Lei}\ \bibnamefont {Wang}}, \bibinfo {author} {\bibfnamefont {Matthias}\ \bibnamefont {Troyer}}, \ and\ \bibinfo {author} {\bibfnamefont {Xi}~\bibnamefont {Dai}},\ }\bibfield  {title} {\enquote {\bibinfo {title} {Topological charge pumping in a one-dimensional optical lattice},}\ }\href {\doibase 10.1103/PhysRevLett.111.026802} {\bibfield  {journal} {\bibinfo  {journal} {Phys. Rev. Lett.}\ }\textbf {\bibinfo {volume} {111}},\ \bibinfo {pages} {026802} (\bibinfo {year} {2013})}\BibitemShut {NoStop}%
\bibitem [{\citenamefont {Walter}\ \emph {et~al.}(2023)\citenamefont {Walter}, \citenamefont {Zhu}, \citenamefont {Gächter}, \citenamefont {Minguzzi}, \citenamefont {Roschinski}, \citenamefont {Sandholzer}, \citenamefont {Viebahn},\ and\ \citenamefont {Esslinger}}]{hubbarad_thouless_pump}%
  \BibitemOpen
  \bibfield  {author} {\bibinfo {author} {\bibfnamefont {Anne-Sophie}\ \bibnamefont {Walter}}, \bibinfo {author} {\bibfnamefont {Zijie}\ \bibnamefont {Zhu}}, \bibinfo {author} {\bibfnamefont {Marius}\ \bibnamefont {Gächter}}, \bibinfo {author} {\bibfnamefont {Joaquín}\ \bibnamefont {Minguzzi}}, \bibinfo {author} {\bibfnamefont {Stephan}\ \bibnamefont {Roschinski}}, \bibinfo {author} {\bibfnamefont {Kilian}\ \bibnamefont {Sandholzer}}, \bibinfo {author} {\bibfnamefont {Konrad}\ \bibnamefont {Viebahn}}, \ and\ \bibinfo {author} {\bibfnamefont {Tilman}\ \bibnamefont {Esslinger}},\ }\bibfield  {title} {\enquote {\bibinfo {title} {Quantization and its breakdown in a hubbard–thouless pump},}\ }\href {\doibase 10.1038/s41567-023-02145-w} {\bibfield  {journal} {\bibinfo  {journal} {Nature Physics}\ }\textbf {\bibinfo {volume} {19}},\ \bibinfo {pages} {1471–1475} (\bibinfo {year} {2023})}\BibitemShut {NoStop}%
\bibitem [{\citenamefont {Ke}\ \emph {et~al.}(2020)\citenamefont {Ke}, \citenamefont {Zhong}, \citenamefont {Poshakinskiy}, \citenamefont {Kivshar}, \citenamefont {Poddubny},\ and\ \citenamefont {Lee}}]{qubit_pumping}%
  \BibitemOpen
  \bibfield  {author} {\bibinfo {author} {\bibfnamefont {Yongguan}\ \bibnamefont {Ke}}, \bibinfo {author} {\bibfnamefont {Janet}\ \bibnamefont {Zhong}}, \bibinfo {author} {\bibfnamefont {Alexander~V.}\ \bibnamefont {Poshakinskiy}}, \bibinfo {author} {\bibfnamefont {Yuri~S.}\ \bibnamefont {Kivshar}}, \bibinfo {author} {\bibfnamefont {Alexander~N.}\ \bibnamefont {Poddubny}}, \ and\ \bibinfo {author} {\bibfnamefont {Chaohong}\ \bibnamefont {Lee}},\ }\bibfield  {title} {\enquote {\bibinfo {title} {Radiative topological biphoton states in modulated qubit arrays},}\ }\href {\doibase 10.1103/PhysRevResearch.2.033190} {\bibfield  {journal} {\bibinfo  {journal} {Phys. Rev. Res.}\ }\textbf {\bibinfo {volume} {2}},\ \bibinfo {pages} {033190} (\bibinfo {year} {2020})}\BibitemShut {NoStop}%
\bibitem [{\citenamefont {Ogino}\ \emph {et~al.}(2021)\citenamefont {Ogino}, \citenamefont {Furukawa}, \citenamefont {Kaneko}, \citenamefont {Morita},\ and\ \citenamefont {Kawashima}}]{four_spin_furukawa}%
  \BibitemOpen
  \bibfield  {author} {\bibinfo {author} {\bibfnamefont {Takuhiro}\ \bibnamefont {Ogino}}, \bibinfo {author} {\bibfnamefont {Shunsuke}\ \bibnamefont {Furukawa}}, \bibinfo {author} {\bibfnamefont {Ryui}\ \bibnamefont {Kaneko}}, \bibinfo {author} {\bibfnamefont {Satoshi}\ \bibnamefont {Morita}}, \ and\ \bibinfo {author} {\bibfnamefont {Naoki}\ \bibnamefont {Kawashima}},\ }\bibfield  {title} {\enquote {\bibinfo {title} {Symmetry-protected topological phases and competing orders in a spin-$\frac{1}{2}$ xxz ladder with a four-spin interaction},}\ }\href {\doibase 10.1103/PhysRevB.104.075135} {\bibfield  {journal} {\bibinfo  {journal} {Phys. Rev. B}\ }\textbf {\bibinfo {volume} {104}},\ \bibinfo {pages} {075135} (\bibinfo {year} {2021})}\BibitemShut {NoStop}%
\bibitem [{\citenamefont {Ogino}\ \emph {et~al.}(2022)\citenamefont {Ogino}, \citenamefont {Kaneko}, \citenamefont {Morita},\ and\ \citenamefont {Furukawa}}]{frustrated_xxz_furukawa}%
  \BibitemOpen
  \bibfield  {author} {\bibinfo {author} {\bibfnamefont {Takuhiro}\ \bibnamefont {Ogino}}, \bibinfo {author} {\bibfnamefont {Ryui}\ \bibnamefont {Kaneko}}, \bibinfo {author} {\bibfnamefont {Satoshi}\ \bibnamefont {Morita}}, \ and\ \bibinfo {author} {\bibfnamefont {Shunsuke}\ \bibnamefont {Furukawa}},\ }\bibfield  {title} {\enquote {\bibinfo {title} {Ground-state phase diagram of a spin-$\frac{1}{2}$ frustrated xxz ladder},}\ }\href {\doibase 10.1103/PhysRevB.106.155106} {\bibfield  {journal} {\bibinfo  {journal} {Phys. Rev. B}\ }\textbf {\bibinfo {volume} {106}},\ \bibinfo {pages} {155106} (\bibinfo {year} {2022})}\BibitemShut {NoStop}%
\bibitem [{\citenamefont {Liu}\ \emph {et~al.}(2012)\citenamefont {Liu}, \citenamefont {Yang}, \citenamefont {Han}, \citenamefont {Yi},\ and\ \citenamefont {Wen}}]{spt_spin_ladder}%
  \BibitemOpen
  \bibfield  {author} {\bibinfo {author} {\bibfnamefont {Zheng-Xin}\ \bibnamefont {Liu}}, \bibinfo {author} {\bibfnamefont {Zhen-Biao}\ \bibnamefont {Yang}}, \bibinfo {author} {\bibfnamefont {Yong-Jian}\ \bibnamefont {Han}}, \bibinfo {author} {\bibfnamefont {Wei}\ \bibnamefont {Yi}}, \ and\ \bibinfo {author} {\bibfnamefont {Xiao-Gang}\ \bibnamefont {Wen}},\ }\bibfield  {title} {\enquote {\bibinfo {title} {Symmetry-protected topological phases in spin ladders with two-body interactions},}\ }\href {\doibase 10.1103/PhysRevB.86.195122} {\bibfield  {journal} {\bibinfo  {journal} {Phys. Rev. B}\ }\textbf {\bibinfo {volume} {86}},\ \bibinfo {pages} {195122} (\bibinfo {year} {2012})}\BibitemShut {NoStop}%
\bibitem [{\citenamefont {Nogrette}\ \emph {et~al.}(2014)\citenamefont {Nogrette}, \citenamefont {Labuhn}, \citenamefont {Ravets}, \citenamefont {Barredo}, \citenamefont {B\'eguin}, \citenamefont {Vernier}, \citenamefont {Lahaye},\ and\ \citenamefont {Browaeys}}]{broywes_geometry}%
  \BibitemOpen
  \bibfield  {author} {\bibinfo {author} {\bibfnamefont {F.}~\bibnamefont {Nogrette}}, \bibinfo {author} {\bibfnamefont {H.}~\bibnamefont {Labuhn}}, \bibinfo {author} {\bibfnamefont {S.}~\bibnamefont {Ravets}}, \bibinfo {author} {\bibfnamefont {D.}~\bibnamefont {Barredo}}, \bibinfo {author} {\bibfnamefont {L.}~\bibnamefont {B\'eguin}}, \bibinfo {author} {\bibfnamefont {A.}~\bibnamefont {Vernier}}, \bibinfo {author} {\bibfnamefont {T.}~\bibnamefont {Lahaye}}, \ and\ \bibinfo {author} {\bibfnamefont {A.}~\bibnamefont {Browaeys}},\ }\bibfield  {title} {\enquote {\bibinfo {title} {Single-atom trapping in holographic 2d arrays of microtraps with arbitrary geometries},}\ }\href {\doibase 10.1103/PhysRevX.4.021034} {\bibfield  {journal} {\bibinfo  {journal} {Phys. Rev. X}\ }\textbf {\bibinfo {volume} {4}},\ \bibinfo {pages} {021034} (\bibinfo {year} {2014})}\BibitemShut {NoStop}%
\bibitem [{\citenamefont {Labuhn}\ \emph {et~al.}(2016)\citenamefont {Labuhn}, \citenamefont {Barredo}, \citenamefont {Ravets}, \citenamefont {de~L{\'e}s{\'e}leuc}, \citenamefont {Macr{\`i}}, \citenamefont {Lahaye},\ and\ \citenamefont {Browaeys}}]{browaeys_ising}%
  \BibitemOpen
  \bibfield  {author} {\bibinfo {author} {\bibfnamefont {Henning}\ \bibnamefont {Labuhn}}, \bibinfo {author} {\bibfnamefont {Daniel}\ \bibnamefont {Barredo}}, \bibinfo {author} {\bibfnamefont {Sylvain}\ \bibnamefont {Ravets}}, \bibinfo {author} {\bibfnamefont {Sylvain}\ \bibnamefont {de~L{\'e}s{\'e}leuc}}, \bibinfo {author} {\bibfnamefont {Tommaso}\ \bibnamefont {Macr{\`i}}}, \bibinfo {author} {\bibfnamefont {Thierry}\ \bibnamefont {Lahaye}}, \ and\ \bibinfo {author} {\bibfnamefont {Antoine}\ \bibnamefont {Browaeys}},\ }\bibfield  {title} {\enquote {\bibinfo {title} {Tunable two-dimensional arrays of single rydberg atoms for realizing quantum ising models},}\ }\href {\doibase 10.1038/nature18274} {\bibfield  {journal} {\bibinfo  {journal} {Nature}\ }\textbf {\bibinfo {volume} {534}},\ \bibinfo {pages} {667--670} (\bibinfo {year} {2016})}\BibitemShut {NoStop}%
\end{thebibliography}%
\end{document}